\documentstyle[epsfig]{mn}

\newcommand{\iras} {{\it IRAS \/}}

\newcommand{\virgo}{Virgo\ }

\newcommand{\be}{\begin{equation}}
\newcommand{\ee}{\end{equation}}

\newcommand{\bea}{\begin{eqnarray}}
\newcommand{\eea}{\end{eqnarray}}

\newcommand{\bc}{\begin{center}}
\newcommand{\ec}{\end{center}}

\newcommand{\ol}[1]{ {\overline{#1}}}

\newcommand{\lu}{\,h^{-1}{\rm Mpc}}

\renewcommand{\vec}[1]{ {\bmath #1} } 

\newcommand{\dd}{{\rm d}}
\newcommand{\e}{{\rm e}}

\newcommand{\bld}[1] {\mathbf{#1}}

\title{Genus Statistics of the \virgo N-body simulations
and the 1.2-Jy Redshift Survey}

\author[V. Springel et al.]
{Volker Springel$^1$, 
Simon D. M. White$^1$, 
J\"{o}rg M. Colberg$^1$,\cr
Hugh M. P. Couchman$^2$,
George P. Efstathiou$^3$,
Carlos S. Frenk$^4$,\cr
Adrian R. Jenkins$^4$,
Frazer R. Pearce$^4$,
Alistair H. Nelson$^5$,\cr
John A. Peacock$^6$,
and Peter A. Thomas$^7$ (The Virgo Consortium)\\
$^1$Max-Planck-Institut f\"{u}r Astrophysik,
Karl-Schwarzschild-Stra\ss{}e 1, 85740 Garching bei M\"{u}nchen, Germany\\
$^2$Dept. of Physics \& Astronomy, Univ. of Western Ontario, London,
Ontario, N6A 3K7, Canada\\
$^3$Dept. of Astrophysics, Nuclear \& Astrophysics Laboratory, Keble
Road, Oxford, OX1 3RH, UK\\
$^4$Dept. of Physics, South Road, Durham, DH1 3LE, UK\\
$^5$Dept. of Physics, College of Cardiff, P.O.Box 913, Cardiff, CF4
3TH, UK\\
$^6$Royal Observatory Edinburgh, Blackford Hill, Edinburgh, EH9 3HJ,
UK\\
$^7$Astronomy Centre, University of Sussex, Falmer, Brighton, BN1, 9QH,
UK
}

\begin{document}
\maketitle

\begin{abstract}

We study the topology of the \virgo 
N-body simulations and compare it
to the 1.2-Jy redshift survey of \iras galaxies
by means of the genus statistic. 
Four high-resolution simulations of variants of the CDM cosmology are considered: a
flat standard model (SCDM), a variant of it with 
more large-scale
power ($\,\tau$CDM), and two low density universes, one
open (OCDM, $\Omega_0=0.3$) and one flat 
($\Lambda$CDM, $\Omega_0=0.3$, 
$\Lambda=0.7$).
In all cases, the initial fluctuation
amplitudes are chosen so that the simulations approximately reproduce
the observed abundance of rich clusters of galaxies at the present day.
The fully
sampled N-body simulations are examined down to strongly nonlinear 
scales, both with spatially fixed smoothing, and with an adaptive
smoothing technique. 
While the $\tau$CDM, $\Lambda$CDM, and OCDM simulations have
very similar genus statistics in the regime accessible to fixed
smoothing, they can be separated with adaptive smoothing at small mass
scales.
In order to compare the N-body models with the 1.2-Jy survey, we extract
large ensembles of mock catalogues from the
simulations. These mock surveys are used to test for various
systematic effects in the genus analysis and to establish the
distribution of errors of the genus curve.
We find that a simple multivariate analysis of the genus measurements
is compromised both by non-Gaussian distributed errors and by noise
that dominates the covariance matrix. We therefore introduce a
principal components analysis of the genus curve. 
With a likelihood ratio test we find that the 1.2-Jy data favours the
$\Lambda$CDM, $\tau$CDM and OCDM models compared to SCDM. When
genus measurements for different smoothing scales are combined, the
SCDM model can be excluded at a 99 per cent confidence level, while the
other three models fit the 1.2-Jy data well.
These results are unlikely to be significantly modified if galaxies
are biased tracers of the mass, provided that biasing preserves a
monotonic relation between galaxy density and mass density.

\end{abstract}


\section{Introduction}

The observed 
large-scale structure of the Universe represents one of the most
important constraints for theories of cosmic structure formation. In
the past, the clustering of galaxies was mainly studied
with statistics 
like  
the two-point correlation function, the power spectrum, 
or the counts-in-cells analysis. These statistical measures 
have been routinely applied to more and more powerful redshift
surveys, leading to
significant advances in our understanding of cosmic
history. Cosmological N-body simulations have played a 
vital part in this development, and they are already strongly
constrained by the available data.
 
However,
the two-point correlation function and the power spectrum
describe the properties of the galaxy distribution only 
to a limited extent. A
full description would involve a hierarchy of three-, four- and higher
correlation functions or, alternatively, information on the phase 
correlations among the
different Fourier modes of the density field. Furthermore,
the low-order statistics are also quite 
insensitive to the geometrical aspects
of the clustering which the human eye can so easily detect in
pictures 
of the matter distribution. For example, it is still 
unclear whether the galaxy distribution is best described as
filamentary, cellular or sheet-like.

In order to develop a measure of the geometrical and morphological
aspects of the galaxy distribution a 
number of measures have been proposed, among
them percolation \cite{Kl88}, level-crossing statistics \cite{Ry88},
genus statistics \cite{Go86}, minimal spanning
tree \cite{Pe95}, shape statistic \cite{Lu95}, 
and Minkowski functionals
\cite{Me94}.

In this paper we focus on the genus statistic
which was first proposed by Gott, Melott \& Dickinson \shortcite{Go86} 
and has become a widely
accepted statistical tool in cosmology since then. 
The genus 
probes the
topology of isodensity surfaces of a smoothed mass density field.
It is therefore sensitive to
global aspects of
the density field that manifest themselves in 
higher order correlations, which reflect 
the connectedness and morphology of the structure in the Universe. 
Such features 
are not revealed by standard measures like the
power spectrum or the two-point correlation function.
For
example, galaxy distributions that are wall-like, bubble-like or
filamentary would all lead to different signatures in the genus
statistic. Thus the topology has potentially strong discriminative power 
and might be used to rule out or support certain models for structure
formation.

A particularly
interesting application of the genus statistic is
a test of the random phase hypothesis for the initial
density fluctuation field.
Because the topology is invariant during the linear growth of structure,
the topology of the present galaxy distribution can be
related to the topology of the initial density field, which in
turn allows a
test of the random phase
hypothesis.
Any departure of the measured topology from the random
phase 
prediction would be
evidence for the presence of phase correlations that might reflect
non-Gaussian initial conditions. 

In contrast to the linear regime, the topology 
on small, strongly non-linear 
scales has hardly been studied to date. In particular, there are 
no theoretical predictions for the genus
statistics in this
regime.
Hence it
is presently unclear whether the genus on small scales
is a useful statistic to discriminate
between different models of structure formation.

In this work we examine the topology of a suite of large, 
high-resolution N-body simulations carried out by the \virgo 
collaboration \cite{Je97}.
The suite of models consists of a standard cold dark
matter model (SCDM), and three variants of the cold dark matter
cosmology, which have more power on large scales.
These are a flat $\Omega_0=1$ model ($\tau$CDM), and two low density
universes with $\Omega_0=0.3$, one open model (OCDM), and one closed
by means of a cosmological constant ($\Lambda$CDM).

In the first of a series of papers,
Jenkins et~al. \shortcite{Je97} measured the correlation function, the
power spectrum and various statistics of the velocity field of these
simulations. 
As Jenkins et~al. \shortcite{Je97} 
demonstrate, all three models with the power spectrum shape
$\Gamma=0.21$ 
can fit the APM
two-point galaxy correlation function reasonably well, if one allows
for a moderate scale-dependent bias. 
However, the differences {\it between}
these models are rather small, at least when 
only the distribution at $z=0$ is
examined with these statistics.

In the second paper of the series, Thomas et~al. \shortcite{Th97} showed that
the internal structure of halos of rich clusters is also similar in
all the models and would be difficult to distinguish in practice.

Here we use the genus statistic to analyse 
these N-body simulations down
to strongly non-linear scales. In this regime, differences
between the models can be detected.
 
In a second thread we compare the 
genus of the N-body models to the 1.2-Jy redshift
survey of \iras galaxies. Our particular aim is to
improve the statistical methodology of such a comparison. 
For this purpose we work with ensembles of mock
catalogues to assess the properties of new smoothing
techniques and to derive accurate estimates for errors and systematic
effects. This Monte-Carlo technique also allows the derivation of formal exclusion
levels for the N-body models.
New larger redshift surveys may be used subsequently to further
tighten these constraints.

This paper is organized as follows. 
In section 2 we briefly review the genus statistic which we apply in
section 3 to the fully sampled \virgo N-body simulations,
both with fixed and adaptive smoothing.
Section 4 describes the 1.2-Jy redshift survey and the construction of
ensembles of mock catalogues. In section 5 we introduce different
methods to compute smoothed density fields from the redshift survey
data, and in section 6 we discuss various systematic effects that
affect the genus statistic.
We then turn in section 7 to the statistical methodology we adopt for the
comparison with the 1.2-Jy redshift survey, 
and we present the results of this comparison
in section 8.
Finally we summarize and conclude in section 9.

\section{Genus statistic}
	
We first review briefly the genus statistic as introduced 
by Gott et~al. \shortcite{Go86}. 
A number of redshift
surveys have been analysed with it, starting with
a compilation of relatively small
samples examined by
Gott et~al. \shortcite{Go89}.
Further surveys that have been studied include 
the SSRS \cite{Pa92}, QDOT \cite{Mo92}, Abell Clusters \cite{Rh94},
CfA \cite{Vo94}, and most recently PSCz \cite{Ca97}.
There have also been 
a number of theoretical studies
of the genus in the mildly non-linear regime \cite{Ma94,Ma96a,Ma96b,Ma96c},
and the genus has been applied in two dimensions to the microwave
background \cite{Co96}
and to slice
surveys \cite{Pa92a,Co97}.

Given an isodensity contour of a smoothed mass density field 
we define the genus as
\be
G=-\frac{1}{4\pi}\int \kappa\; \dd A ,
\ee
where 
\be
\kappa=\frac{1}{r_1 r_2}
\ee 
is the local Gaussian curvature. Here $r_1$ and $r_2$ denote the principal
radii of curvature,
and the integration extends over the
whole surface. 
The Gauss-Bonnet theorem shows that this definition of
genus makes $G$ equal to the 
number of topological holes (like the one in a doughnut) 
minus the number of isolated regions of the surface.

Assuming an ergodic universe, we define a 
genus 
\be
g=\frac{G}{V}
\ee
per unit volume, where $V$ is finite, but large enough to be a
representative patch of the
universe. The genus depends on the density threshold 
used to construct the isodensity surface. As a function of threshold we
therefore obtain a {\it genus curve}, which is the central object of
this investigation.

We parameterize the genus curve
by the fraction $f$ of the volume above the density 
threshold value or by the quantity 
\be
\nu=\sqrt{2}\; {\rm erf}^{-1}(1-2f)
\label{C11}
\ee
derived from it. Here ${\rm erf}^{-1}$ denotes the inverse of the error
function ${\rm erf}(x)=\frac{2}{\sqrt{\pi}}\int_0^x \e^{-t^2} \dd t$. 
We will stick to the usual convention and present 
genus curves in the form $g=g(\nu)$.  

The definition (\ref{C11}) is chosen such that 
for a Gaussian random field 
the quantity  $\nu=\delta_{\rm t}/\sigma$ just
measures the
threshold value $\delta_{\rm t}$ in units of the dispersion $\sigma$.
However, we always define $\nu$ in terms of the volume
fraction via equation (\ref{C11}), because this definition has the advantage
of making the genus curve invariant under arbitrary monotonic
one-to-one 
transformations of the density field. For example, a simple linear
biasing transformation would not affect it. Also, it is insensitive
to the skewness of the one-point probability distribution function, 
that quickly develops in the
mildly non-linear regime.

There is a theoretical prediction for the 
expected genus curve of a Gaussian random field \cite{Ha86}. 
It is given by
\be
g(\nu)=N\,(1-\nu^2)\exp\left(-\frac{\nu^2}{2}\right) , 
\label{GeRP}
\ee
where the amplitude
\be
N=\frac{1}{(2\pi)^2}\left(\frac{\left<k^2\right>}{3}\right)^{\frac{3}{2}}
\ee
is determined by the second moment 
\be
\left<k^2\right>=\frac{\int k^2 P(k) \dd ^3k}{\int P(k)\dd ^3k}
\ee
of the (smoothed) power spectrum.
Interestingly, only the amplitude of the genus curve depends on the
shape of the power spectrum. Apart from that it exhibits a universal, 
symmetric w-shape that we will use as a benchmark to detect
non-Gaussian features of the density field.

For a given threshold value we compute the genus with the algorithm
proposed by Gott et~al. \shortcite{Go86}.
The method tesselates space in small cubes that allow the isodensity
surface to be defined as the boundary between the volume elements 
above and
below threshold. If the cubes are sufficiently small this
approximation does not change the topology of the smooth isodensity
surface. The curvature of the resulting
polygonal surface 
is compressed into the vertices of the cubes. 
This property allows a computer
to rapidly sum up the appropriate angle deficits and to compute the
genus per unit volume.
The method also allows arbitrarily shaped survey volumes. 
Here one just counts those vertices that are surrounded by eight
volume elements
that all lie inside the actual survey region.

Based on Weinberg's \shortcite{We88} code {\small CONTOUR} for computing the genus
we have written a new version in C that is optimized for a high
execution speed, since we need to compute several thousand
genus curves in this work. 
A simple 
sorting of the density field prior to the genus computation
led already to a major speed-up because it 
is then possible to instantly
find the threshold value corresponding to a desired volume
fraction. This also allows the efficient 
computation of 
high resolution genus curves.

\section{Fully sampled Virgo simulations}
	
\subsection{The N-body models}

\begin{table}
\bc
\caption{Parameters of the examined CDM models. The simulations have
been done by the 
\virgo collaboration.\label{modelparameters}
}
\begin{tabular}{l|c|c|c|c|}
\multicolumn{1}{l|}{ }& SCDM & $\tau$CDM & $\Lambda$CDM & OCDM\vspace{0.1cm}\\ 
\multicolumn{1}{l|}{Number of particles } & $256^{3}$ & $256^{3}$ & $256^{3}$ & $256^{3}$ \\
\multicolumn{1}{l|}{Box size$[\lu]$ } & $239.5$ & $239.5$ & $239.5$ & $239.5$ \\
$z_{\rm start}$ & $50$ & $50$ & $30$ & $119$ \\
$\Omega_{0}$ & $1.0$ & $1.0$ & $0.3$ & $0.3$ \\
$\Omega_{\Lambda}$ & $0.0$ & $0.0$ & $0.7$ & $0.0$ \\
 \multicolumn{1}{l|}{Hubble constant $h$} & $0.5$ & $0.5$ & $0.7$ &$0.7$\\
$\Gamma$ & 0.5 & 0.21 & 0.21 & 0.21 \\
$\sigma_{8}$ & $0.60$ & $0.60$ & $0.90$ & $0.85$ \\

\end{tabular}
\ec
\end{table}

\begin{figure*}
\bc
\resizebox{8cm}{!}{\includegraphics{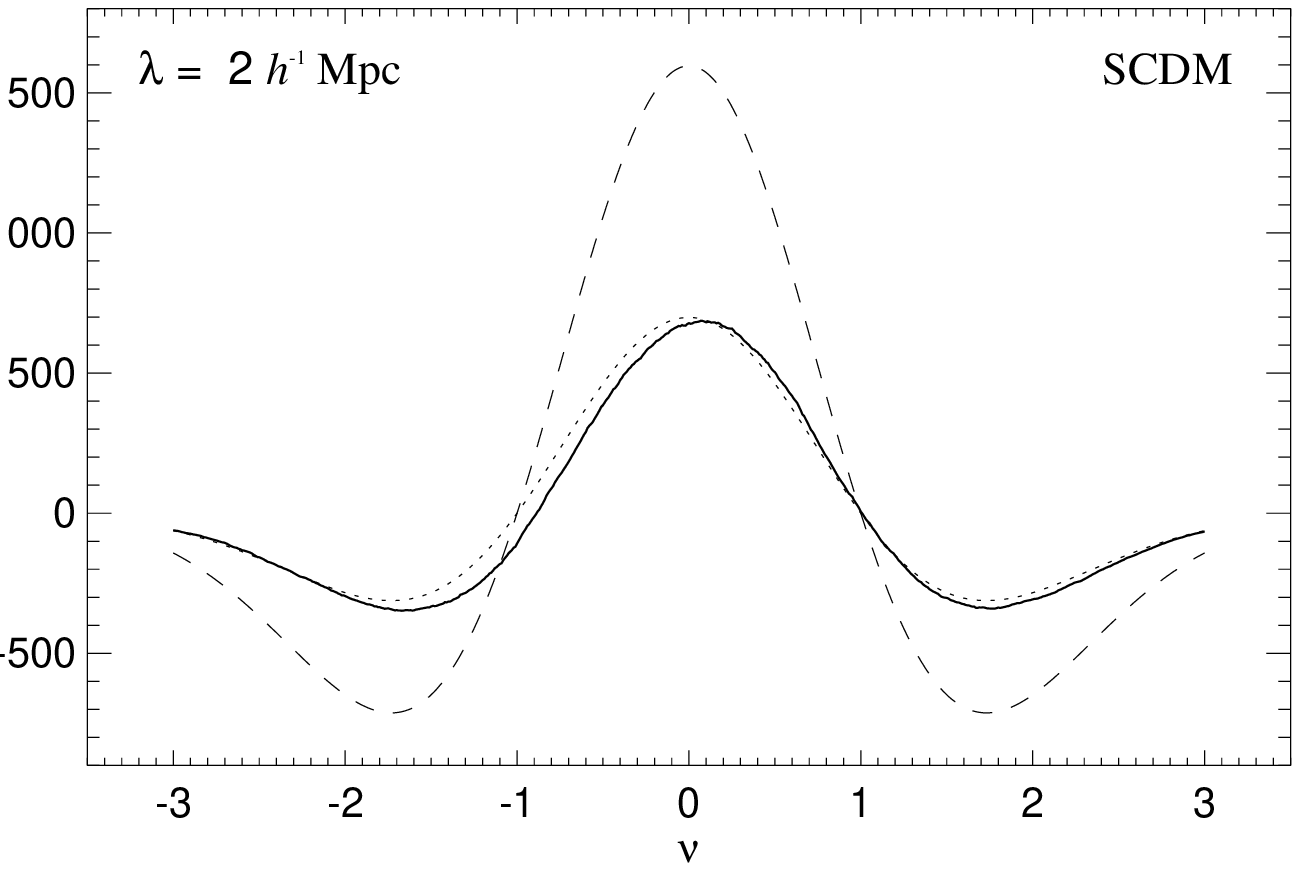}}
\hspace*{0.4cm}\resizebox{8cm}{!}{\includegraphics{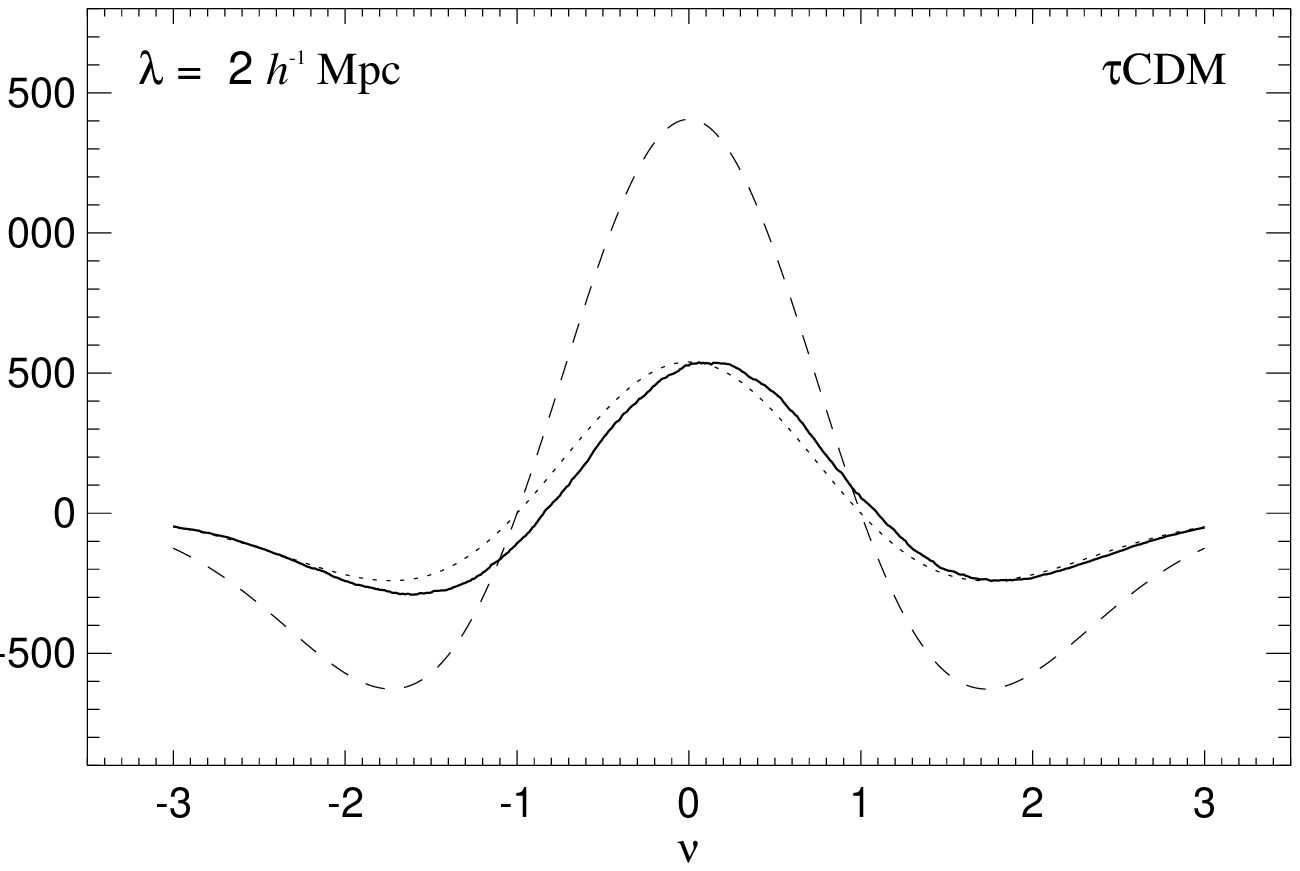}}
\resizebox{8cm}{!}{\includegraphics{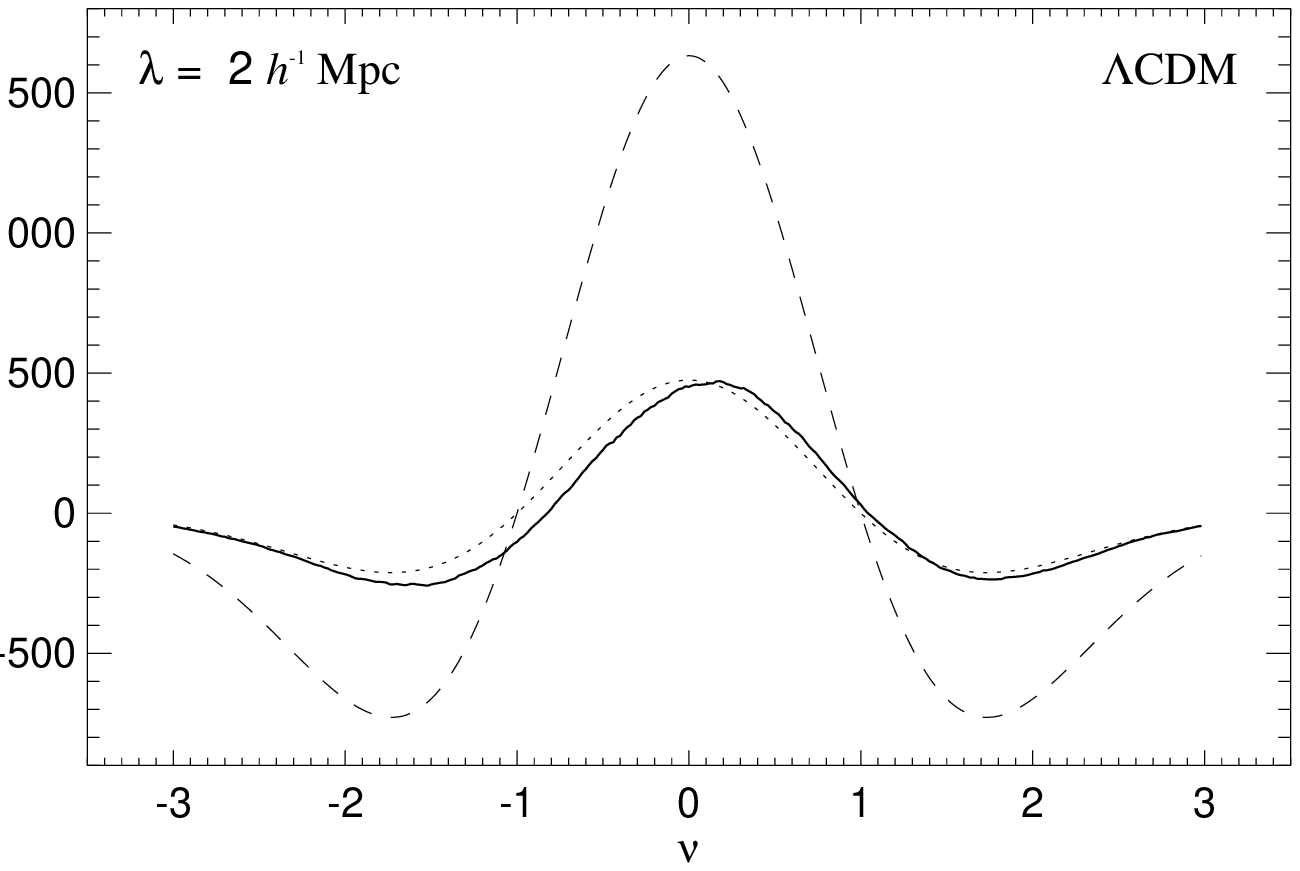}}
\hspace*{0.4cm}\resizebox{8cm}{!}{\includegraphics{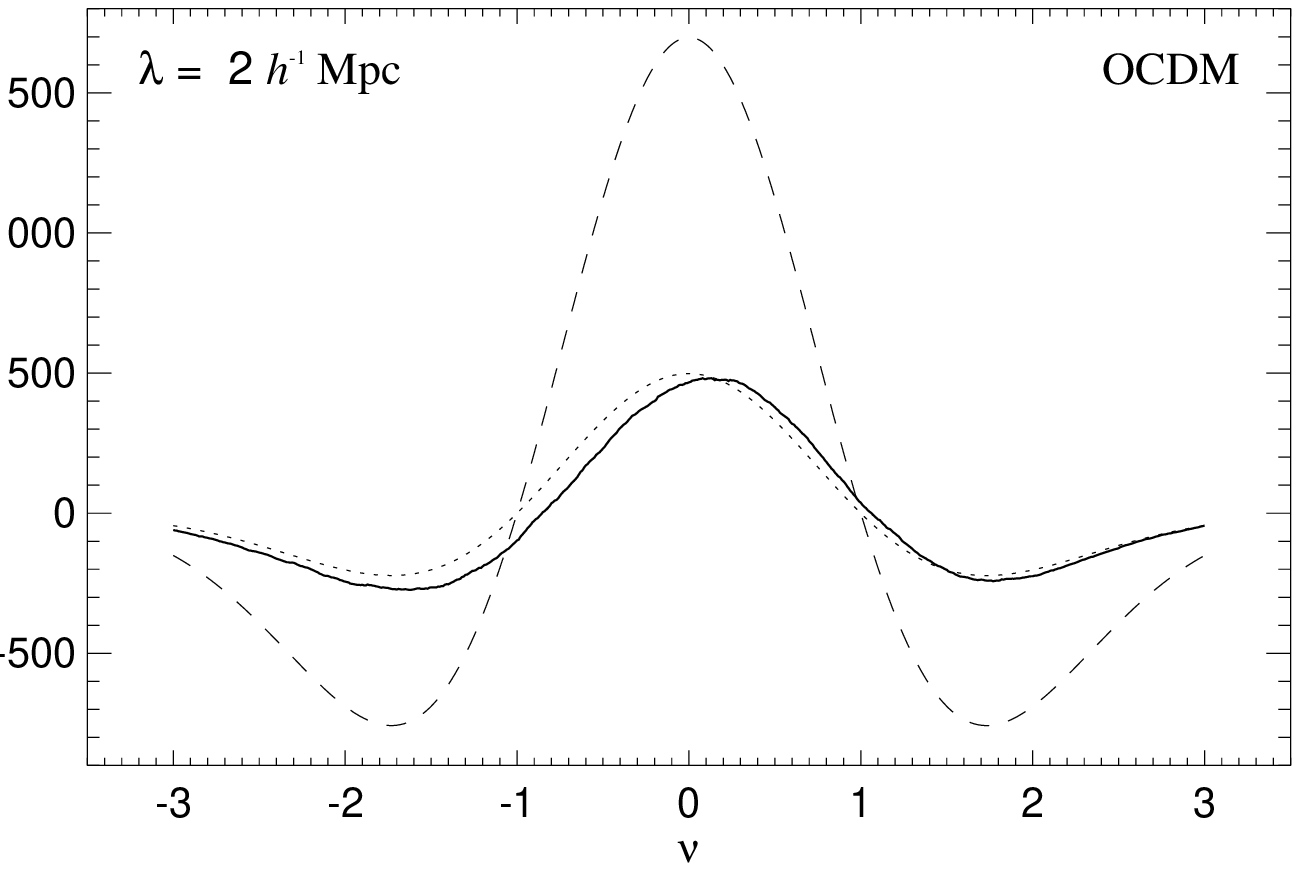}}
\caption{Genus curves of four variants of the CDM
cosmology.
The solid lines in each panel show the genus of the
evolved density fields of the \virgo simulations for a smoothing scale of $2\lu$.
The dotted line is a fit to the random phase genus curve, while the
dashed curve gives the genus of the corresponding {\it Gaussianized} field.
\label{genusvirgo}
}
\ec
\end{figure*}

In this section we compute genus curves for four fully sampled N-body
simulations of variants of the CDM cosmology. These simulations, 
part of the \virgo project \cite{Je97} to examine
structure formation
in the universe 
at very high
resolution, follow
$256^3$ dark matter particles in boxes of comoving size of $239.5\lu$. 
The simulations have been carried out with the parallel adaptive grid
code {\small HYDRA} \cite{Pe97}.

The linear theory power spectrum used to generate the initial
conditions
of the models
is parameterized by the generic fitting form
\be
\label{PowerModels}
P(k)=\frac{Bk}{\left(1+\left[ak+(bk)^{3/2}+(ck)^2\right]^\nu\right)^{2/\nu}} ,
\ee
where $a=6.4\,\Gamma^{-1} \lu$, $b=3.0\,\Gamma^{-1} \lu$, 
$c=1.7\,\Gamma^{-1} \lu$, $\nu=1.13$, and $\Gamma$ is a shape
parameter \cite{Ef92}. 

While the SCDM model has the shape parameter $\Gamma = 0.5$, the other
three models have the same linear power spectrum with $\Gamma=0.21$.
All four models are normalized so as to give 
the observed abundance of rich clusters
of galaxies at the present day. Further simulation parameters are listed
in table \ref{modelparameters} 
and may be found in Jenkins et~al. \shortcite{Je97}.

\subsection{Fixed smoothing}

We start with the ordinary genus statistic, i.e.\ we assume a
spatially constant smoothing kernel.
In order to construct a smooth density field we 
bin the 16.7 million particles of one simulation 
onto a mesh using the
cloud-in-cell assignment, and we smooth the resulting density
field with a Fast Fourier
convolution. We use a
Gaussian kernel of the form 
\be
W(\vec{x})=\frac{1}{\pi^{3/2}\lambda^{3}}\,\exp\left({-\frac{\vec{x}^{2}}{\lambda^{2}}}\right).
\label{EQ1}
\ee
Note that this differs from an ordinary normal distribution 
by a factor of $\sqrt{2}$ in the definition of the smoothing scale. We stick
to this convention which is used in the majority of the literature on
the subject.

Typically we employ a $128^3$ grid to represent the density field.
Only for smoothing lengths below  $5\lu$ 
we do find that a smaller mesh is indicated. We then use a $256^3$ grid.
For test purposes, we repeated one of our calculations on a $512^3$ mesh.

\subsection{Results}

In figure \ref{genusvirgo} we show the genus curves 
of the four simulations at
a smoothing scale of $2\lu$, the smallest scale considered here. We
will come back to the question of what happens on 
still smaller ones later on.
It is evident that the genus curves retain 
their universal w-shape,
even at this small
smoothing scale where the density fields are already fairly non-linear.
Only a small {\it bubble} shift to the right 
has developed, which seems somewhat weaker for the
SCDM model than for the other cosmologies.

\begin{figure}
\bc
\resizebox{8cm}{!}{\includegraphics{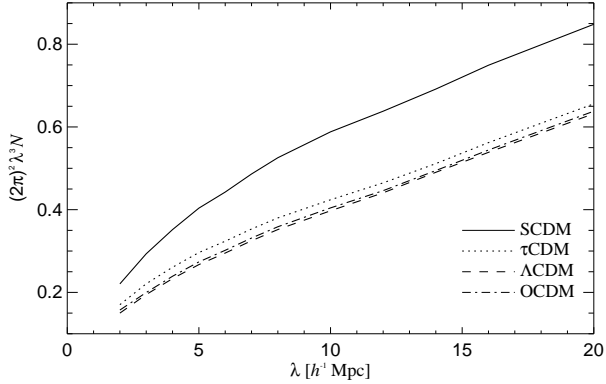}}
\caption{Genus amplitude of the \virgo simulations as a function
of smoothing scale (fixed smoothing). The dimensionless 
vertical axis gives the
genus amplitude $N$ times the factor $(2\pi)^2\lambda^3$. 
\label{figAmp}
}
\ec
\end{figure}

We obtain qualitatively similar genus curves for larger smoothing
scales. For $\lambda>8\lu$ the small bubble shift has vanished
for all models and the shape
of the genus curves is fit perfectly  by the random phase form of
equation (\ref{GeRP}). In figure \ref{figAmp} we show the measured genus
amplitudes as a function of smoothing scale. 
The SCDM model exhibits a
significantly higher amplitude, 
reflecting the different shape of its power
spectrum, while the other three models show very similar amplitudes.
This demonstrates that the genus amplitude in the linear and mildly
non-linear regime is determined by the shape of the power spectrum
alone.

\begin{figure}
\bc
\resizebox{8cm}{!}{\includegraphics{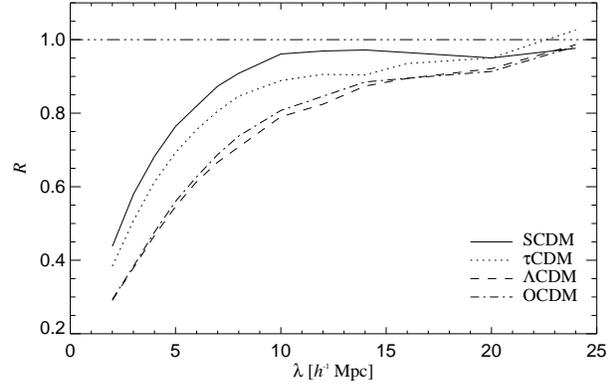}}
\resizebox{8cm}{!}{\includegraphics{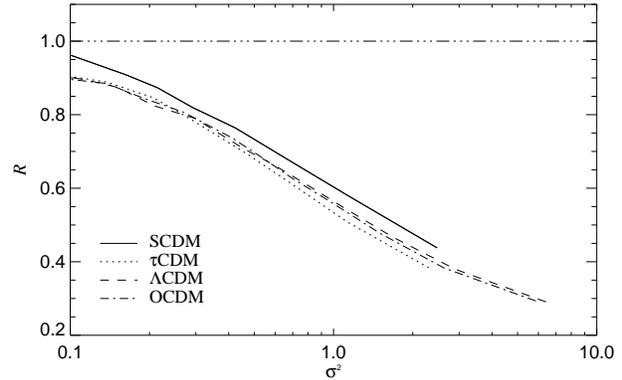}}
\caption{Amplitude drop for the \virgo simulations, i.e. the ratio $R$
of the genus amplitude $N$ to the corresponding amplitude of the
Gaussianized density field.
The top panel shows the amplitude 
drop versus the smoothing scale, while
the lower panel displays it against the variance of the smoothed
fields.
\label{fig8}
}
\ec
\end{figure}

Even if the genus curve is well described by equation (\ref{GeRP}), the
underlying density field does not have to be Gaussian.
In fact,
Vogeley et~al. \shortcite{Vo94} and Canavezes et~al. \shortcite{Ca97} pointed out
that the amplitude of the genus curve is
suppressed on small scales compared to the expected amplitude based on
the power spectrum alone. This amplitude drop is a direct consequence
of phase correlations that develop during the nonlinear growth of
density perturbations. The phase drop may be measured for a periodic
N-body simulation by {\it Gaussianizing} the evolved density field,
i.e. by taking it to Fourier space, randomizing the phases of all
modes constrained by the reality condition
$\delta_{\vec{k}}=-\delta_{\vec{k}}^\star$,
and transforming back to real space. 
In this way a Gaussian field with
identical power spectrum as the evolved density field is obtained, and
a measurement of its genus allows an estimate of the amplitude drop.

\begin{figure}
\bc
\resizebox{8cm}{!}{\includegraphics{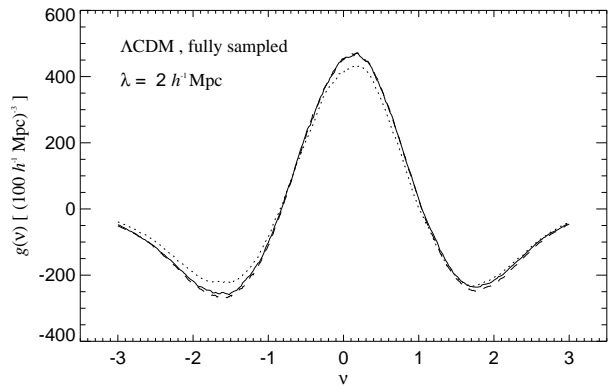}}
\caption{Finite grid size effect. 
The lines are genus curves for the $\Lambda$CDM simulation,
smoothed with $\lambda=2\lu$. The solid line is computed with a
$512^3$ grid, the dashed line with $256^3$, and the dotted line with $128^3$.
\label{fig16}
}
\ec
\end{figure}

We have measured the genus amplitude drop in this way and show the results in
figure \ref{fig8}. Here we reach 
smoothing scales as small as $2\lu$, thereby extending the work of Canavezes 
et~al. \shortcite{Ca97}.
Note that the 
amplitude drop becomes very substantial
at small scales, showing
that the genus is indeed quite sensitive to
higher order correlations in this regime. If we plot the amplitude
drop against the
variance of the smoothed density fields, we can approximately take out the 
differences between the models due to their slightly different
normalizations.
As the lower panel in figure \ref{fig8} shows, we again find that the 
models with the same shape
of the power spectrum exhibit a nearly degenerate behaviour. 
However, the horizontal axis of this plot is
affected by biasing. Since the bias required to match the observed
two-point correlation function \cite{Je97} is different for the four 
models, the 
amplitude drop may still be a useful measure 
to discriminate between different
CDM variants.
Future 
galaxy redshift surveys should allow an accurate
measurement of the amplitude drop by combining genus statistics with
an independent measure of the power spectrum or clustering strength as
outlined by Canavezes et~al. \shortcite{Ca97}.
Comparing these measurements with the  
results
obtained in figure \ref{fig8} for the dark matter may then allow
a direct measure of the bias $b^2=\sigma^2_{\rm gal}/\sigma^2_{\rm DM}$.

We now examine systematic limitations of the genus analysis performed
above. We will
consider finite
grid size effects first and then 
examine the resolution limit of the \virgo
simulations in terms of the genus statistic.

\subsubsection{Finite grid size}

Our method for computing the genus curve relies on an approximate
representation of isodensity contours as polygonal surfaces that are
made up of faces
of small cubes used to tesselate space. As Hamilton
et~al. \shortcite{Ha86} show 
the genus curve is expected to be unaffected
by this approximation, if $d/\lambda \ll 1$, where $d$ is the size of
the cubes. Typically we use $128^3$-grids, and for $\lambda\le 5\lu$
$256^3$ grids. We find that for $d\le 0.5\lambda$ there is hardly any
finite grid size effect. For example, in figure \ref{fig16}
we show a
comparison of the genus
curves for the $\Lambda$CDM simulation, smoothed at $\lambda=2\lu$,
using $128^3$, $256^3$, and $512^3$ grids. 
There is only a
small depression of the amplitude and the minima of the genus curve, 
when the $128^3$
mesh is used, but for $256^3$ the asymptotic behaviour is clearly
reached. It is therefore not necessary to use costly 
computations at $512^3$ resolution.

\subsubsection{Resolution limit}

\begin{figure}
\bc
\resizebox{8cm}{!}{\includegraphics{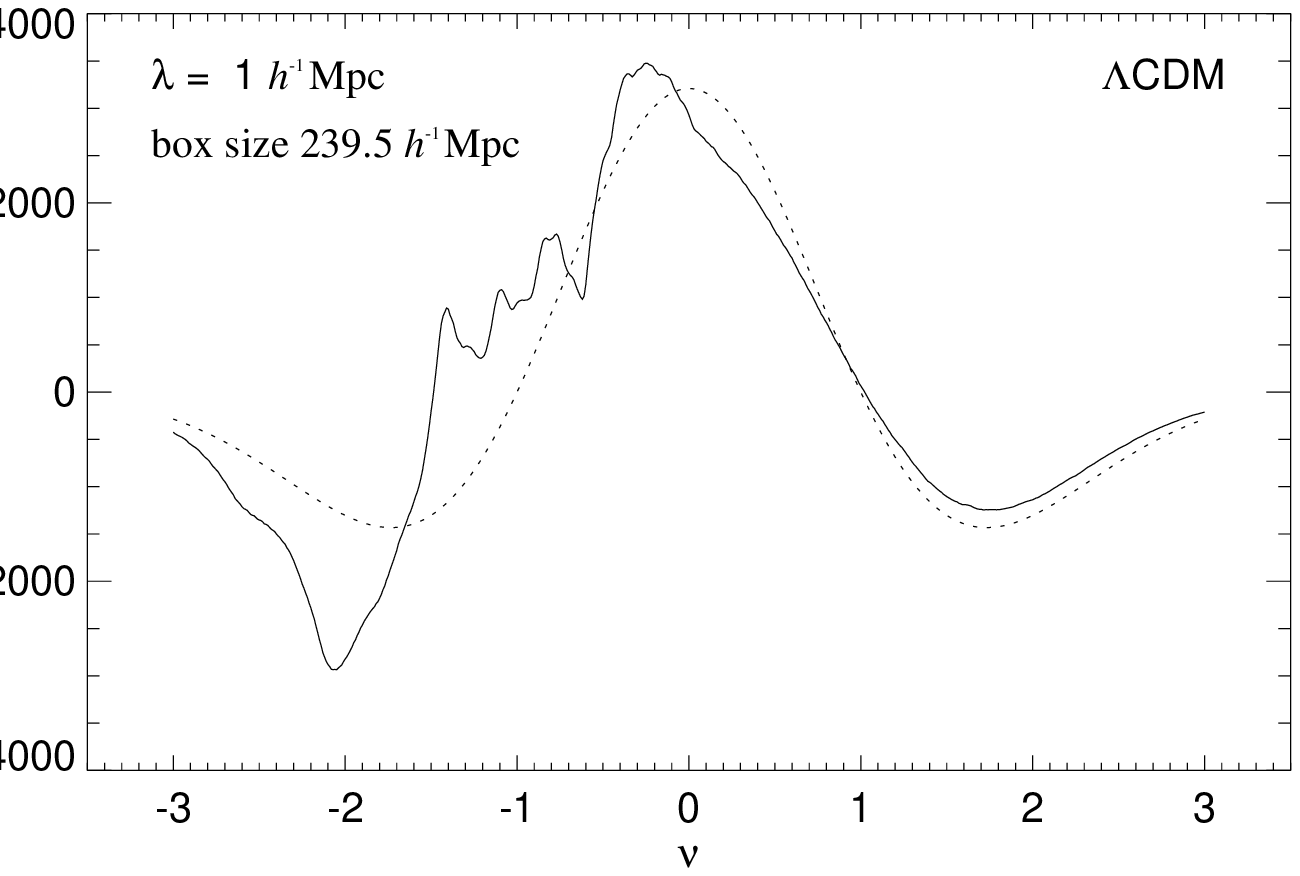}}
\resizebox{8cm}{!}{\includegraphics{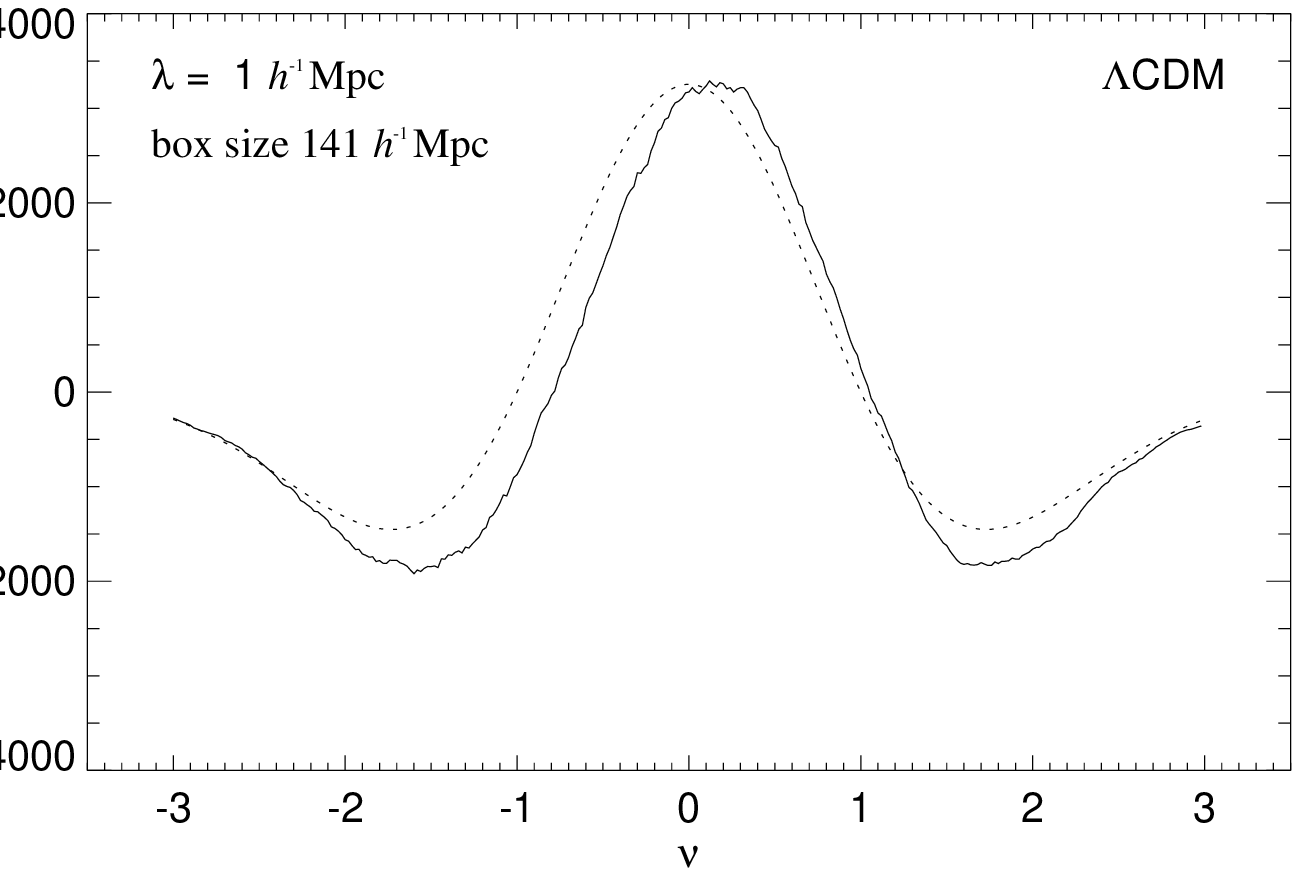}}
\caption{Resolution limit of the \virgo simulations. The solid line of
the top panel shows
the genus of the $\Lambda$CDM model smoothed with $\lambda=1\lu$, and
the dotted curve shows a best-fit random phase genus curve. In this case
the smoothing reaches the level of the inter-particle separation and
the features in the genus curve are in fact artifacts. 
We show that
this is the case 
by computing the genus for a second $\Lambda$CDM simulation with
smaller box size of $141\lu$, but with the same number of particles and
hence significantly higher spatial resolution. As seen in the bottom panel, the
genus curve remains featureless.
\label{fig788}
}
\ec
\end{figure}

If the smoothing length is reduced below $2\lu$ 
one suddenly starts
to see features in the genus curve, as exemplified in figure
\ref{fig788}, where
we show the genus for the $\Lambda$CDM simulation at a smoothing scale
of $1\lu$.
However, the `ringing' on the low density side is
just an artifact due to the fact that 
the resolution of the \virgo simulations is limited. We demonstrate
that this is indeed the case 
by analysing yet another simulation of the \virgo consortium, 
the same $\Lambda$CDM model, but with 
a smaller box size of $141\lu$. 
This model has roughly twice the spatial resolution of the original simulation.
As the bottom panel of figure \ref{fig788} demonstrates, it
still gives a `normal' genus curve for $\lambda=1\lu$.
 
The smoothing scale of $1\lu$ is already close to the mean
inter-particle separation of $\approx$ 0.94$\lu$ for the large-box
\virgo runs. Due to the strong
clustering of matter, the voids apparently contain too few particles to
prevent discreteness effects becoming visible when a fixed smoothing
kernel with $\lambda=1\lu$ is used. 
On the other hand, the regions with high particle 
density allow a much higher resolution in principle. In order to take full
advantage of this spatially varying resolution an adaptive smoothing
technique needs to be developed. We will introduce such a scheme in the
next section.

\subsection{Adaptive Smoothing}

\begin{figure}
\bc
\resizebox{8cm}{!}{\includegraphics{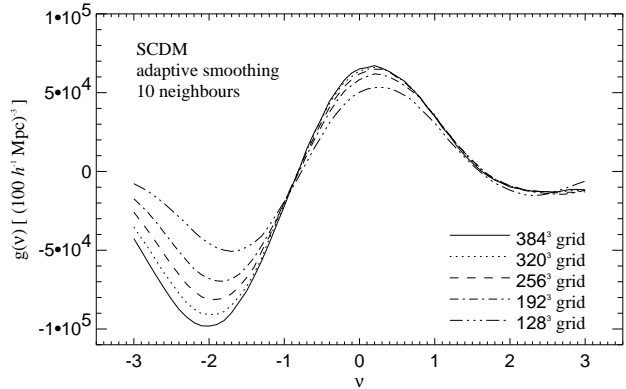}}
\caption
{Finite grid size effects in the adaptive smoothing technique.
Shown are five genus curves for one of the SCDM subvolumes, 
computed at different grid resolutions. 
For 10 neighbours, the 
number considered here, a $256^3$ mesh is sufficient to
resolve accurately 
the genus of the density field, at least in the regions of
positive genus and for $\nu>0$. However, the minimum on the negative
side is still not fully resolved by a grid as fine as $384^3$.
Comparing the size distribution of the isolated regions at the 
two minima of the
genus curve, we find that this is due
to a larger relative abundance 
of very small regions 
at the minimum on the negative side.
This population of very small structures is difficult to resolve;
those regions with volume smaller than a mesh cell can
be lost, leading to a suppression of the genus amplitude.
\label{figgrideffad}}
\ec
\end{figure}

\begin{figure*}
\bc
\resizebox{8cm}{!}{\includegraphics{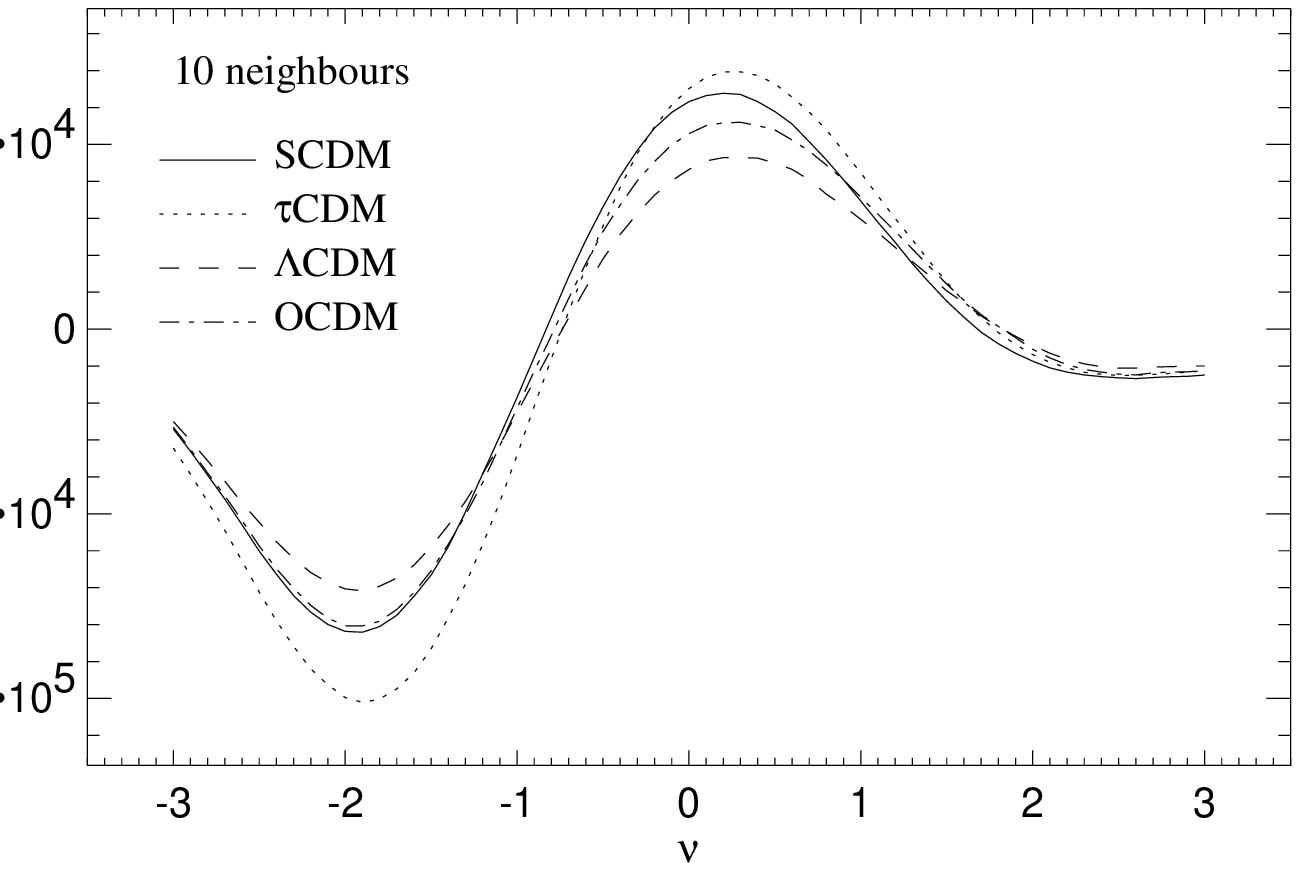}}
\hspace{0.4cm}\resizebox{8cm}{!}{\includegraphics{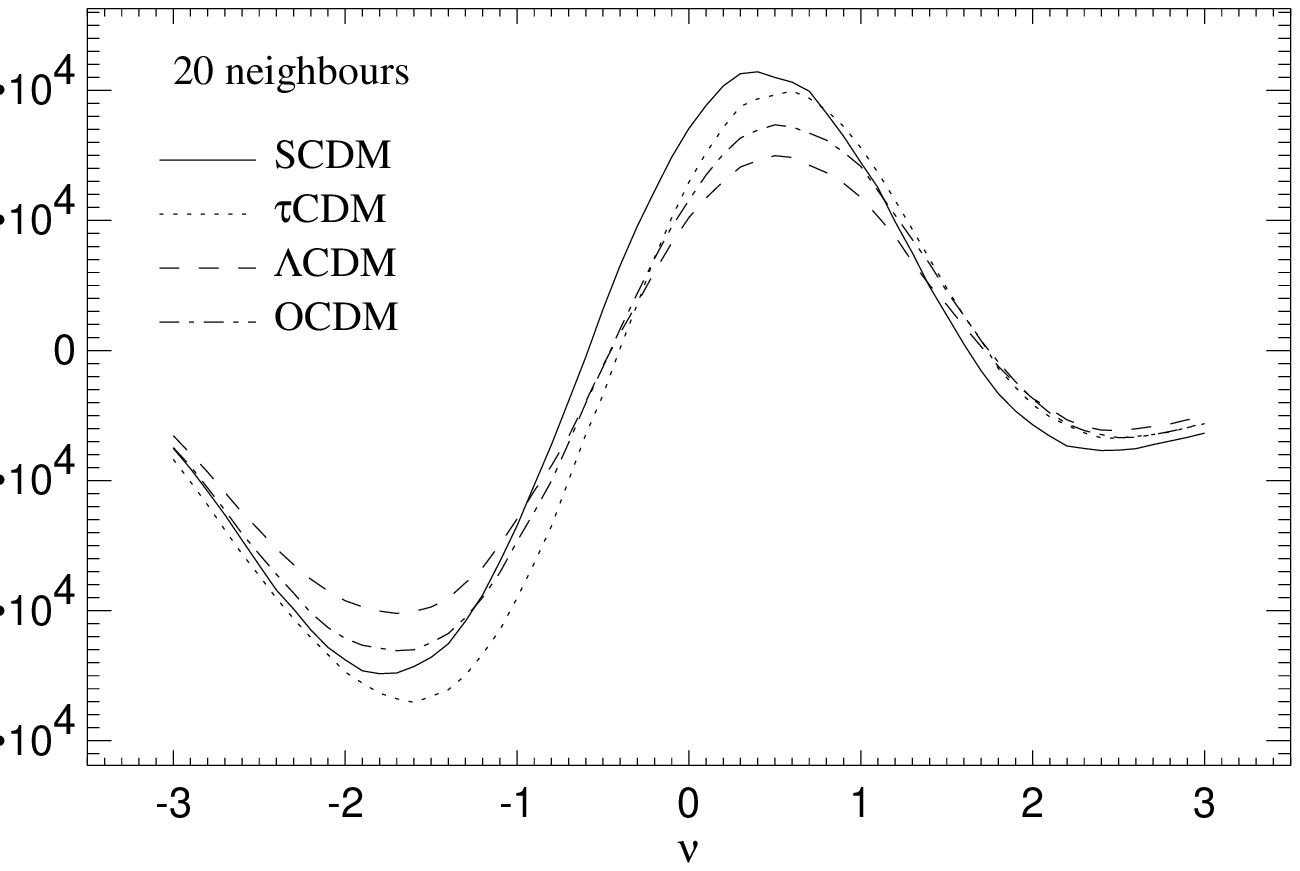}}
\resizebox{8cm}{!}{\includegraphics{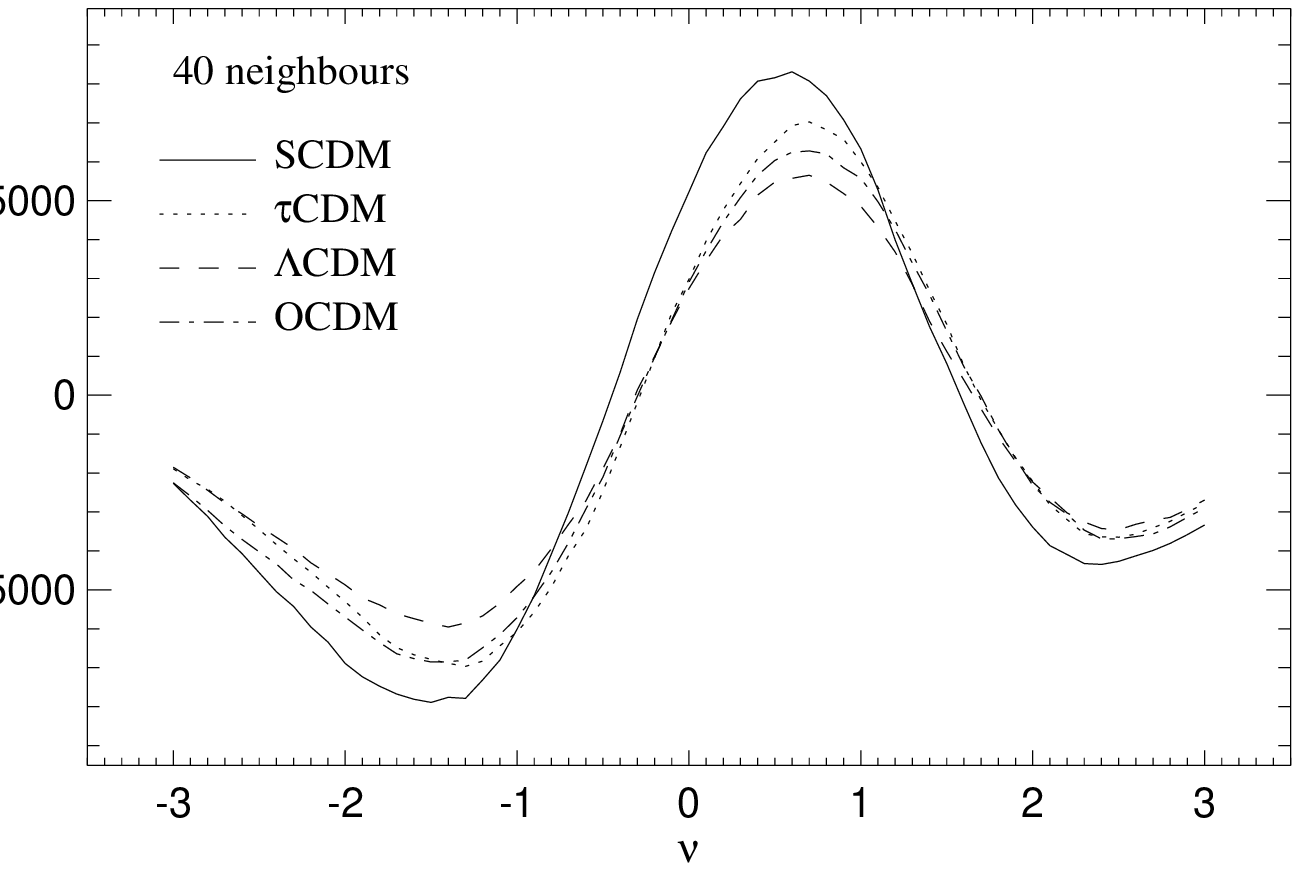}}
\hspace{0.4cm}\resizebox{8cm}{!}{\includegraphics{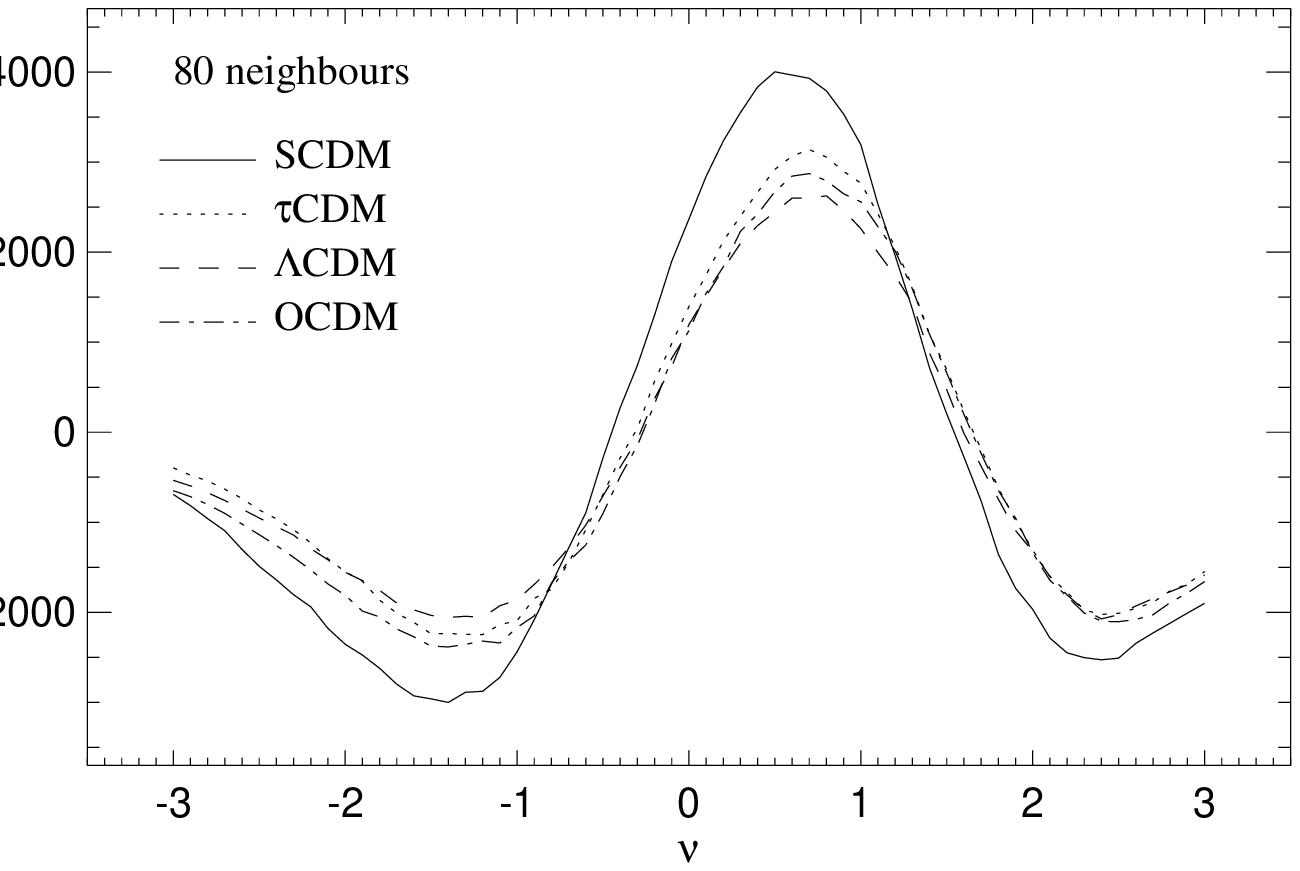}}
\caption{Genus curves for the adaptively smoothed \virgo simulations.
The four panels show results when 10, 20, 40, and 80 neighbours are used to
determine the local smoothing scale. Each curve is the average of 4
subvolumes, each being 64 times smaller than the full simulation
volume. Since the variance between these curves is quite small, it is
not necessary to extend the computation to a larger fraction of the
total volume.
\label{figadmo}}
\ec
\end{figure*}

As we saw above, a spatially fixed smoothing is not able to take full
advantage of the information content on small scales. 
Borrowing an idea from 
smoothed particle hydrodynamics (SPH),
we can improve on this by
following the particles in a Lagrangian sense and varying the smoothing
scale with local density. 
This application of SPH-smoothing 
to the dark matter has first been proposed
by Thomas \& Couchman \shortcite{Th92}.

Of course, a difficulty with such an approach is that the resulting
density
field cannot be studied analytically. In particular, the power
spectrum of the smoothed field is not related to that of 
the underlying field
in a simple way. Because of that, it is also not obvious 
what one
can expect for the genus statistic. However, 
a comparison between theory and observation is always possible
if
ensembles of mock surveys are used to calibrate the adaptively smoothed genus 
statistic.

As Hernquist \& Katz \shortcite{He89} point out there are two
different approaches to defining a smoothed density estimate with
variable smoothing scale. In the {\it scatter} approach the mass of
each particle is distributed in space, and the density estimate at a
particular point $\vec{x}$ follows from the overlap of the individual
smoothing spheres. Alternatively, one can define a smoothing radius for
each point $\vec{x}$ and weight all particles in its neighbourhood by
the resulting kernel. Only for a fixed smoothing scale does this {\it gather}
approach become identical to the scatter method.

We want to compute the
density field on a fine mesh at a large number of points. Because 
the
number of tracer particles is also large, the gather approach is 
computationally less costly in this case. 
Hence we will adopt it in the following.

As a smoothing kernel we choose the spherically symmetric
spline kernel
\be
W(r;h)=\frac{8}{\pi h^3}
\left\{
\begin{array}{ll}
1-6\left(\frac{r}{h}\right)^2 + 6\left(\frac{r}{h}\right)^3, &
0\le\frac{r}{h}\le\frac{1}{2} ,\\
2\left(1-\frac{r}{h}\right)^3, & \frac{1}{2}<\frac{r}{h}\le 1 ,\\
0 , & \frac{r}{h}>1 , 
\end{array}
\right.
\ee
which is frequently used in SPH calculations \cite{Mo85}. This kernel
has the advantage of compact support, which again simplifies the
computational task.

For simplicity, we choose the smoothing scale $h(\vec{x})$ as the
distance to the $N$-th nearest neighbour. In this way the smoothing
scale is set to a fixed fraction of an estimate 
of the local mean inter-particle
separation.
We then estimate the density
field as
\be
\rho(\vec{x})=\sum_i W(\vec{r}_i-\vec{x};h(\vec{x})),
\ee
where the sum extends over all particles. 
Due to the compact kernel only
the nearest $N$ particles give a nonzero contribution. 
Hence the smoothing scheme heavily relies on a fast algorithm for
finding the nearest neighbours for arbitrary points in space.
We utilize a coarse search grid
and link-lists to match this requirement.

Nevertheless,
the computational
cost for adaptive smoothing is much higher than for fixed
smoothing, because 
Fourier techniques can no longer be applied and one has to work in
real space.
Furthermore, the adaptive smoothing will also be able to
resolve very small structures that demand a fine mesh in order to
allow a proper resolution of their topology.

Because of this it is not possible to consider the whole simulation
box in one pass. Instead, we compute adaptively smoothed density
fields for subvolumes of the simulation box of size one
quarter of the total box size. In this way, we divide the simulation
in
64 subvolumes, each containing still $64^3$
particles on average. For the small scales we are considering here, we
anticipate that each of these subvolumes will already contain so many
structural elements that the cosmic variance between them will be 
small.
Hence we assume that we will have to consider at most a few of the
subvolumes in order to accurately reproduce the genus of the whole
simulation box.

We compute adaptively smoothed density fields on a regular
grid for each of the subvolumes. In the smoothing process,
we also consider the particles that
lie outside the subvolume, hence 
boundary smoothing effects are not present.
In figure \ref{figgrideffad}
we show the
dependence of the result on the chosen grid resolution for the extreme
case of $N=10$ neighbours. 
A $128^3$ grid clearly leads to
a systematic underestimate of the genus, showing that 
small features are missed due to the coarseness of the grid. 
On the other hand, 
the result for the $256^3$ grid is already very close to
the finer meshes,
at least for $\nu\ge-1$.
However, the minimum of the genus on the negative side is seen to be very
difficult to resolve accurately. 
By comparing the distributions of the sizes of the
isolated regions at the two minima, we find
that the number density of very small regions is much larger at the
minimum on the negative side than on the positive side.
Since these very small voids are so abundant, the loss of the smallest
of them due to the finite grid size leads to a suppression of the
genus.
Fortunately this effect is less severe at larger smoothing scales. 
As a compromise between computational cost and accuracy, we decided to
use $256^3$ grids for $N=10$ neighbours, $192^3$ for $14\le N \le 20$, and
$128^3$ for $N\ge28$.
This ensures that the genus density is accurately resolved
for $\nu\ge-1$. However, the depth of the minimum on the negative side may
be slightly diminished due to finite grid size effects.

\begin{figure}
\bc
\resizebox{8cm}{!}{\includegraphics{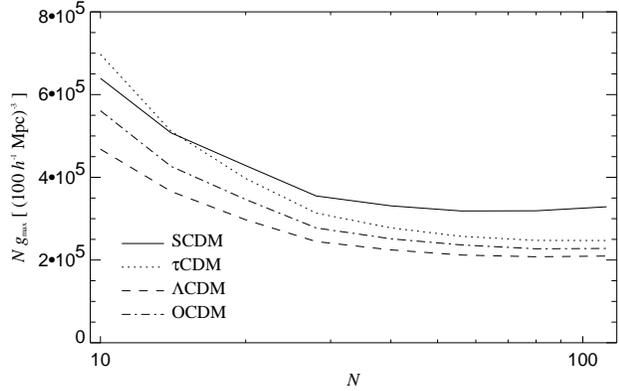}}
\caption
{Peak genus densities in the \virgo models
as a function of the number $N$ of neighbours used
in the adaptive smoothing scheme. We have measured 
the genus for $N=10$, 14, 20, 28, 40, 56, 80, and 112
neighbours, and plot the maximum genus density 
times the number $N$ of smoothing neighbours.
\label{figampladap}}
\ec
\end{figure}

\begin{figure}
\bc
\resizebox{8cm}{!}{\includegraphics{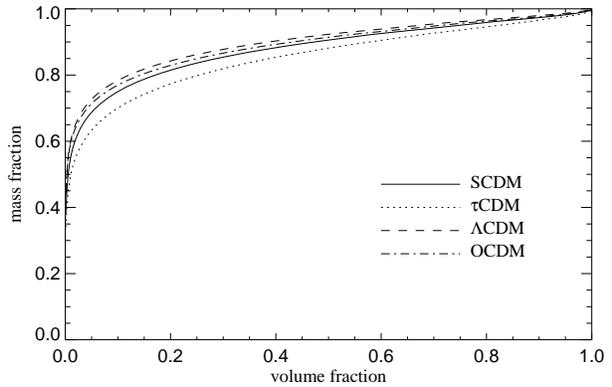}}
\caption
{Mass fraction above threshold versus volume fraction
above threshold for the 
adaptively smoothed
\virgo simulations. Here $N=10$ neighbours have been used
in the smoothing process.
At a given volume fraction, the $\Lambda$CDM model contains a higher
fraction of its mass in the high density regions than the other three
models.
\label{figmfvf}}
\ec
\end{figure}

In figure \ref{figadmo} we show genus curves for adaptive smoothing
with $N=10$, 20, 40, and 80 neighbours. In each panel we plot the
results for SCDM, $\tau$CDM, $\Lambda$CDM, and OCDM.
Interestingly, there are marked differences between the models,
particularly at small smoothing scales. The SCDM model shows
usually the highest genus amplitude, as expected for the larger
amount of small-scale power in this model. 
Only on the smallest scale
considered here, it is surpassed by the $\tau$CDM model.
While the genus amplitudes
of the three models $\tau$CDM, $\Lambda$CDM, and OCDM 
were practically degenerate in the
regime accessible to fixed smoothing, 
they can now be used to discriminate between
the models.

Note that the number density of structure elements resolved with
adaptive smoothing is really much
larger than the one accessible to fixed
smoothing. Compared to the $\lambda=2\lu$ fixed smoothing, 
the $N=10$ adaptive scheme
reaches a
density of structure elements which
is
approximately two orders of magnitude larger. This really opens up a
new regime for topological analysis.

In figure \ref{figampladap} 
we plot the peak genus densities as a function 
of the number of neighbour particles used for the adaptive smoothing.
It is interesting that the relative difference between $\tau$CDM,
$\Lambda$CDM, and OCDM
grows with decreasing smoothing length. 
$\Lambda$CDM consistently shows the smallest amplitude, which
demonstrates that its structure has a higher degree of coherence
with fewer small-scale features than the other models. 

However, for the $N=10$ smoothing, 
all four simulations give roughly the same genus signal at
the minimum on the positive side 
of the genus curves. This minimum
occurs at $\nu\simeq 2.4$, corresponding to a volume fraction above
threshold of
only 0.0082. 
For a Gaussian random field, the
minimum occurs at $\nu=\sqrt{3}$, equivalent to 
the much larger volume fraction of 0.042.
At the minimum the genus is completely dominated by the
number density of very dense, isolated clumps. 
While this number density of roughly  $0.012\,h^{3}{\rm Mpc}^{-3}$ is similar for the four models,
the mass fraction contained in
the clumps is different. 
For the $N=10$ smoothing, 
we plot in figure \ref{figmfvf} the
mass fraction above threshold as a function of 
the volume fraction above
threshold. 
Note the very high concentration of the mass in these adaptively
smoothed density fields.
In the SCDM model
50 per
cent of the mass
is contained in 
only 0.65 per cent of the volume.
The $\tau$CDM model requires almost twice the volume
fraction,
1.23 
per cent,
to include half of its mass.
On the other hand, 
the $\Lambda$CDM and OCDM simulations have half-mass
volume fractions of 0.33 and 0.31 per cent, respectively.
At high densities
a given volume fraction therefore
contains a higher fraction of their mass than in the high density models.
This reflects the normalization of the models which ensures that they
all have the same abundance of virialized clusters of any given mass.

In contrast to the minimum on the positive side, the minimum on the
negative side of the genus curve 
is not shifted significantly. However, here the models show
stronger differences in their genus signal. Again, at the minimum the genus is 
dominated by a high
number density of isolated regions. 
These voids are significantly more abundant
in the $\tau$CDM model
than in the
$\Lambda$CDM model. This means that the underdense regions
are choppier in the
$\tau$CDM model than in $\Lambda$CDM. The latter model has voids which
are more coherent and
larger on the average. 
Presently it is unclear, whether
this can be understood merely as a consequence of the smaller amount of mass 
left in the $\Lambda$CDM model to fill the voids.

The maxima of the genus densities occur at a smaller
volume fraction than 0.5, i.e.\ the adaptive smoothing results in a
substantial bubble-shift to the right.
At the maxima, the isodensity
surfaces have the topology of a sponge, with an interlocking high
density region and a complex system of tunnels and voids. The smaller
amplitude of $\Lambda$CDM can again be interpreted as a larger degree
of coherence; there are not so many topological holes as in $\tau$CDM,
for example, and the typical size of tunnels and cavities is expected
to be larger.

It should be noted that the two low-$\Omega_0$ models 
show only small 
differences when their two-point
correlation functions or their velocity fields are considered
\cite{Je97}. 
This demonstrates that the genus can indeed reveal
additional information about the morphological properties of the
matter distribution. Here we conclude 
that the voids in the $\Lambda$CDM
model are in some sense `emptier' than in the OCDM model.

We also note that only the adaptive smoothing allowed an extension of
the genus statistics essentially down to the mass resolution of the
\virgo simulations. Future large redshift surveys feature a large
number of galaxies, possibly in the range $10^6$ for the Sloan and 2dF
surveys.
Adaptive smoothing techniques should prove very 
powerful for this kind of data.
We now work out a first test of this idea for the 1.2-Jy
redshift survey.

\section{The 1.2-Jy redshift survey and Virgo mock catalogues}

\subsection{The 1.2-Jy redshift survey data}

The data of the 1.2-Jy redshift survey \cite{St90} of 
\iras galaxies has been
published \cite{Fi95} and can be retrieved electronically from the
Astronomical Data Center (ftp://adc.gsfc.nasa.gov).
The 5321 galaxies of the survey are selected from the PSC catalogue above a
flux limit of 1.2$\,$Jy in the 60$\,\mu$m band. The sky coverage is 87.6
per cent,
excluding only the zone of avoidance for $|b|<5^{\rm o}$ and 
a few unobserved or contaminated patches at higher latitude.  

There have been numerous studies of the 
1.2-Jy survey, essentially covering all standard methods
for analysing large-scale structure.
Very recently the catalogue has also been examined with topological
methods. Yess, Shandarin \& Fisher \shortcite{Ye97} used a percolation
analysis, Kersher et~al. \shortcite{Ke97} employed Minkowski
functionals, and Protogeros \& Weinberg \shortcite{Pr97} used genus
statistic. 

Our approach is similar to that of the latter authors. We
also work with mock catalogues to derive the statistical properties of
the genus statistics. However, in contrast to Protogeros \& Weinberg
\shortcite{Pr97} we do not volume-limit our catalogues but use instead
a selection function weighting to derive the density fields. 
Additionally,
we introduce adaptive smoothing techniques. 
We also differ in some conclusions; e.g.\ we find no 
confirmation of their claim of a finite volume bias, and we 
explicitly show that the errors of the genus curve are not
multivariate normally distributed.

In our analysis of the 1.2-Jy survey
we convert the redshifts to
the Local Group frame and use them to infer distances without
further corrections for peculiar velocities. Redshift space
distortions have only a negligible effect on the genus, as has been
shown in a number of studies.

We will assume an Einstein-de-Sitter model for the
background cosmology throughout. The results will not be sensitive to
this choice because the 1.2-Jy density field
maps only the very local Universe.

We define 
the selection function $S(z)=\left<m(\vec{r})\right>$ of the survey 
as the mean expected comoving number density of sources at redshift
$z=|\vec{r}|$. We employ the fitting form
\be
S(z)=\frac{\psi}{z^{\alpha}\left[
1+\left(\frac{z}{z^{\star}}\right)^{\gamma}\right]^\frac{\beta}{\gamma}},
\label{EQ10}
\ee
and adopt the parameters (table \ref{tabSel})
determined by
Springel \& White \shortcite{Sp97}. 
Note that the selection function
includes a correction for the strong evolution seen in \iras galaxies.

\subsubsection{Depth of maps}

\begin{table}
\bc
\caption{\label{tabSel}
Parameters of the selection function of the 1.2-Jy survey.}
\begin{tabular}{ccc}
$\alpha$  & $\beta$ & $\gamma$ \\
$0.741^{+0.128}_{-0.135}$  
&  $4.210^{+0.419}_{-0.344}$ 
& $1.582^{+0.237}_{-0.214}$ \\
\\
$z^{\star}$ & $\psi \;\;[h^3{\rm Mpc}^{-3}]$ & \\
$0.0184^{+0.00213}_{-0.00167}$ 
&$(486.5\pm 13.0 ) \times 10^{-6}$ & \\
\end{tabular}
\ec
\end{table}

The galaxy density of a flux limited sample falls off quickly
with distance. As a consequence, 
the uncertainty in the density estimate grows
rapidly with redshift. 
It is desirable, of course, to use a
survey volume that is as large as possible 
in order to beat down statistical noise and cosmic variance.
According to a useful rule of thumb \cite{We87} 
discreteness effects are approximately negligible  
if 
\be
\lambda\ge d=S^{\,-\frac{1}{3}} \, ,
\label{C15}
\ee
where $d$ is the mean inter-galaxy separation. 
Adopting this criterion we choose a 
maximal radius $R_{\rm max}$ 
by 
$\lambda= S(R_{\rm max})^{-\frac{1}{3}}$
and use it to delimit 
the usable survey volume $V_{\rm s}$. This choice ensures that at the far
edge of the survey volume the sampling condition is just met, and in the
remainder of the volume the sampling is denser.

\begin{figure}
\bc
\resizebox{8cm}{!}{\includegraphics{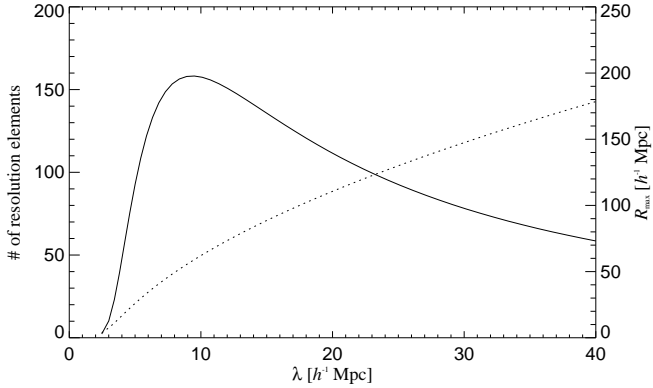}}
\caption
{The number of resolution elements (solid) for the 1.2-Jy survey when the maximal
survey volume is used. Also shown is the radius (dashed) of the 
usable survey volume.\label{fig1}}
\ec
\end{figure}

\begin{table}
\bc
\caption{The smoothing lengths adopted for the topological analysis of the 1.2-Jy
survey. Listed are the chosen survey depth $R_{\rm max}$, the
resulting number
$N_{\rm res}$ of resolution elements and the number $N_{\rm gal}$ 
of galaxies inside the survey volume.
\label{tab2}
}
\begin{tabular}{c|c|r|r|r}
\vspace*{0.2cm}$\lambda\;\;[\lu]$  & $R_{\rm max}\;\;[\lu]$  &
$N_{\rm res}$ & $N_{\rm gal}$  \\
           5 &   26.00 &    92.6 &     1030\\
           7 &   42.23 &   144.7 &     1783\\
          10 &   62.07 &   157.6 &     2820\\
          14 &   83.73 &   140.9 &     3582\\
          20 &  110.63 &   111.5 &     4199\\

\end{tabular}
\ec
\end{table}

\subsubsection{Resolution elements}

The notion of number of resolution elements provides 
a useful way to compare roughly the statistical power of genus
measurements.
Because the 
smoothing extends over an effective volume
$V_{\rm sm}=\pi^{3/2}\lambda^3$
the number of independent structures that can be present in a finite
survey volume is limited. This number is of order
\be
N_{\rm res}=\frac{V_{\rm s}}{V_{\rm sm}}=\frac{ \omega
R_{\rm max}^3}{3\pi^{3/2}\lambda^3},
\ee
where $\omega$ is the solid angle
covered by the survey.

The number $N_{\rm res}$ indicates the power of a data set used for
topological analysis.
With the QDOT survey Moore et~al. \shortcite{Mo92} reached a maximum of about
$N_{\rm res} =80$ whereas the CfA
survey allowed 
Vogeley et~al. \shortcite{Vo94} to achieve
$N_{\rm res}=260$ for their best subsample. 
The most powerful dataset examined so far is PSCz, making it possible 
for Canavezes
et~al. \shortcite{Ca97} to reach $N_{\rm res}=415$.
For the 
1.2-Jy redshift survey we have to be content with 
$N_{\rm res} = 158$.
This already indicates that one can hardly expect very tight constraints
from the genus statistics of this data set. Instead, 
the sparse sampling of the
density field can be expected to severely limit the power of the genus
test.

For the genus analysis we have examined the smoothing lengths
5, 7, 10, 14, and $20\lu$ in approximately logarithmic spacing. 
Table \ref{tab2} lists some relevant parameters for the different
cases.

\subsection{Construction of mock surveys}

We use the \virgo
N-body simulations
to obtain artificial redshift surveys 
that mimic the statistical properties of the 1.2-Jy survey with
respect to sky coverage, selection function
and luminosity distribution. The suites of mock catalogues are then
analysed in exactly the same way as the 1.2-Jy redshift survey
data. In this way sampling noise, cosmic variance, and systematic
biases can be reliably modeled, which allows a fair assessment of the
viability of the models, even if strong biases in our analysis existed.

To construct a mock catalogue 
we first select an arbitrary observer position in the periodic
simulation box, which we replicate periodically in order to allow the
construction of catalogues with sufficient depth. For computational
convenience, we adopt a depth of 239.5$\lu$, corresponding to the size
of the simulation box.
We do
not attempt to restrict the observers to positions 
that match certain properties of the 
immediate neighbourhood of the solar system because we want to
allow for a realistic degree of cosmic variance.
Because we have only one simulation at our disposal, not all the mock
catalogues are independent. However, 
the survey volumes considered 
are much smaller than the simulation volume itself.

We identify every dark matter particle with a possible galaxy site, i.e.
we construct only unbiased galaxy catalogues. We then compute the
cosmological redshifts
$z_i$
of the potential galaxy sites with respect to the observer and draw
uniformly distributed random numbers $x_i\in [0,1]$ for them.
Only those galaxies with
$x_i\le S(z_i)$ are kept,
where $S(z)$ is the selection function scaled such that
$S(z_0)=1$ for some very small $z_0$.
This selection results in a catalogue with selection function
proportional to $S(z)$.

We finally discard the galaxies behind the angular mask of the 1.2-Jy
survey and degrade the source density randomly such that the mock
surveys contain the same number of sources (5083) 
as the 1.2-Jy survey
in a sphere of radius
$239.5\lu$.

In a second step (optional for this work) we assign 
observed fluxes to the galaxies according to the luminosity
function that results as a consequence of the adopted selection
function \cite{Sp97}.
For this purpose we draw
a random number $q_i \in [0,1]$
from a uniform distribution,
compute a 
maximum redshift 
\be
z_{\rm max}=S^{-1}\left((1-q_i)S(z_i)\right),
\ee
and assign an observed flux 
\be
f_i=f_{\rm min}
\frac{r_{\rm max}^2}{r_i^2}\left(\frac{1+z_{\rm max}}{1+z_i}\right)^{1-\alpha}
\ee
for each source. Here $S^{-1}$ denotes the inverse selection function, and
$f_{\rm min}=1.2\,{\rm Jy}$ is the flux limit. Note 
we here assumed 
a straight
K-correction with $\alpha=-2$,  
a non-evolving luminosity function, and
an Einstein-de-Sitter universe. For $\Omega_0\neq 1$, 
the latter may be replaced by the appropriate cosmology.

We usually 
neglect peculiar velocities in the construction of 
mock catalogues. As pointed out above, redshift space distortions 
are generally found to have only a minor effect on the genus
statistics, at least on the relatively large scales accessible so far.
We test that this is indeed the case by producing an additional 
set of mock catalogues in redshift space, where we adopt the velocities
of the dark matter particles themselves as peculiar velocities of the
mock galaxies.

\subsubsection{The mock samples}

We construct two main suites of mock catalogues; one is drawn from the
SCDM simulation, the other from the $\Lambda$CDM simulation. Each of
them contains 500 1.2-Jy mock surveys, that feature 5083 galaxies
on 87.6 per cent of the sky, flux limited at 1.2-Jy and volume limited
at a depth of 239.5$\lu$.

These samples are used to develop the statistical methodology for the
comparison of the 1.2-Jy data to the N-body models. The large number
of mock catalogues is necessary to determine the statistical
distribution of the genus reliably. 
We restrict ourselves to the SCDM and $\Lambda$CDM simulations 
because we do not
expect the genus of such a sparse sample as 1.2-Jy to be able
to discriminate between $\Lambda$CDM, $\tau$CDM, and OCDM.

For the adaptive smoothing technique that we outline below, 
we fill the masked regions of the sky with 
a small set of fake galaxies.
For simplicity, we construct for each catalogue 733 points with 
random angular positions 
in the masked regions, and with a redshift distribution sampled from
the catalogue itself.

To examine a number of systematic effects
we construct additional mock catalogues.
For example, we use 100 mock catalogues in redshift space to assess
the influence of redshift space distortion on the genus.
We further compute additional catalogues that are full-sky 1.2-Jy
surveys, i.e. which have no angular mask. These are used to test the
influence of the mask.
Finally, we compute a number of different survey realizations for
fixed observer positions in order to separately determine the relative
importance of sampling noise and cosmic variance.

\section{Smoothing techniques}

We now outline our procedures to construct smoothed 
density fields from
the 1.2-Jy redshift survey and the mock catalogues. These fields are
then used as input to the genus computation.

\subsection{Fixed smoothing}

Assuming a universal luminosity function an unbiased estimate of 
the galaxy density field can
be obtained by weighting the discrete point distribution $m(\vec{r})$ 
of the
observed galaxies with the inverse of the selection function $S(r)$:
\be
\rho(\vec{r}) \propto \frac{m(\vec{r})}{S(r)}.
\label{EQ0}
\ee
We obtain an estimate of the density field smoothed on some scale
$\lambda$ by convolving with a filter $W(\vec{r})$, which we 
choose as the Gaussian of equation (\ref{EQ1}).

However, due to the absence of galaxies in the regions 
of the angular mask, the density would be systematically 
underestimated at locations close to unobserved patches 
of the sky if the smoothing were just done by a straightforward
use of the kernel of equation (\ref{EQ1}).
In order to avoid this problem we employ the ratio method proposed by Melott \&
Dominik \shortcite{Me93}, who 
have shown in a systematic study that a smoothing according to 
\be
\hat{\rho}(\vec{r})=\frac{\int W(\vec{r}-\vec{r}') \rho(\vec{r}')
\,\dd\vec{r}'}
{\int W(\vec{r}-\vec{r}'') M(\vec{r}'')
\,\dd\vec{r}''},
\label{EQ2}
\ee
leads to the smallest loss or distortion of topological information
compared to a number of alternative schemes that treat the
mask differently. Here $M(\vec{r})$ is a mask field 
defined to be equal to 0 for
$\vec{r}$ lying behind the angular mask and to be 1 otherwise. 
For this choice the 
denominator of equation (\ref{EQ2}) essentially renormalizes the
smoothing kernel to the survey volume visible from the reference point
$\vec{r}$.

In the actual 
computation of the genus curve we only 
use the volume with $M(\vec{r})=1$ which is not hidden by the
mask. Additionally, we restrict the genus computation to a sphere
carved out of the smoothed density field. Note that there is no boundary
smoothing effect due to the outer surface of this sphere since we also
include the sources outside this final region in the smoothing
process.

We compute the convolutions that appear in the numerator and denominator of
equation (\ref{EQ2}) with the help of a Fast Fourier Transform (FFT) on a
 $128^3$ mesh. We choose a grid size of $b=\lambda/8$, which
ensures that the genus is free of finite mesh size
effects \cite{Ha86}, as we will demonstrate below.
The final depth $R_{\rm max}$ of the 
density field we use for the topological analysis
is always small enough to avoid wrap around effects due to the periodic
FFT smoothing.

\subsection{Adaptive smoothing}

As we saw in the analysis of the \virgo simulations, the smoothing
scale is limited by the 
poor
sampling of underdense regions.
To make use of the 
additional resolution in high density regions 
a variable
smoothing length can be employed.
We have already implemented
an adaptive 
smoothing scheme 
for the fully sampled N-body simulations.
We now define such a scheme for the analysis of flux
limited redshift surveys as well. Again, the hope is
that the overall effect of adaptive smoothing
is an increase of the number of structure elements visible in a given
density field.

\subsubsection{Spherically symmetric kernel}

We start by making the kernel a function of position, i.e. the
smoothed density field is computed as
\be
\hat{\rho}(\vec{r})=\int \rho(\vec{r}') W(\vec{r}-\vec{r}';\vec{r}')
\dd \vec{r}'.
\ee
This definition corresponds to the {\it scatter} approach of SPH.
Since this scheme strictly conserves
`mass' (if $W$ is properly normalized), it seems more natural to use 
for a small number of tracer particles than the {\it gather}
formulation, which we employed for the densely sampled N-body simulations.

In a first variant of the adaptive smoothing, we stay with a spherical
Gaussian and allow only the smoothing scale to vary with the local
density.
Later we will generalize the technique to triaxial Gaussians.

For a given survey volume $V_{\rm s}$ and 
a prescribed smoothing scale $\lambda_0$
we first compute the average mass $M_0$ in a Gaussian sphere of radius
$\lambda_0$,
that is
\be
M_0\equiv \frac{\pi^{\frac{3}{2}}\lambda^3}{V_{\rm s}} 
\int_{V_{\rm s}} \rho(\vec{r})\dd\vec{r}
=\pi^{\frac{3}{2}}\lambda_0^3,
\label{Eqas}
\ee
where the last equality 
holds (at least on average) 
due to our normalization of the selection function.

We then compute for every galaxy site $\vec{r}_i$ 
an individual smoothing radius
$\lambda_i$ such that 
\be
\int \rho(\vec{r}) \exp\left[ -\frac{(\vec{r}-\vec{r}_i)^2}{\lambda_i^2}\right]=M_0.
\ee
This definition implies that $\lambda_i$ will be smaller than $\lambda_0$
if the density around $\vec{r}_i$ is higher than the mean, and
it will be larger if the local density is lower than the mean.

\subsubsection{Triaxial kernel}

Up to now we have only varied the volume of the kernels. In an attempt
to improve the flexibility of the smoothing we may also allow the shape
of the kernel to vary. 
For this purpose we adopt
triaxial Gaussians
\be
W(\vec{x};\vec{r})=\frac{1}{\pi^{\frac{3}{2}} ({\rm det}
\bld{\Lambda})^{\frac{1}{2}}}\exp\left(-\vec{x}^{T}{\bld\Lambda}^{-1}\vec{x}\right)
\ee
as kernels, where the quadratic form ${\bld\Lambda}$ is a function of
$\vec{r}$.
We now need to specify the matrices ${\bld\Lambda}_i$ for every particle.
For simplicity, we set ${\bld\Lambda}_i$
proportional to the local moment of inertia tensor around the site
$\vec{r}_i$, i.e.
\be
{\bld\Lambda}_i\propto \int (\vec{r}-\vec{r}_i) 
(\vec{r}-\vec{r}_i)^{T} \rho(\vec{r})
\exp\left[-\frac{(\vec{r}-\vec{r}_i)^2}{\lambda^2_i} \right]
\dd \vec{r},
\ee
and we keep the original smoothing volume fixed by requiring
$({\rm det}{\bld\Lambda}_i)^{\frac{1}{2}}=\lambda_i^3$. 
Once the matrices ${\bld\Lambda}_i$ are determined we `only' need to compute 
\be
\hat{\rho}(\vec{r})=\sum_i m_i W(\vec{r}-\vec{r}_i;{\bld\Lambda}_i)
\ee
with $m_i=[S(\vec{r}_i)]^{-1}$
to arrive at the triaxially smoothed density field.

Because the adaptive smoothing has to be carried out in real space 
it requires much more CPU time than the fixed smoothing described
above. For this reason we construct the density fields only in spheres
of radius $R_{\rm max}$, each of them inscribed in a $64^3$ mesh.
For simplicity we deal with the mask by filling 
the empty region with fake galaxies as described above.
Note that the 
masked volume is not used in the calculation of the genus curve.

The adaptive smoothing schemes
we employ here are by no means unique and many other variants 
are conceivable. In fact, we have tried a
number of alternatives ourselves.  However, the suggested procedure is
fairly intuitive and allows a study of the effects of volume and shape
adaptivity separately.
Currently we see no alternative to 
Monte-Carlo experiments in determining
the performance and properties
of such adaptive smoothing techniques.

\section{Systematic effects}
	
Before turning to the results for the 1.2-Jy
survey and the \virgo mock catalogues, we first examine various
systematic effects
that can affect the genus measurements. In particular, we are
interested in the amount of bias present in the genus 
curve derived from a 1.2-Jy-like sample compared to the genus of the
underlying density field.

\subsection{Finite volume effect}

Recently, Protogeros \& Weinberg \shortcite{Pr97} have claimed that the
genus curve is severely biased high if it is computed from subvolumes
carved out of large N-body simulations or out of Gaussian random
fields. In particular, for a roughly spherical volume of radius
$R_{\rm max}=60\lu$ and smoothing scale $\lambda=10\lu$ 
they detected an increase of the genus amplitude by a
whopping factor of 1.5-2. 
They further found that this volume effect 
becomes somewhat smaller for larger volumes, yet it seems to be present
independently of the shape of the survey volume.

Their result is puzzling since the volumes used in their analysis 
seem large enough 
that any correlation of the curvature between adjacent points on the 
surface of the volume is expected to average out, i.e. the mean curvature of
the subvolume should allow an unbiased estimate of the genus of the
full volume.

\begin{figure}
\bc
\resizebox{8cm}{!}{\includegraphics{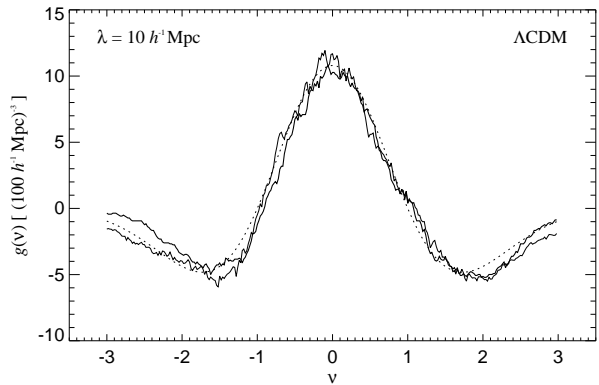}}
\caption{Finite volume effect. The thin line shows the average genus
curve of 10 spherical subvolumes of radius $60\lu$ extracted from the
SCDM simulation, while the thick line gives the genus of the full
simulation volume. The genus curves have been computed for a
resolution of $\Delta\nu=0.01$.
Comparing the two curves we find no evidence for
the strong amplitude bias claimed by Protogeros \& Weinberg \protect\shortcite{Pr97}. 
They claim an increase of
the genus amplitude for the finite volume curves by a factor of
1.5-2.0. 
\label{fig15}
}
\ec
\end{figure}

We have searched for this volume effect ourselves, but we could not
find it. For example, in 
figure \ref{fig15} we compare the average genus curve for 10
spherical subvolumes of radius $R_{\rm max}=60\lu$ 
extracted from the SCDM simulation with the genus of
the full simulation box. Reassuringly, the average genus curve
of the
subvolumes
appears to be largely 
unbiased; there is no trace of the claimed strong 
finite volume effect.

\subsection{Mask effects}

We dealt with the limited sky coverage of the 1.2-Jy survey 
by employing the ratio
smoothing method of Melott \& Dominik \shortcite{Me93}. Here we look for systematic
biases inflicted on the genus curve because of that.
For this purpose we have constructed an additional  
suite of 100 mock catalogues
for the SCDM model. These catalogues 
have full sky coverage, with 5816 galaxies to a depth of 239.5$\lu$.

\begin{figure}
\bc
\resizebox{8cm}{!}{\includegraphics{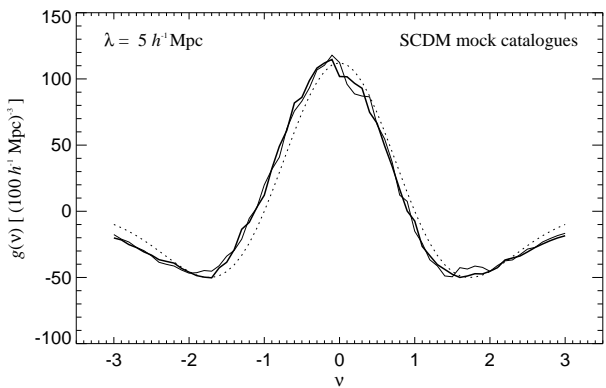}}
\caption{Effects due to the angular mask. The thick curve shows the
average genus curve for 100 SCDM mock catalogues before they are
subjected to the 1.2-Jy mask, i.e.\ their genus is computed for an
all-sky 
density field. The thin line shows the genus for the masked mock
catalogues, where the genus computation involves the ratio method.
The curves shown are for fixed smoothing with $\lambda=5\lu$. Larger
smoothing scales or the adaptive smoothing techniques result in
similar small effects.
\label{fig35}
}
\ec
\end{figure}

In figure \ref{fig35} 
we 
compare the
average genus curve of these full sky mock catalogues to the 
average curve  of the masked catalogues, where the computation
involves the ratio method.
The good agreement between the two results gives us confidence that
any bias due to our treatment of the mask is very small. 
The result
shown is for fixed smoothing with $\lambda=5\lu$, but we
observe  
a similar small influence
of the mask
for
other smoothing
scales and for the adaptive smoothing techniques.

\subsection{Genus curves}

\begin{figure}
\bc
\resizebox{8cm}{!}{\includegraphics{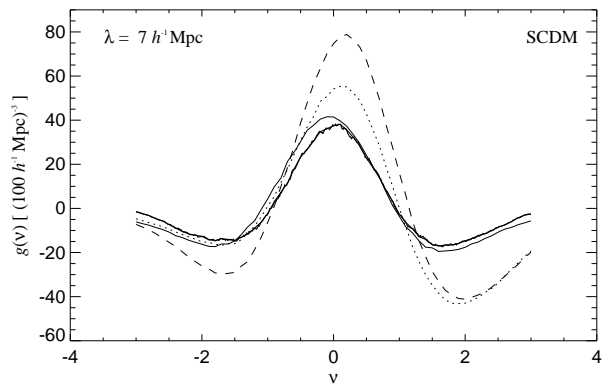}}
\caption{Genus curves for the SCDM simulation, smoothed at $7\lu$ with
different smoothing techniques. The thick solid line gives the result
for the fully sampled SCDM simulation, while the thin line gives the
corresponding average genus curve with a fixed smoothing kernel 
for the SCDM mock catalogues.
The dotted line shows the result for the spherical adaptive smoothing,
and the dashed curve is for triaxial adaptive smoothing.
\label{figGcomp}
}
\ec
\end{figure}

Figure \ref{figGcomp}
shows the average genus curve of the SCDM suite of 
mock catalogues compared to the genus of the fully sampled simulation.
Obviously the genus curve derived from the mock catalogues is
biased high compared to full sampling. This can be explained by the
influence of
shot noise on the genus amplitude, as demonstrated by
Canavezes et~al. \shortcite{Ca97}.
Furthermore, the average curve exhibits a slight {\it meatball} bias, that is a
shift of the peak to the left. This effect can also be understood
as a discreteness error that results from the sparse sampling of the
density field.

Also shown in figure \ref{figGcomp} 
are the mean genus curves resulting from the
adaptive smoothing techniques we tried. Both of them show a strong
enhancement of the genus amplitude. 
The spherically symmetric smoothing also results in a strong
asymmetry between the minima of the genus curve, with the minimum on the
positive side being lowered strongly, while the minimum on the negative
side remains practically at the level obtained for the fixed smoothing
technique. Note that the genus curves for the different smoothing
techniques are expected to be different since the smoothing
procedures are sensitive to different properties of the density field.

Interestingly, the triaxial smoothing technique gives a genus 
very close to the spherical smoothing around the minimum on the
positive side, while it increases the genus signal for smaller values
of $\nu$. Presumably the amplitude of the minimum on the positive side
is just set by the number density of individual, isolated
clumps. Apparently, the triaxial technique does not change the topology
of these isolated regions.

For fixed smoothing, 
the measured genus is biased compared to the fully
sampled simulations, albeit by a small amount.
Instead of trying to correct for it as attempted by Canavezes et~al. 
\shortcite{Ca97}, we compare the 1.2-Jy
measurements only to the results obtained for the mock catalogues, and
not to the fully sampled simulations themselves. This is a viable 
procedure, even if strong biases
are present.

\section{Statistical methodology}

Ultimately we want to use the genus statistic to compare theory with
observation, i.e. to quantify the level of agreement of
the 1.2-Jy survey with the \virgo N-body models.
A prerequisite to derive 
formal exclusion levels is a precise
understanding of the distribution of errors
of
the genus measurement.

Perhaps the most general method to assess 
random and systematic errors 
is to work with
ensembles of mock galaxy surveys that mimic the statistical properties
of the actual observed data set.
When the mock catalogues and the redshift survey 
are analysed in the same way, systematic biases 
that might be present in the adopted analysis                      
enter in the same way.

\subsection{Distribution of errors \label{sec71}}

For a suite of $n$ mock catalogues 
we measure the genus curve 
at $k$ values $\nu_1,\nu_2,\ldots,\nu_k$ of the filling
factor. In what follows, we compute the genus with spacing $\Delta\nu=0.1$
in the range $[-3.0,3.0]$, i.e. at $k=61$ positions.
We now use these measurements to estimate the distribution of
errors in the genus. 
The mean genus curve and its covariance matrix may be estimated as
\be
\ol{\vec{g}}=\frac{1}{n}\sum_{l=1}^{n}\vec{g}^{(l)}
\ee
and
\be
{\bld V}={\rm cov}(g_i,g_j)=\frac{1}{n-1}\sum_l
(\vec{g}^{(l)}-\ol{\vec{g}})
(\vec{g}^{(l)}-\ol{\vec{g}})^T ,
\label{chi2matrix}
\ee
where
$\vec{g}^{(l)}=(g^{(l)}_1,\ldots,g^{(l)}_k)$ denotes the measured 
genus curve 
for the catalogue $l$.

\begin{figure*}
\begin{minipage}{160mm}
\bc
\resizebox{5.2cm}{!}{\includegraphics{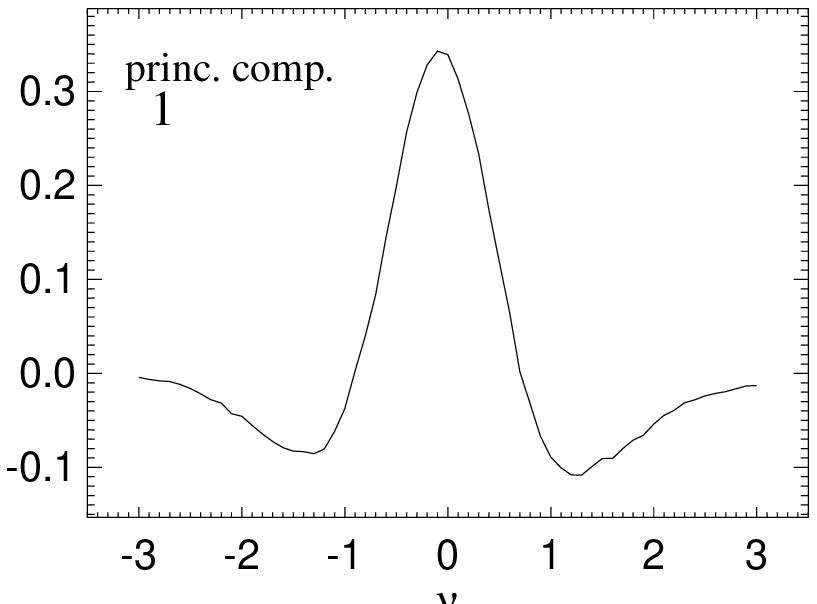}}
\resizebox{5.2cm}{!}{\includegraphics{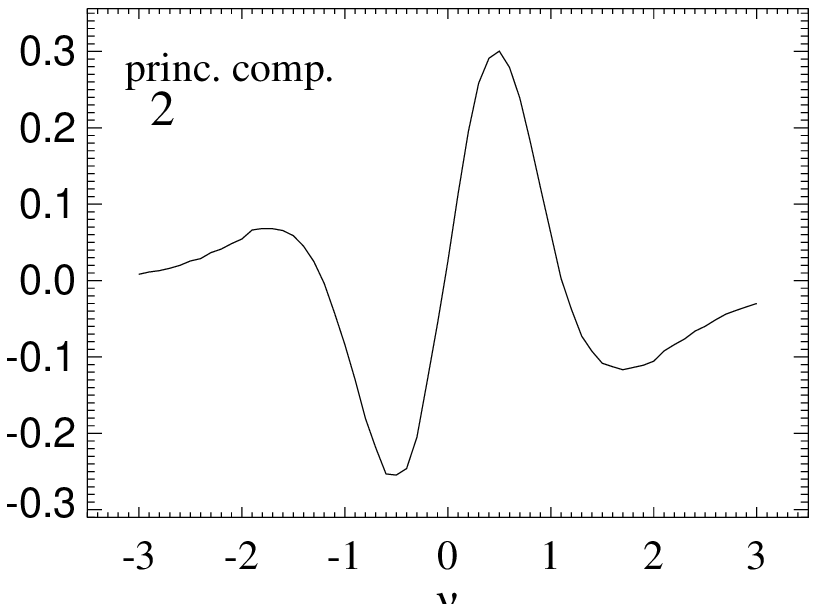}}
\resizebox{5.2cm}{!}{\includegraphics{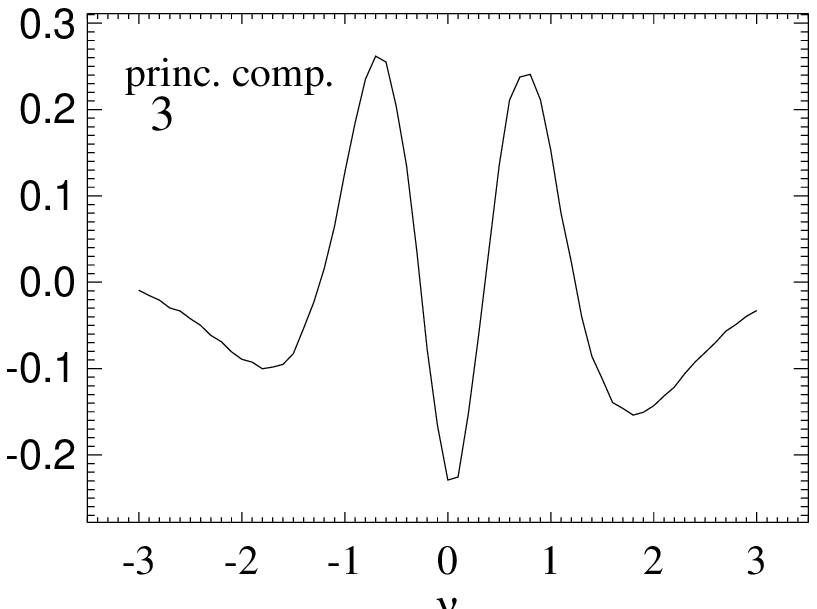}}

\resizebox{5.2cm}{!}{\includegraphics{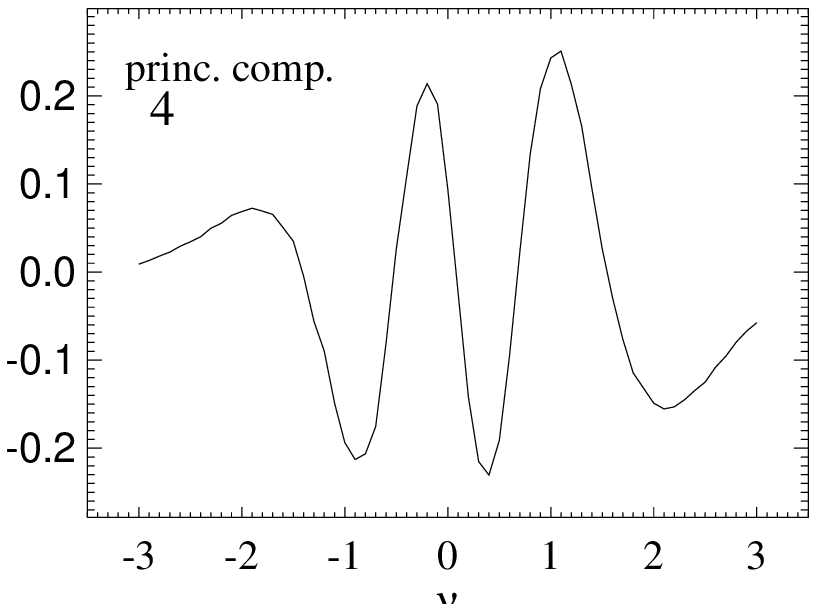}}
\resizebox{5.2cm}{!}{\includegraphics{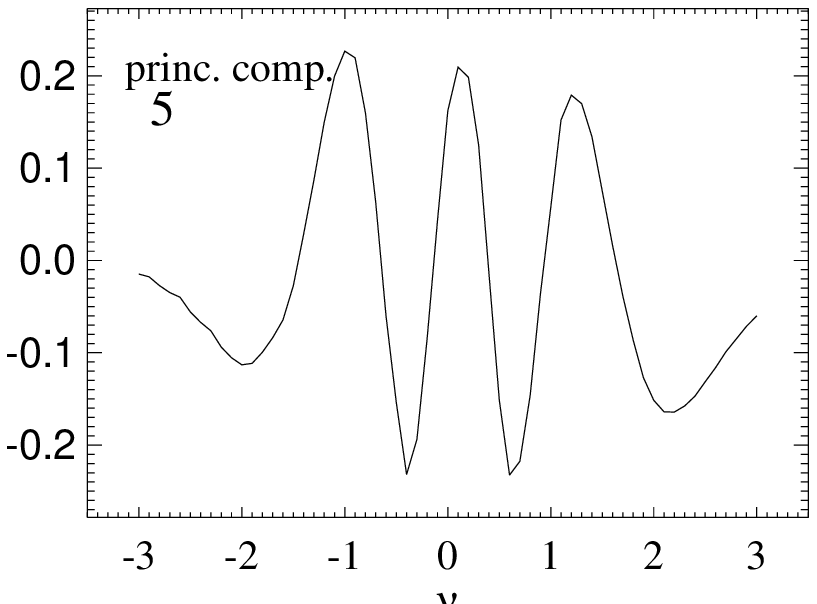}}
\resizebox{5.2cm}{!}{\includegraphics{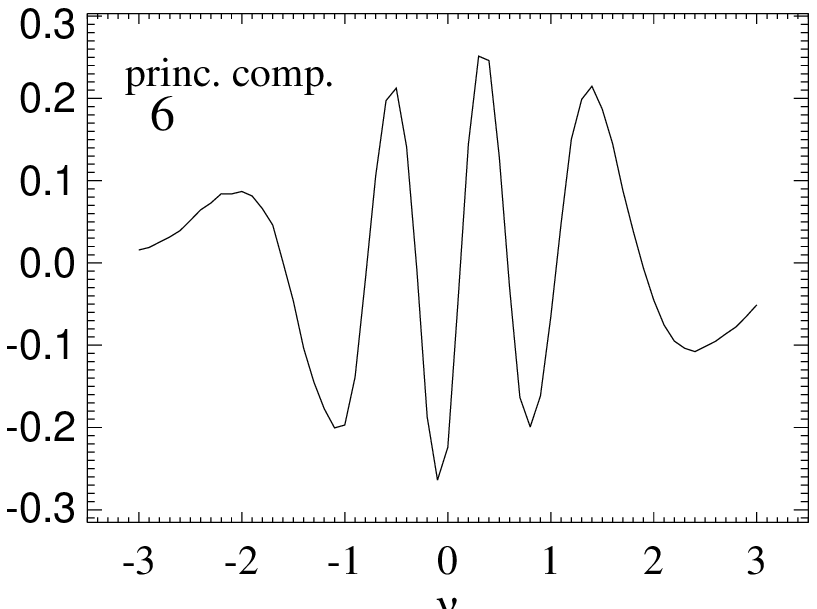}}
\caption{Principal components of the genus measurement.
The displayed 
curves are based on an average of covariance matrices corresponding to
different smoothing scales. The matrices have been scaled
to a common highest eigenvalue before the averaging. The resulting
curves are similar to the ones obtained for a single
covariance matrix, but they are smoother due to the reduction of noise
by the averaging. We use these curves as generic
principal components in the PCA of all examined samples.
Note that the
principal components are normalized to unity and mutually orthogonal
to each other.
\label{figPC}
}
\ec
\end{minipage}
\end{figure*}

Recently 
Protogeros \& Weinberg \shortcite{Pr97} conjectured that the 
distribution of errors is well described by a
multivariate Gaussian
\be
f(\vec{g})=\frac{1}{(2\pi)^\frac{k}{2} |{\rm
det}\bld{V}|^{\frac{1}{2}}}
\exp\left[-\frac{1}{2}(\vec{g}-\ol{\vec{g}})^T {\bld V}^{-1}
(\vec{g}-\ol{\vec{g}}) \right].
\ee
Then the quantity
\be
\chi^2(\vec{g})=(\vec{g}-\ol{\vec{g}})^T {\bld V}^{-1}
(\vec{g}-\ol{\vec{g}})
\label{AAA001}
\ee
would exhibit a $\chi^2$ distribution with $k$ degrees of freedom, 
and 
one could  
estimate the 
probability 
\be
p=1-
\frac{1}{2^{\frac{k}{2}}\Gamma\left(\frac{k}{2}\right)}\int_{\chi^2}^{\infty}x^{\frac{k}{2}-1}\e^{-\frac{x}{2}}
\dd x
\label{EQchi2pdf}
\ee
of finding a mock survey in the ensemble that differs from 
the mean of the mock
catalogues
by more than
a particular observation with $\chi^2=\chi^2(\vec{g}^{({\rm obs})})$.
In this way a formal
exclusion level could be derived.

This procedure seems attractive, yet we find 
that it can fail 
miserably
in
practice. As we show in the appendix, this failure is 
partly due to the fact
that
the errors are at best approximately distributed as a
multivariate Gaussian. 
More importantly,
we will use a principal components analysis (PCA)
to demonstrate that there
are only a small number of principal components that can be determined
with some confidence. 
The rest of them are dominated heavily by noise;
it is therefore not a good idea to invert the noisy $61\times 61$
covariance matrix of equation (\ref{chi2matrix}). Instead we will
regularize the problem by means of a PCA.
Note that a simple smoothing of the genus curve can remove some of
the noise. However, as a side effect this will make the covariance matrix
close to singular. Of course, this causes
trouble if one naively goes
ahead and tries to compute $V_{ij}^{-1}$, so a PCA cannot be avoided
in this way.

\subsection{Principal components analysis \label{sec72}}

The principal components analysis (PCA, Murtagh \& Heck 1987) 
is frequently applied in
astronomy 
to extract the most relevant features from data sets that
may be strongly contaminated by noise. 
Since PCA is a linear method it works best for uncorrelated noise, a
situation not really valid for the genus. However, we can still
expect that it allows the construction of a clean multivariate
analysis of the measured genus curves.

Each measured genus curve can be viewed as a point in a 61-dimensional
space.
The idea of PCA is to transform to a new set of coordinates which
correspond to the directions of maximum extension of the cloud of
measured genus points. These
principal axes are just the eigenvectors of the covariance
matrix (\ref{chi2matrix}). It is convenient to 
order these components in descending order of
their eigenvalues, i.e.\ the first principal component shows
the highest variance. 
Usually one then considers only the
first few principal components, 
which are the ones that describe the most prominent features of the
signal.
In this way the method 
allows to efficiently filter out the
noisy contributions to the signal and to concentrate on its essential
features.

The principal axes and their
eigenvalues can be conveniently found by a singular value
decomposition (SVD) of the covariance matrix. 
Having ordered the principal
components in descending order we  
consider only the first $m$ of them.
We can then construct a $m\times61$ matrix $\bld{P}$ that
contains the $m$ eigenvectors in its rows and that
projects a genus curve $\vec{g}$ onto new coordinates $\vec{h}=\bld{P}
\vec{g}$. One can also transform back to the original space, giving
rise to a PCA-filtered genus curve
\be
\vec{g}_{\rm PCA}(m)={\bld P}^T\bld{P}\vec{g}.
\ee

In figure \ref{figPC} we display the first six 
principal components of the genus measurement. 
Actually these
principal components have been derived from an
averaged 
covariance
matrix, obtained by adding up matrices corresponding to several
smoothing scales which we have scaled to a common highest eigenvalue.
The curves resulting from just one suite of genus curves look very
similar, although they are not quite so smooth. We employ the averaging
procedure to establish a generic set of smooth orthogonal principal
components that we subsequently apply to all the different samples on
an equal footing. Of course, 
depending on the covariance matrices used
in the averaging procedure the derived principal components may differ
slightly in detail. However, in all cases 
the first few
components (we will use $m=6$ of them) 
span very nearly the same region in the full 61-dimensional
space of measured genus curves. Hence the specific choice of 
covariance matrices is uncritical as long as all the information of these
principal components is used in a multivariate analysis.

Some of the principal components shown in figure \ref{figPC} 
are easy to interpret. The first
clearly measures the amplitude, 
while
the next two can be seen to be sensitive to a horizontal shift and a
broadening of the genus curve. Hence the first three principal
components
are similar in
meaning to the genus meta-statistics introduced by
Vogeley et~al. \shortcite{Vo94}.
However, here these measures are not postulated in an ad-hoc way, but
they suggest themselves naturally 
as the most relevant features of the measured genus curves.

\begin{figure}
\bc
\resizebox{8cm}{!}{\includegraphics{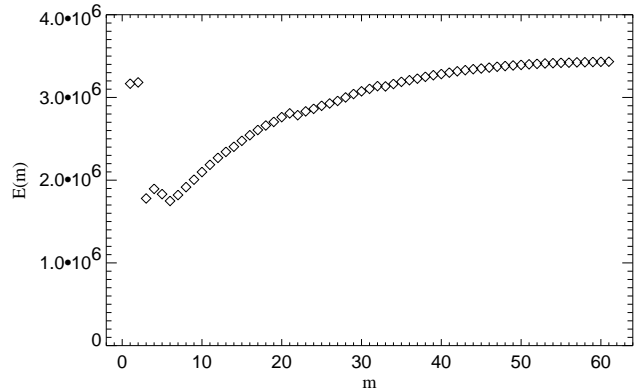}}
\caption{Global approximation error $E$
as a function of the number $m$ of included principal components.
The points shown are for the SCDM mock catalogues, smoothed with
$7\lu$. Other smoothing scales show similar results. For the adaptive
smoothing techniques the minimum occurs also at 
$m\approx 6$. However, the higher
components lead to a somewhat smaller increase of $E$.
\label{figPCAError}
}
\ec
\end{figure}

How many principal components should we reasonably take? To
answer this question we examine a global error function
\be
E(m)=\sum_l ( \vec{g}_{\rm {PCA}}^{(l)}(m) - \ol{\vec{g}})^2 ,
\ee
where the sum is over the mock ensemble and  $\vec{g}_{\rm
{PCA}}^{(l)}(m)$
denotes the genus curve of catalogue $l$, treated with a PCA filter of
order $m$. Figure \ref{figPCAError} shows a minimum of $E(m)$, when $m\approx 6$
principal components are used. Higher principal components lead to 
additional noise, in the sense that 
the reconstructed genus curves differ more and more from the ensemble
average.
For this reason we
will restrict ourselves to the first 6 principal components.

\begin{figure}
\bc
\resizebox{8cm}{!}{\includegraphics{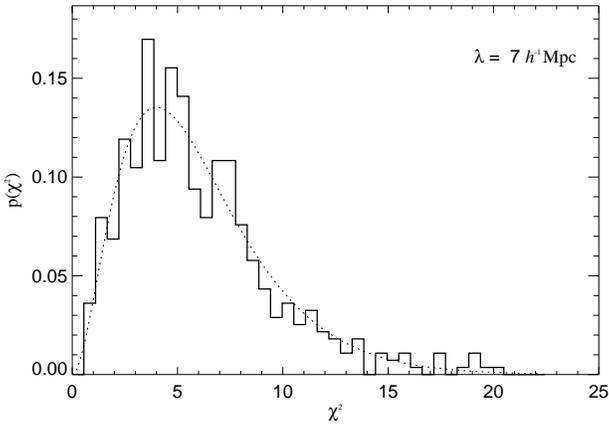}}
\caption{Distribution of $\chi^2(\vec{h})$, when only the first 6 principal
components
are included. The variance of the distribution is 12.66, close to the
expected value of 12.0 for a $\chi^2$ distribution with $m=6$ degrees of
freedom (dotted line).
The data shown here are for the SCDM catalogues with fixed smoothing 
of $7\lu$. All the other cases show equally good agreement
with the theoretical $\chi^2$ distribution. Note that the test curve 
$\vec{g}=2\ol{\vec{g}}$, which demonstrates the failure of the method
applied in the appendix (figure \ref{figchi2dist}), 
here gives a $\chi^2$ of 32.4, indicating
correctly a terrible fit.
\label{chi26}
}
\ec
\end{figure}

\begin{figure*}
\begin{minipage}{160mm}
\bc
\resizebox{5.2cm}{!}{\includegraphics{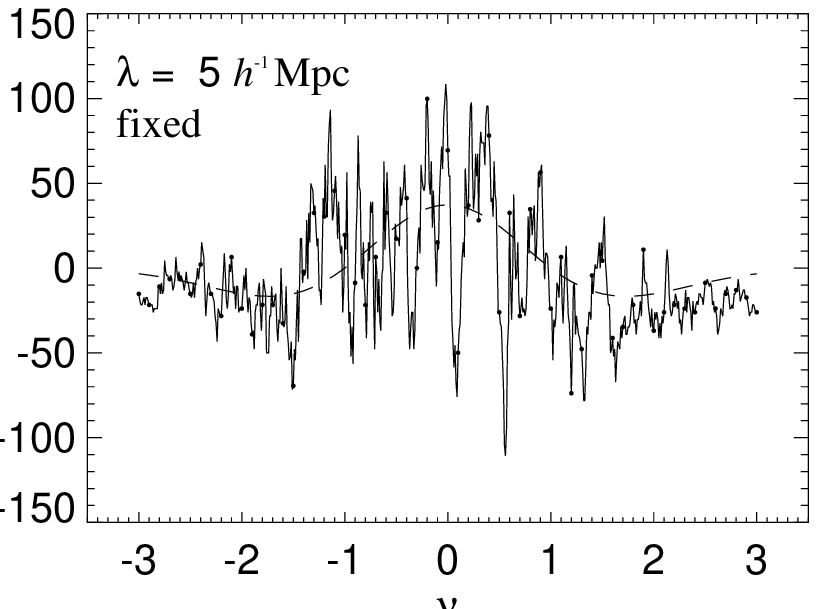}}
\resizebox{5.2cm}{!}{\includegraphics{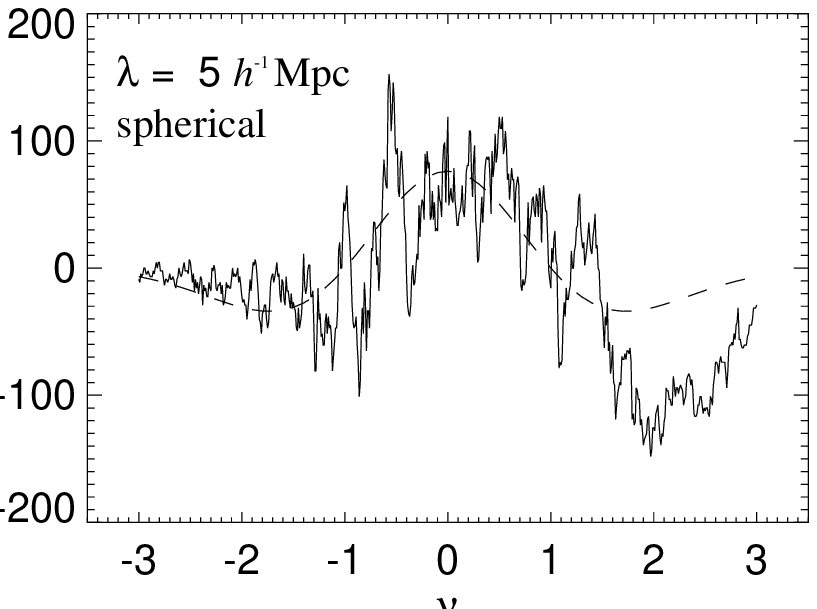}}
\resizebox{5.2cm}{!}{\includegraphics{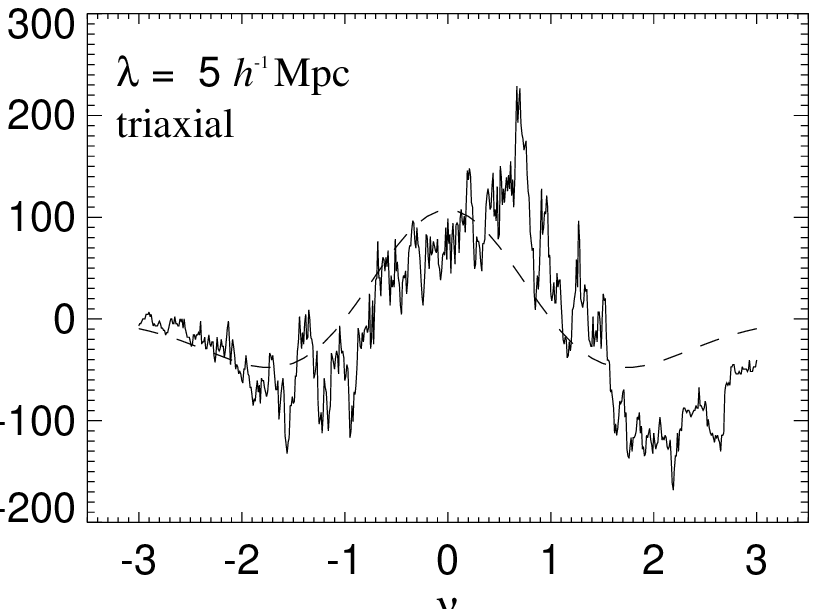}}
\resizebox{5.2cm}{!}{\includegraphics{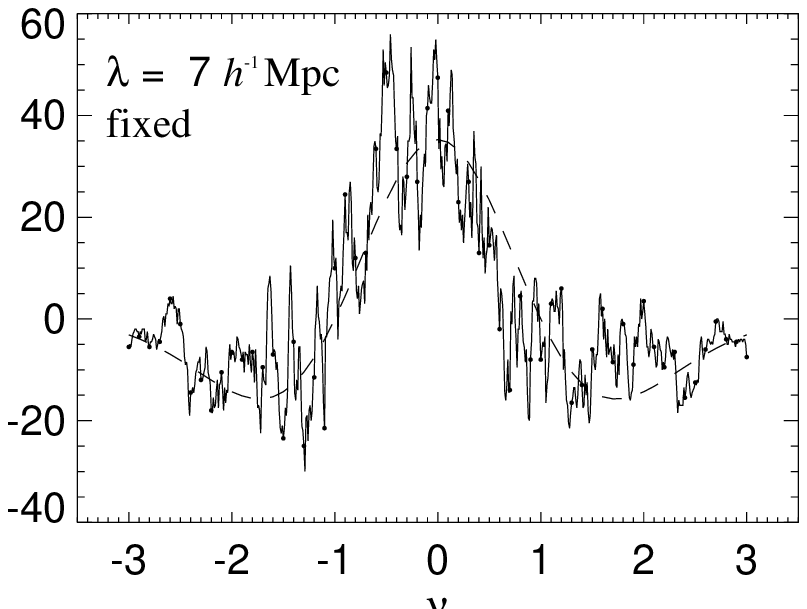}}
\resizebox{5.2cm}{!}{\includegraphics{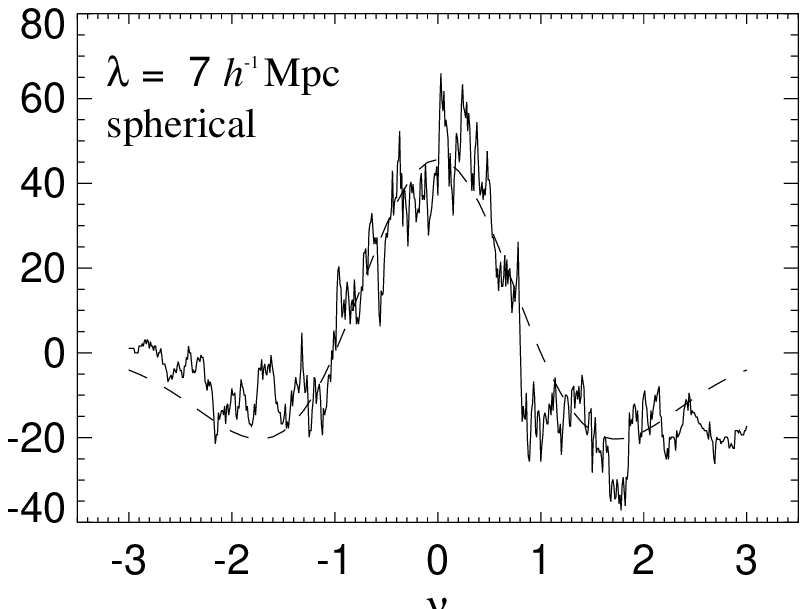}}
\resizebox{5.2cm}{!}{\includegraphics{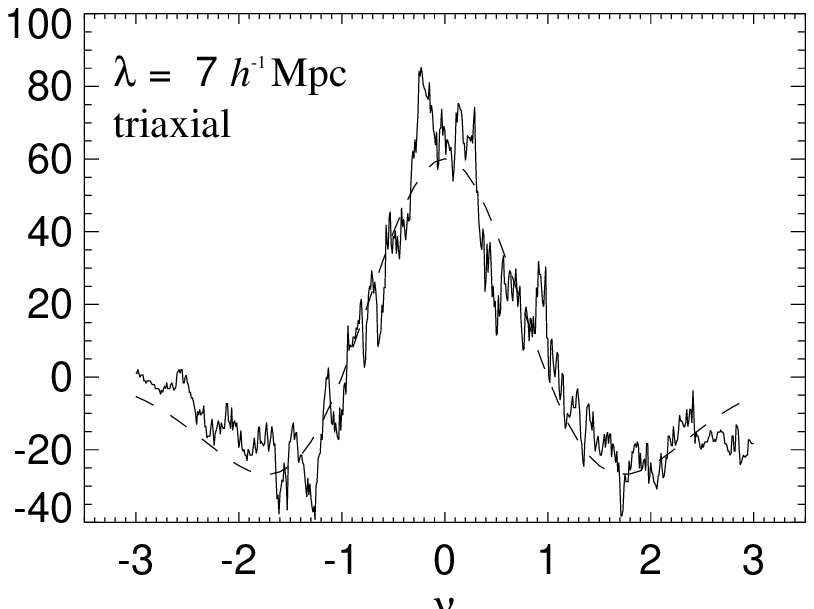}}
\resizebox{5.2cm}{!}{\includegraphics{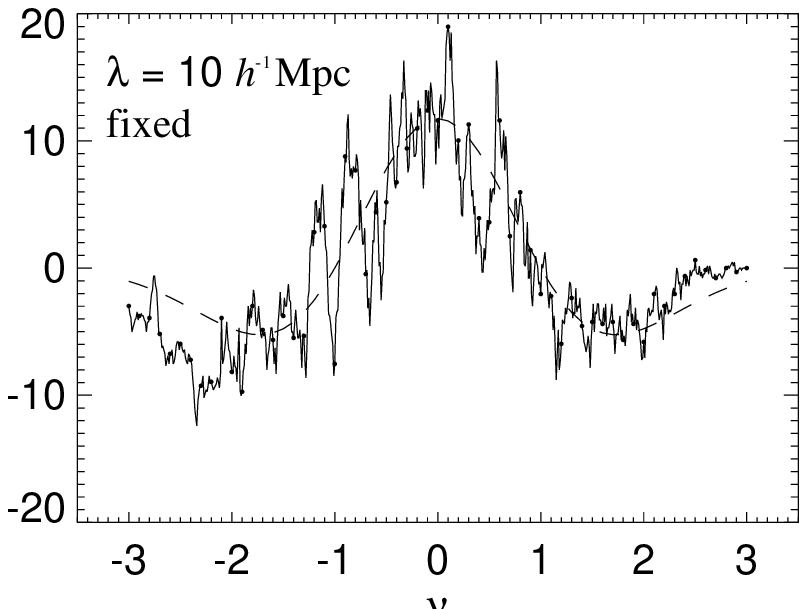}}
\resizebox{5.2cm}{!}{\includegraphics{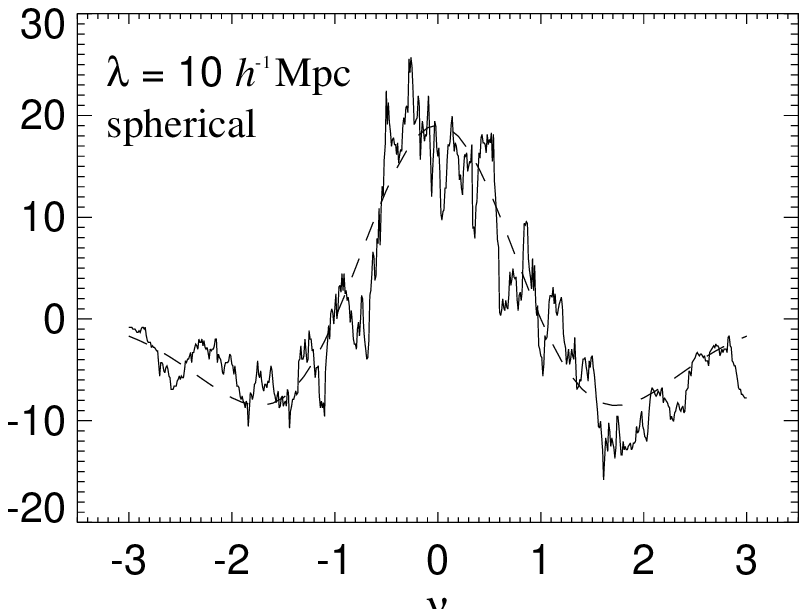}}
\resizebox{5.2cm}{!}{\includegraphics{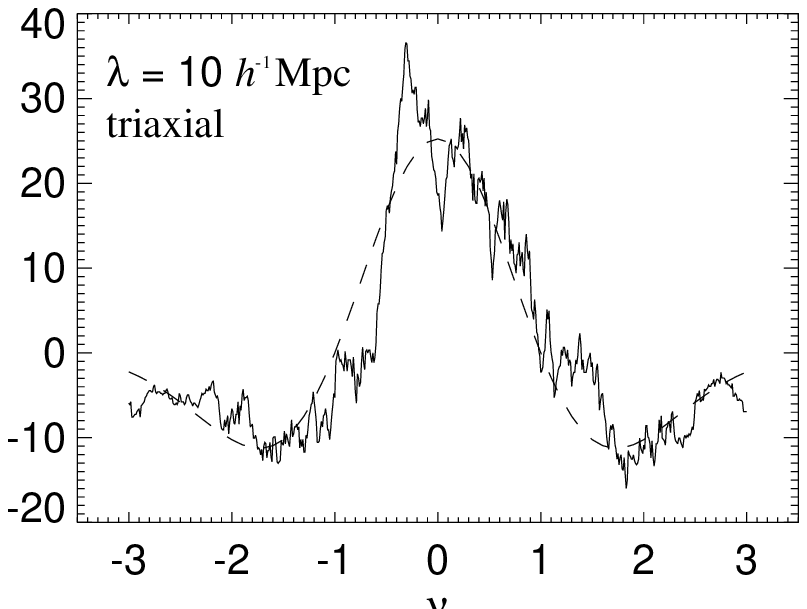}}
\resizebox{5.2cm}{!}{\includegraphics{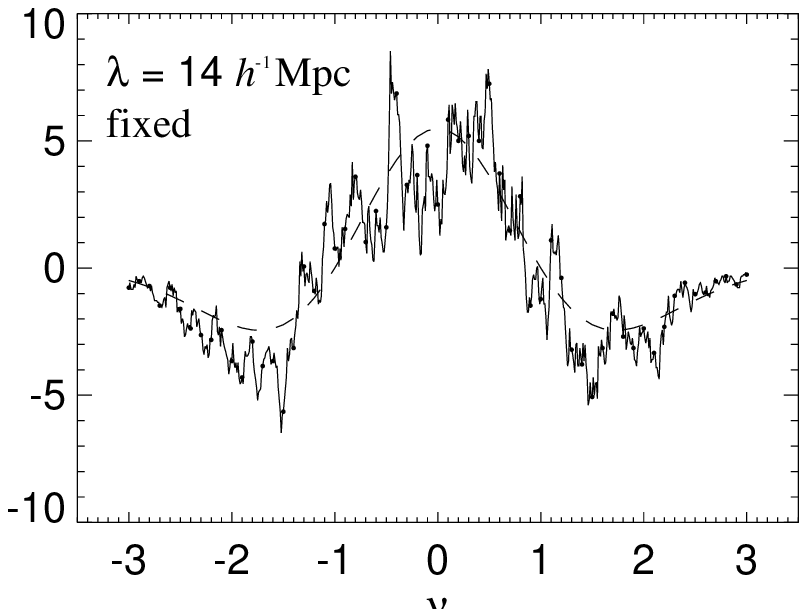}}
\resizebox{5.2cm}{!}{\includegraphics{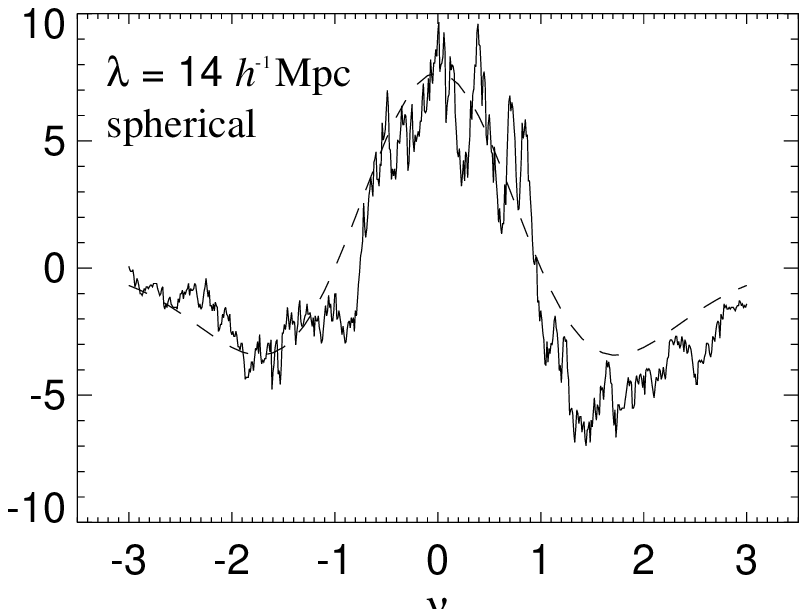}}
\resizebox{5.2cm}{!}{\includegraphics{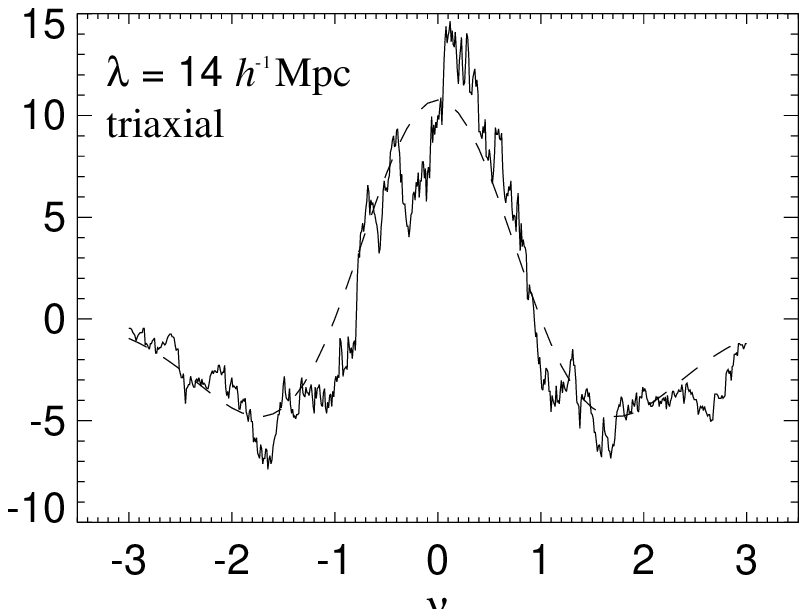}}
\resizebox{5.2cm}{!}{\includegraphics{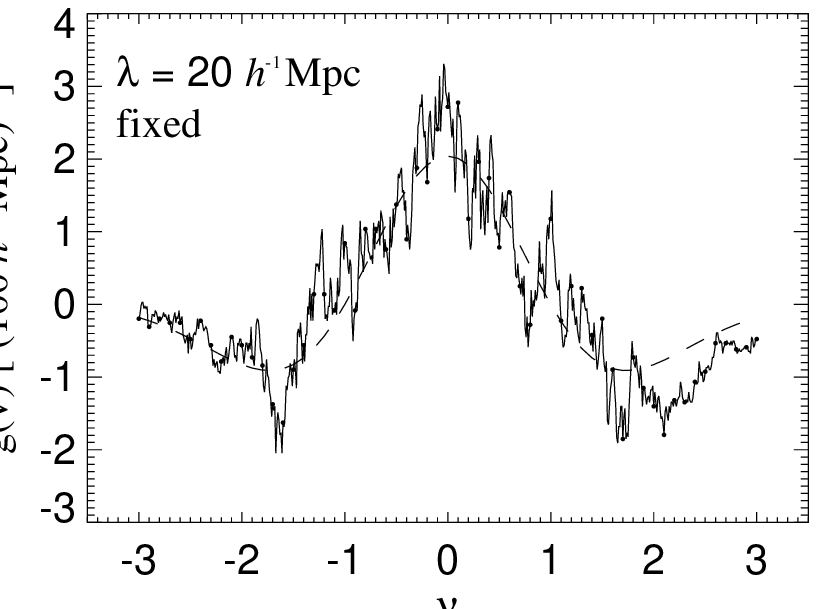}}
\resizebox{5.2cm}{!}{\includegraphics{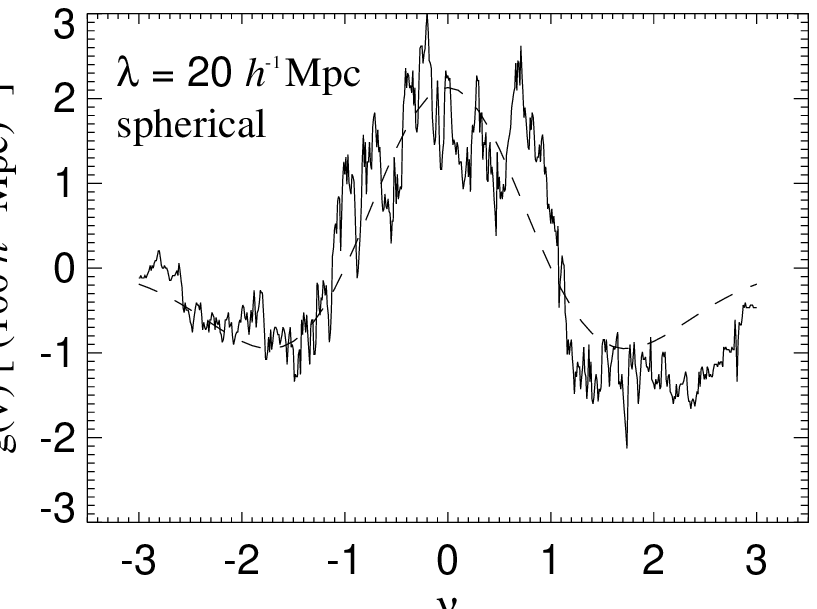}}
\resizebox{5.2cm}{!}{\includegraphics{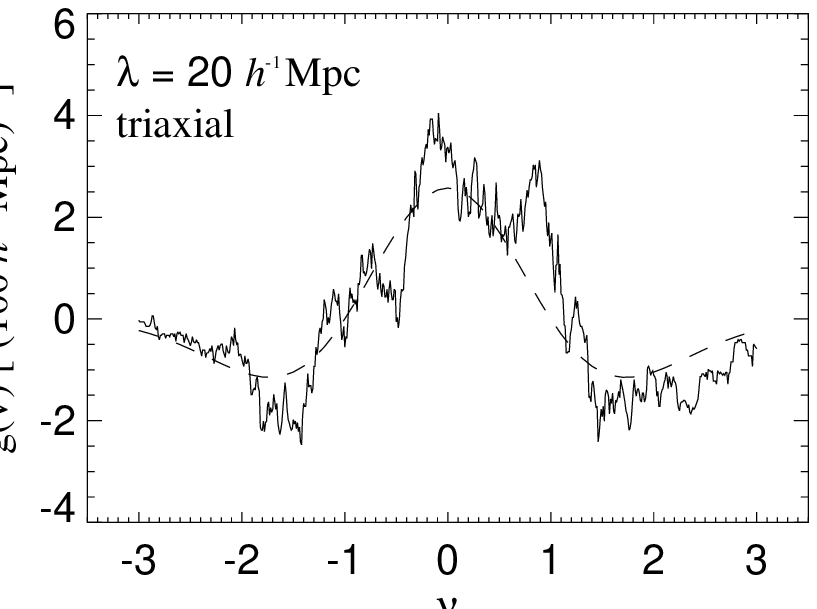}}
\caption{Genus curves of the 1.2-Jy redshift survey of \iras galaxies.
The left column shows high resolution genus curves obtained with fixed
smoothing, while the middle and right columns give the 
results for the two adaptive smoothing schemes. In each row the same
smoothing scale is considered, rising from $5\lu$ to $20\lu$. Note however that the vertical scale is
different for individual panels. The adaptively smoothed genus curves
exhibit a considerably higher genus amplitude. The dashed lines are
random phase genus curves with amplitudes determined from the first
principal component of the measured genus curves. In the case of
adaptive smoothing, $\lambda$ sets the mass per smoothing volume via
equation (\ref{Eqas}). 
\label{fig7}
}
\ec
\end{minipage}
\end{figure*}

As is seen in figure \ref{figPC} the first six principal components
give only small weight on the genus at low and high values of
$\nu$, where we know that the errors are not normally distributed. 
Because of that, one can hope that 
the distribution of the principal components $\vec{h}$  is
consistent with a Gaussian, at least if we only
use the first $m=6$ of them. A Kolmogorov-Smirnov test (appendix A) 
reveals that this is indeed the
case. 
We will therefore attempt an ordinary multivariate analysis
based on the 
fixed set of principal components
displayed in figure \ref{figPC}.
Since we keep them fixed for
all suites of mock catalogues, we do not expect that the correlations
between the different principal components will exactly vanish, although
they should be small.
For this reason we compute the covariance matrix ${\bld C}$ of the
measured values of $\vec{h}$, and consider the statistic
\be
\chi^2(\vec{h})=(\vec{h}-\ol{\vec{h}})^T {\bld C}^{-1}
(\vec{h}-\ol{\vec{h}}) .
\label{AAA002}
\ee

In figure \ref{chi26} we show an example of the distribution of this
quantity. It is indeed very well fit by a $\chi^2$ distribution with
$m=6$ degrees of freedom. Note that for this statistic the
pathological result obtained in appendix A for the curve
$\vec{g}=2\ol{\vec{g}}$ is gone; here 
this curve gives $\chi^2=32.4$, implying a terrible disagreement, as it
should.

\section{Comparison of the 1.2-Jy survey with \virgo}
	
\subsection{1.2-Jy genus curves}

\begin{figure}
\bc
\resizebox{8cm}{!}{\includegraphics{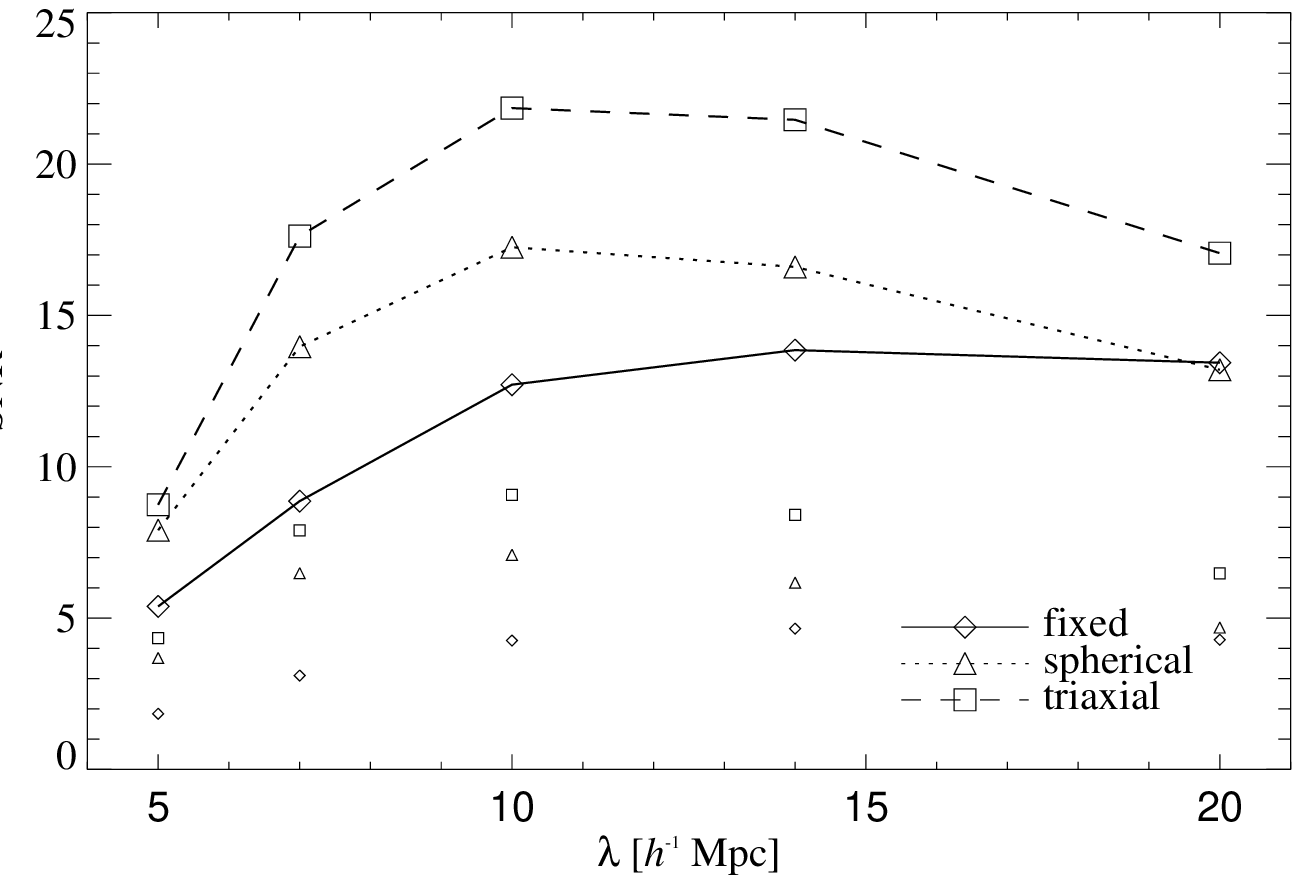}}
\caption{Average signal-to-noise ratio
of the measured genus curves for different smoothing techniques as a
function of smoothing scale.
The results are for the SCDM mock catalogues and PCA-filtered ($m=6$)
genus curves. The small symbols show the corresponding SNR for the raw
genus curves. The $\Lambda$CDM catalogues give a similar result.
\label{figsnr}
}
\ec
\end{figure}

\begin{figure*}
\begin{minipage}{160mm}
\bc
\resizebox{5.2cm}{!}{\includegraphics{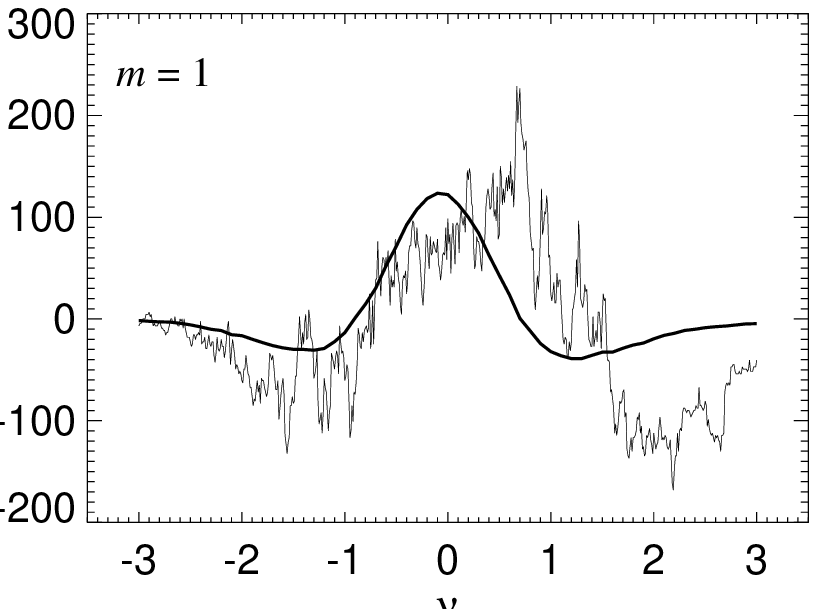}}
\resizebox{5.2cm}{!}{\includegraphics{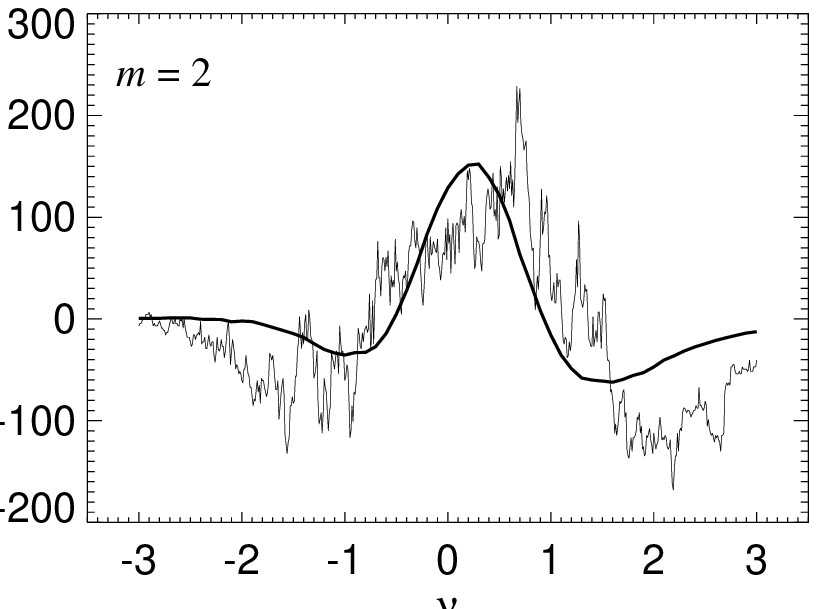}}
\resizebox{5.2cm}{!}{\includegraphics{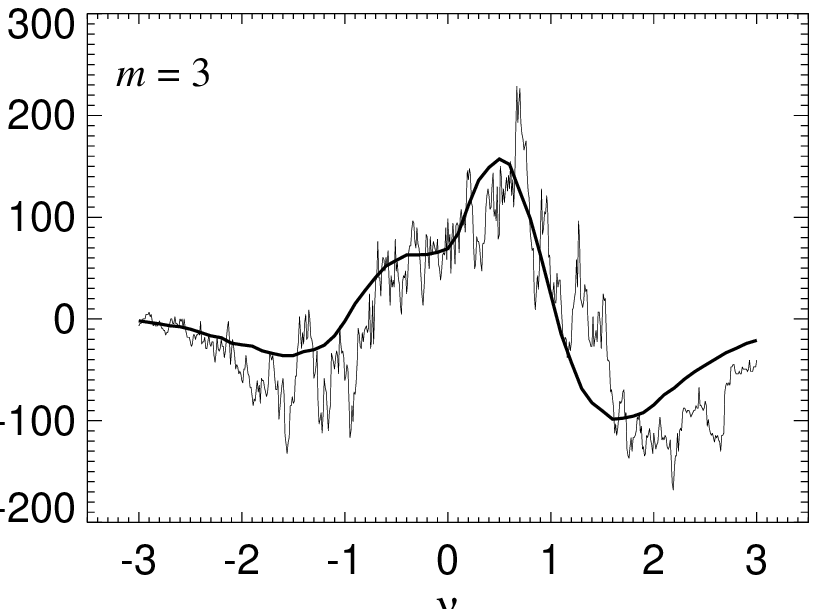}}
\resizebox{5.2cm}{!}{\includegraphics{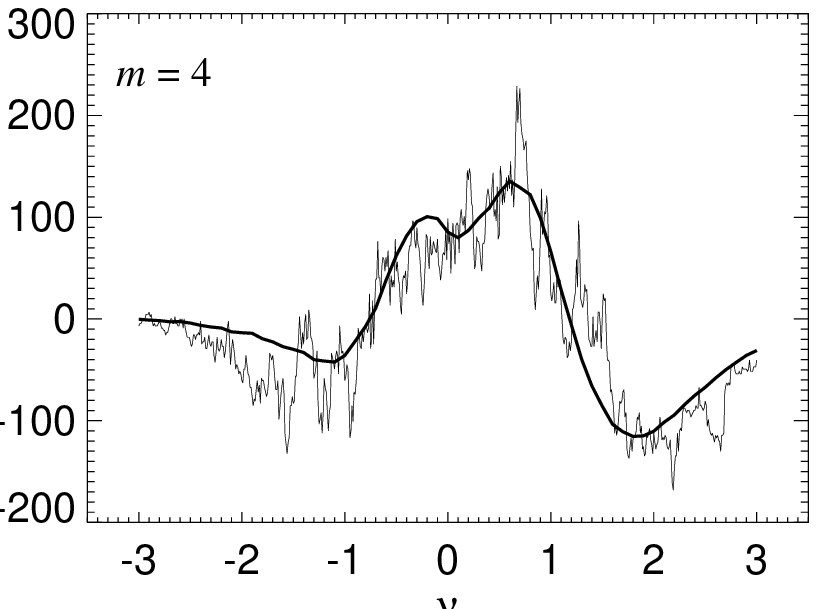}}
\resizebox{5.2cm}{!}{\includegraphics{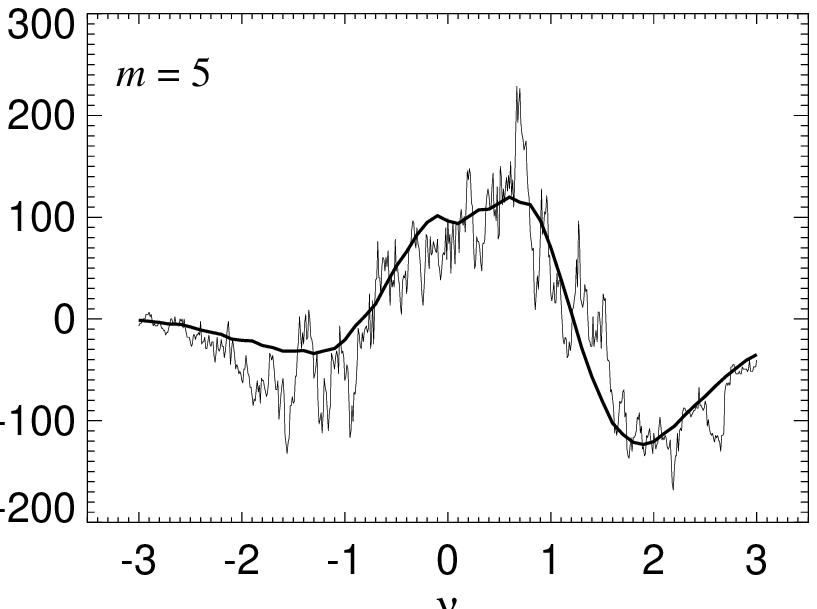}}
\resizebox{5.2cm}{!}{\includegraphics{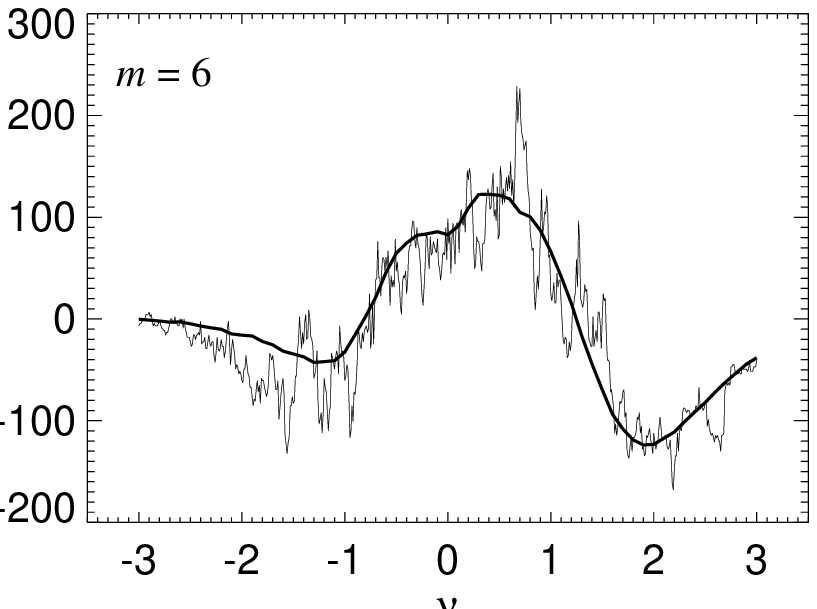}}
\caption{Example for a PCA-filtered genus curve
as a function of the number $m$ of included principal components.
In each panel the thin
line shows the 
1.2-Jy genus curve resulting for triaxial adaptive smoothing with 
$\lambda=5\lu$. The thick line gives the corresponding PCA-filtered
curve for the specified order $m$ of the filter.
\label{figexpca}
}
\ec
\end{minipage}
\end{figure*}

Figure \ref{fig7} shows the genus curves we obtain for the 1.2-Jy
redshift
survey with
three different smoothing schemes. The left column displays results
for fixed smoothing, while 
the middle and right columns give the
corresponding curves
for the two variants of adaptive smoothing. The genus curves
in the figure are computed with a uniform
spacing of 
$\Delta \nu=0.01$ 
in the interval $-3\le\nu\le3$. Note
that the genus curves that are used for the comparison with the \virgo
simulations
represent only a subsampling of this data with a spacing of $\Delta\nu=0.1$.

The large amount of jitter in the curves indicates that there is
substantial noise in the measurement. Interestingly, the adaptively
smoothed genus curves show less scatter and seem to be smoother than
with fixed smoothing. Clearly, the triaxial smoothing technique 
produces the smoothest genus curves. It might therefore be hoped that
this reflects an increase of the number of resolved structural
elements 
which in turn allows a 
measurement of the 
genus
curve with less
error. 

In order to test this expectation we define an average signal-to-noise
ratio (SNR)
\be
\left<{\rm SNR}\right>
\equiv\left<\frac{\ol{g}^2}{\sigma^2}\right>
=\frac{1}{k}\sum_{i=1}^{k}\frac{\ol{g}_i^2}{\sigma^2(g_i)}
\ee
of the measured genus curves. 
Here $\ol{g}_i$ denotes the mean genus density at each of the $k=61$
measured positions, and $\sigma^2({g}_i)$ is the variance of the
corresponding measurements.

In figure \ref{figsnr} we show the SNR for the three different smoothing
techniques as a function of smoothing scale. As expected, the adaptive
smoothing techniques can significantly improve the SNR compared to the
fixed smoothing scheme. Also, the triaxial method clearly performs better
than the spherical adaptive smoothing.
Note that figure \ref{figsnr} 
shows the SNR for genus curves that are filtered with a PCA-filter of
order $m=6$. The SNR for the raw genus curves is also plotted; it is
generally smaller, showing that the PCA-filter can indeed reduce the
contamination with noise.

A further illustration of the effect of the PCA-filtering is given in 
figure \ref{figexpca}. Here we show the 1.2-Jy genus curve resulting
for the triaxial adaptive smoothing technique with $\lambda=5\lu$
together with PCA-filtered versions of it. As the number $m$ of
included principal components is increased, more features of the
measured genus curve can be faithfully reproduced by the filtered
curve. However, adding in more principal components will eventually only
lead to a reproduction of the noise inherent in the measured curve.

\subsection{Comparison with \virgo}

\begin{figure*}
\begin{minipage}{160mm}
\bc
\resizebox{5.2cm}{!}{\includegraphics{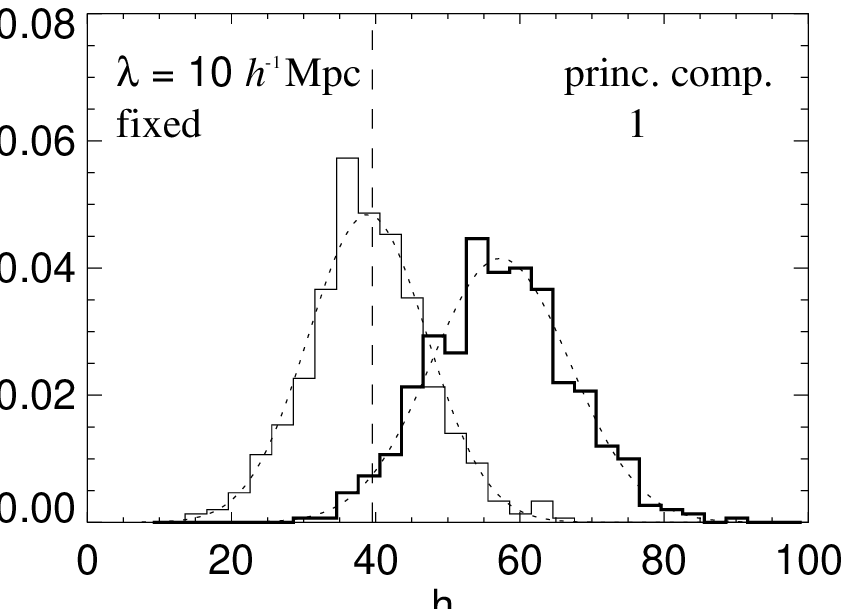}}
\resizebox{5.2cm}{!}{\includegraphics{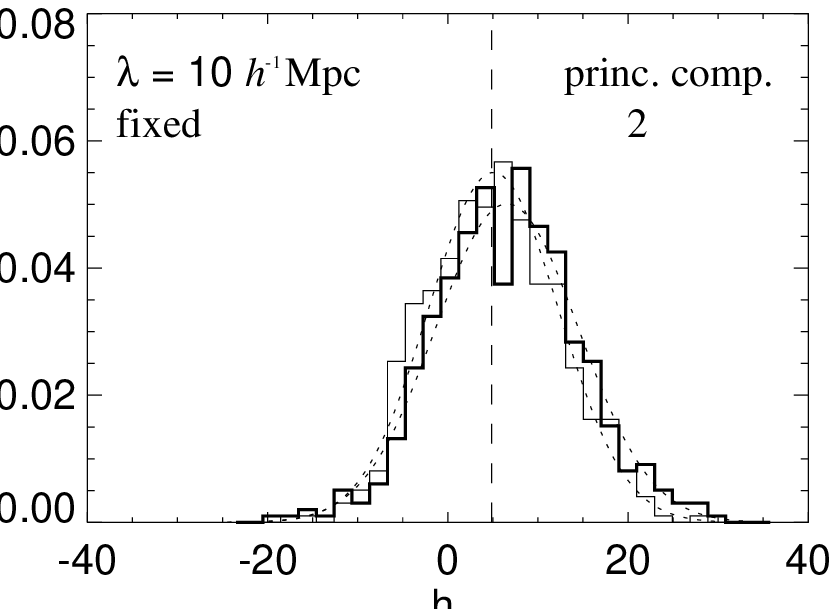}}
\resizebox{5.2cm}{!}{\includegraphics{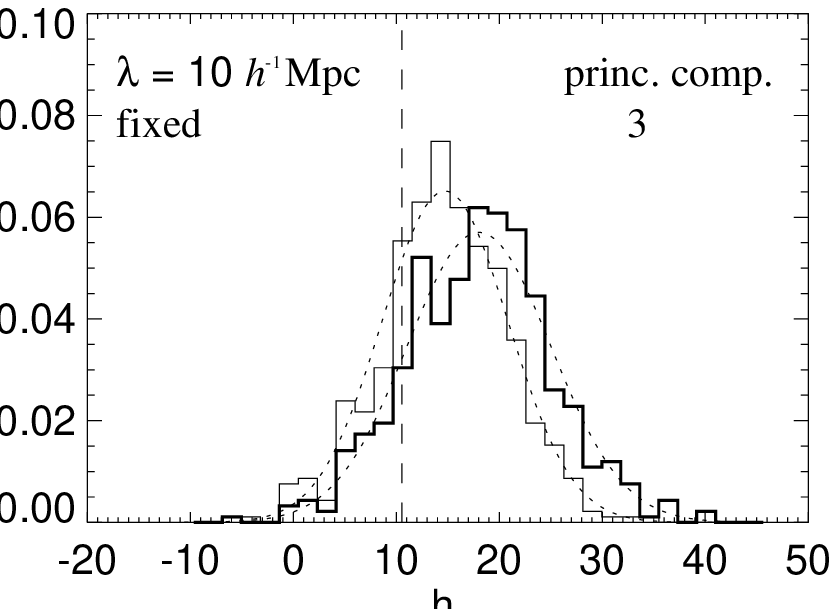}}
\vspace{0.35cm}\resizebox{5.2cm}{!}{\includegraphics{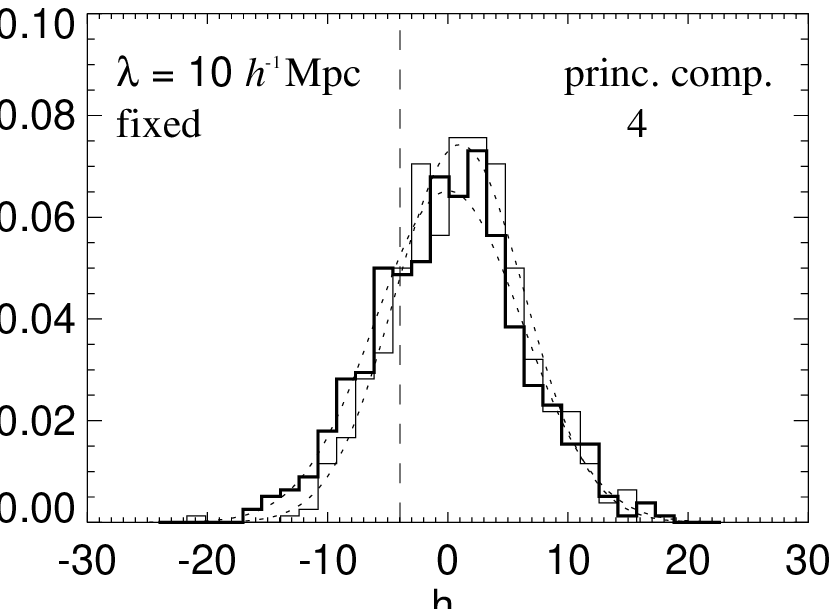}}
\resizebox{5.2cm}{!}{\includegraphics{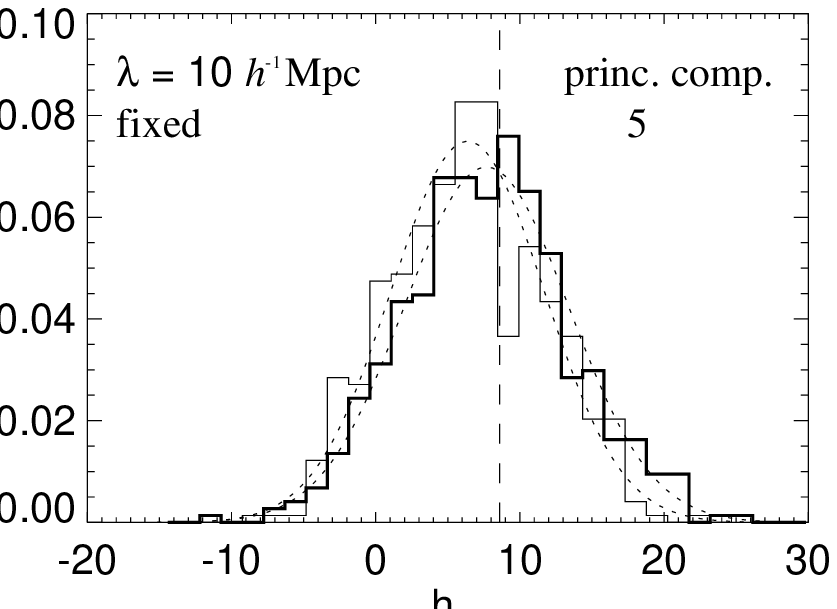}}
\resizebox{5.2cm}{!}{\includegraphics{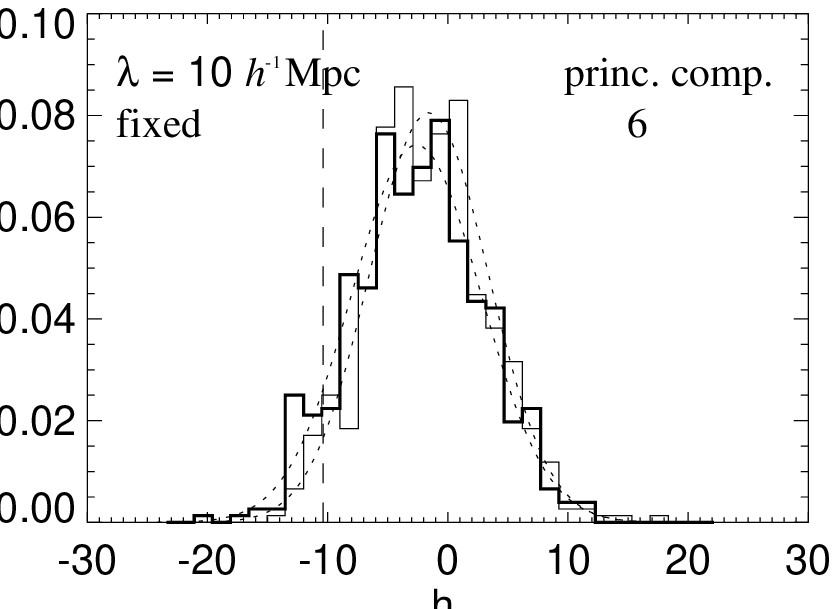}}

\resizebox{5.2cm}{!}{\includegraphics{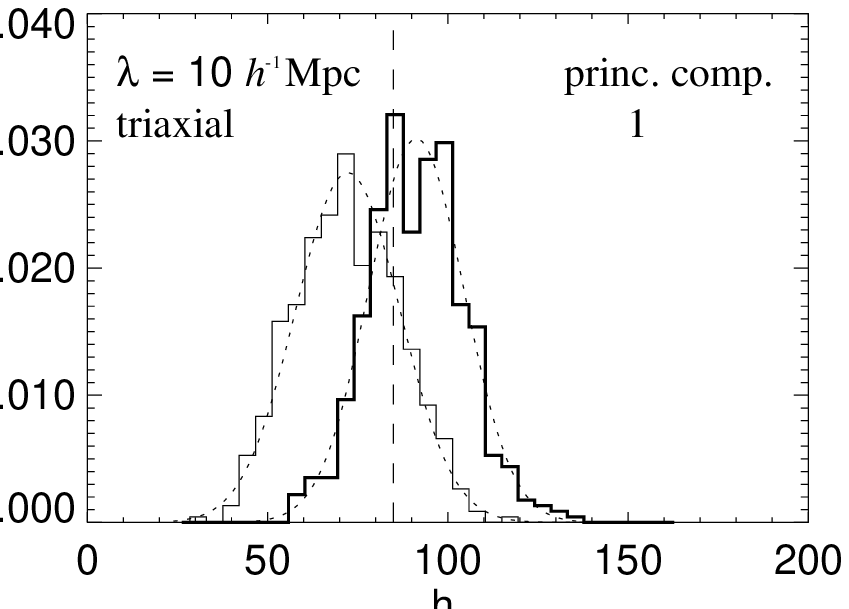}}
\resizebox{5.2cm}{!}{\includegraphics{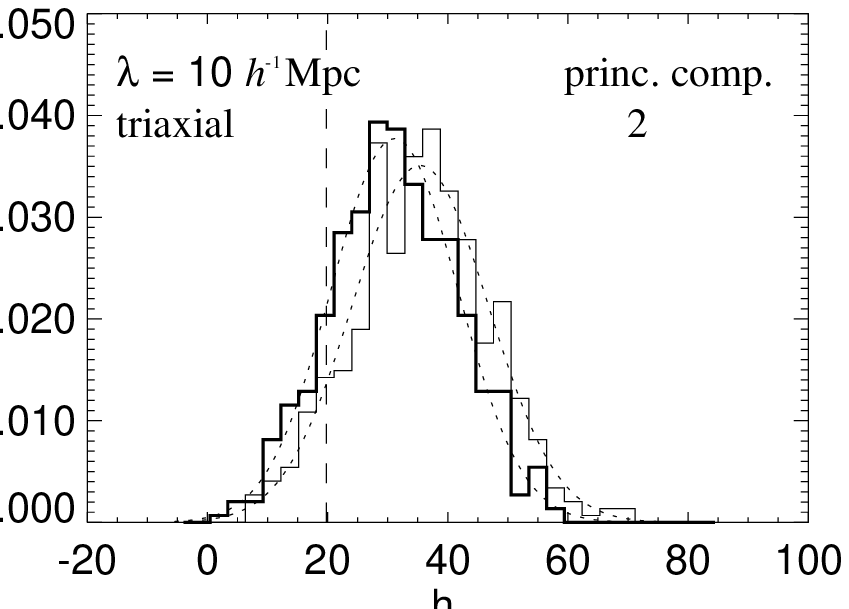}}
\resizebox{5.2cm}{!}{\includegraphics{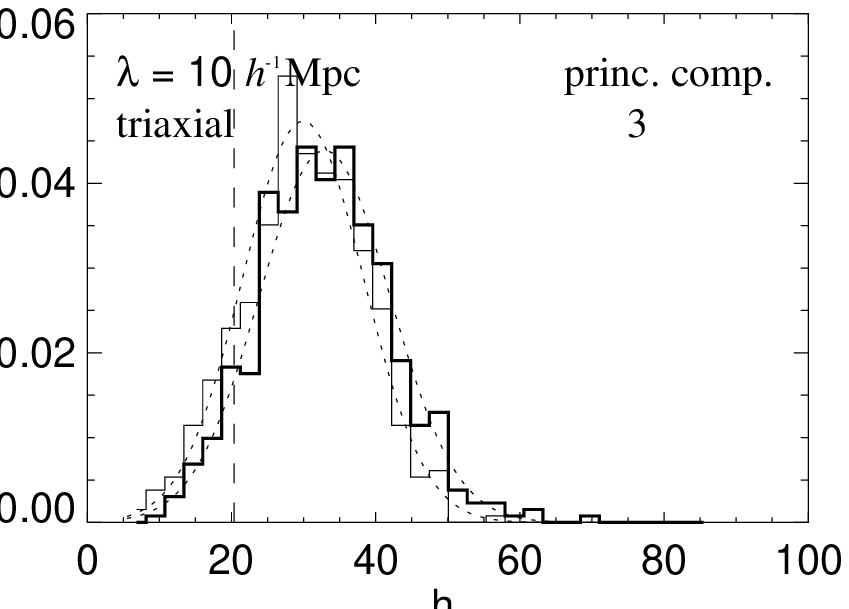}}
\resizebox{5.2cm}{!}{\includegraphics{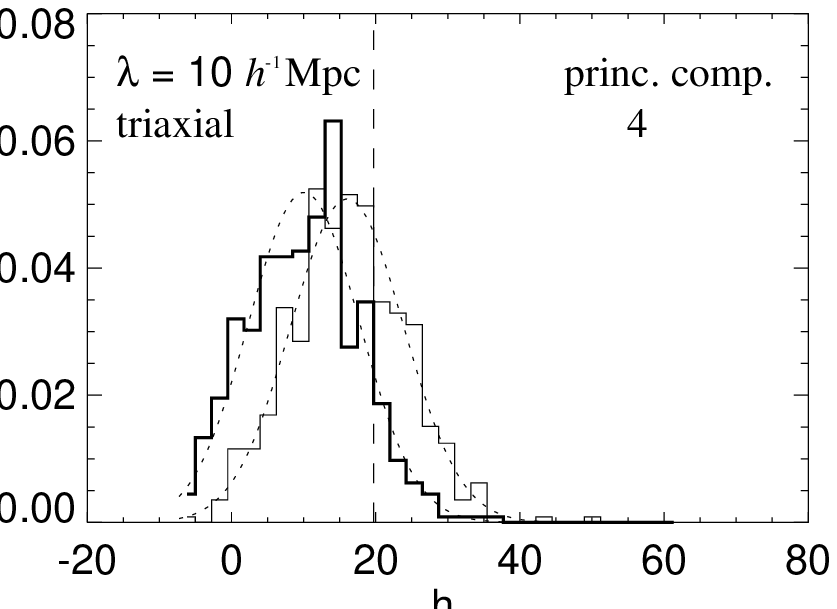}}
\resizebox{5.2cm}{!}{\includegraphics{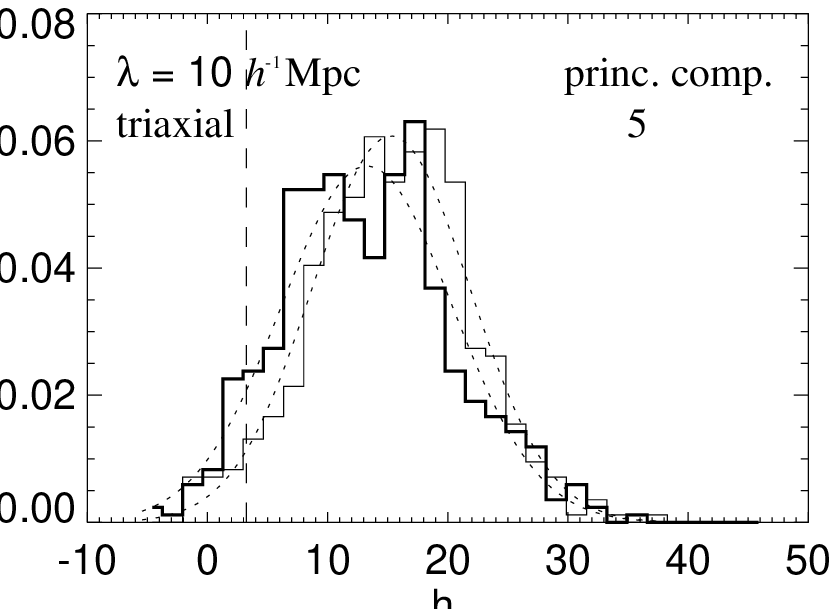}}
\resizebox{5.2cm}{!}{\includegraphics{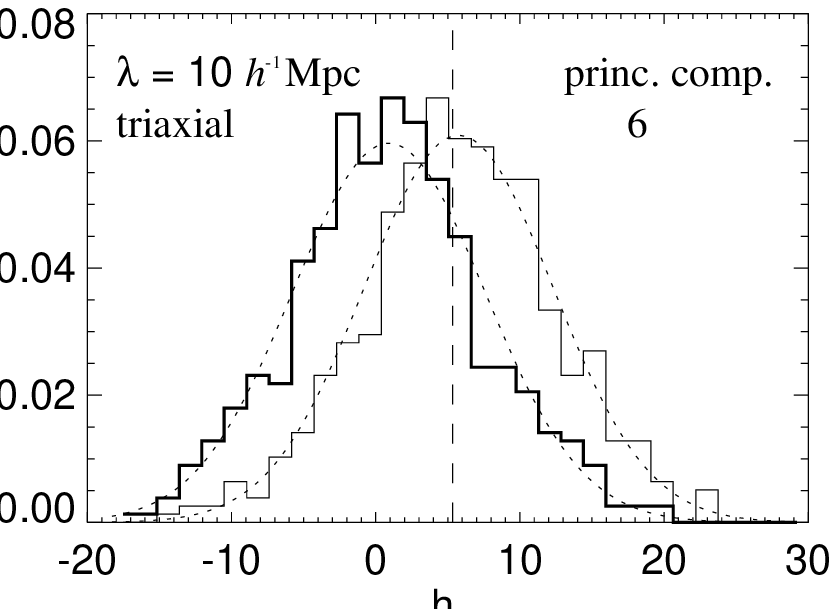}}
\caption{Distribution of the measured principal components for the
SCDM (thick histogram) and 
$\Lambda$CDM (thin histogram) models. In each panel, the dashed vertical line
marks the result for the 1.2-Jy survey. The dotted lines are normal
distributions with the mean values and variances of the
individual histograms. The data shown compares the distribution of the
principal components for the fixed and triaxial adaptive smoothing
techniques with $\lambda=10\lu$. Other smoothing scales show
qualitatively similar results.
\label{figPChisto}
}
\ec
\end{minipage}
\end{figure*}

As outlined in section 7 we base our statistical comparison between
\virgo and the 1.2-Jy survey on 6 principal
components of the measured genus curves. The amplitudes of the 
principal components
are just the projections of the measured genus on the curves shown in figure
\ref{figPC}.

In figure \ref{figPChisto} we show examples of
the distribution of the principal components for fixed smoothing with
$\lambda=10\lu$.
In each panel, 
the solid histogram shows the result for the SCDM suite of mock
catalogues, while the thin line gives the $\Lambda$CDM distribution. The
dotted lines show the normal distributions that result from the mean
and the variances of these histograms, and
the dashed vertical line marks the measurement for the 1.2-Jy survey
itself. The panels of the bottom half
of figure \ref{figPChisto} show the corresponding plots for the
triaxial adaptive smoothing technique.

It is immediately apparent, that the discrimination between SCDM and 
$\Lambda$CDM
is poor with a dataset as sparse as 1.2-Jy. The  amplitude $h_1$
proves to be the most sensitive measure of differences between the
models. However, the large 
uncertainties in the genus measurement  
manifest themselves in broad, overlapping distributions of the principal components.
Nevertheless we can still compute the exclusion probability of the
1.2-Jy measurements for each of the models.
For this purpose we calculate the $\chi^2(\vec{h})$ 
value of the 1.2-Jy
measurement with respect to the distributions of the SCDM and
$\Lambda$CDM models, and the probability $p$ that a mock catalogue
shows a higher $\chi^2$ than the 1.2-Jy survey. Here we assume that the principal
components are multivariate normally distributed.

Additionally we give the
likelihood ratio
\be
{\cal L} \left({\scriptstyle\frac{\Lambda{\rm CDM}}{ {\rm SCDM}}}\right) 
=
\frac{|{\rm det}\,{\bld C}_{\rm S}|^{\frac{1}{2}}}
{|{\rm det}\,{\bld C}_{\Lambda}|^{\frac{1}{2}}}
\exp\left[ -\frac{1}{2}\left( \chi^2_{\Lambda} - \chi^2_{\rm
S} \right)
\right]
\ee
between the $\Lambda$CDM and SCDM models. 
Here $\chi^2_{\rm S}$ and $\chi^2_{\Lambda}$ denote the $\chi^2$-values of
the 1.2-Jy data with respect to the SCDM and $\Lambda$CDM samples, and
${\bld C}_{\rm S}$ and ${\bld C}_{\Lambda}$ are the $6\times 6$
covariance matrices of the measured principal components.
We list a summary of our
results in table \ref{tabfix}.

\begin{table}
\bc
\caption{Comparison of SCDM and $\Lambda$CDM with the 1.2-Jy survey
assuming a multivariate normal distribution for the principal
components.
Listed are the $\chi^2$ values ($m=6$ degrees of freedom) for the
1.2-Jy survey when it is compared to either the SCDM or the
$\Lambda$CDM suite of mock catalogues, and the resulting exclusion
levels $p$.
We also compute the relative likelihood of $\Lambda$CDM compared to SCDM.
\label{tabfix}
}

\begin{tabular}{c|c|c|c|c|c}
\multicolumn{6}{c}{{\bf Fixed smoothing}}\vspace*{0.1cm}\\
 & \multicolumn{2}{c}{SCDM} & \multicolumn{2}{c}{$\Lambda$CDM} & \vspace*{-0.08cm} \\
$\lambda\;\;[\lu]$   & $\chi^2$ & $p$ &  $\chi^2$ & $p$ &
${\cal L} \left(\frac{\Lambda{\rm CDM}}{ {\rm SCDM}}\right) $  
\vspace*{0.2cm}
\\
           5 &    7.22 &    0.30 &    3.91 &    0.69 &   8.122\\
           7 &    7.06 &    0.32 &    5.43 &    0.49 &   4.316\\
          10 &    8.05 &    0.23 &    4.64 &    0.59 &  10.790\\
          14 &    5.46 &    0.49 &    5.15 &    0.52 &   2.429\\
          20 &    9.04 &    0.17 &    7.30 &    0.29 &   4.465\\

\end{tabular}
\vspace*{0.4cm}

\begin{tabular}{c|c|c|c|c|c}
\multicolumn{6}{c}{{\bf Adaptive smoothing (spherical)}}\vspace*{0.1cm}\\
 & \multicolumn{2}{c}{SCDM} & \multicolumn{2}{c}{$\Lambda$CDM} & \vspace*{-0.08cm} \\
$\lambda\;\;[\lu]$   & $\chi^2$ & $p$ &  $\chi^2$ & $p$ &
${\cal L} \left(\frac{\Lambda{\rm CDM}}{ {\rm SCDM}}\right) $  
\vspace*{0.2cm}
\\
           5 &    5.52 &    0.48 &    2.10 &    0.91 &   5.208\\
           7 &   12.97 &    0.04 &   10.72 &    0.10 &   2.860\\
          10 &    5.76 &    0.45 &    6.98 &    0.32 &   0.637\\
          14 &    4.98 &    0.55 &    6.82 &    0.34 &   0.382\\
          20 &    3.62 &    0.73 &    3.90 &    0.69 &   0.911\\

\end{tabular}
\vspace*{0.4cm}

\begin{tabular}{c|c|c|c|c|c}
\multicolumn{6}{c}{{\bf Adaptive smoothing (triaxial)}}\vspace*{0.1cm}\\
 & \multicolumn{2}{c}{SCDM} & \multicolumn{2}{c}{$\Lambda$CDM} & \vspace*{-0.08cm} \\
$\lambda\;\;[\lu]$   & $\chi^2$ & $p$ &  $\chi^2$ & $p$ &
${\cal L} \left(\frac{\Lambda{\rm CDM}}{ {\rm SCDM}}\right) $  
\vspace*{0.2cm}
\\
           5 &    4.87 &    0.56 &    2.25 &    0.90 &   3.478\\
           7 &   11.31 &    0.08 &    8.97 &    0.18 &   2.852\\
          10 &    7.26 &    0.30 &    8.14 &    0.23 &   0.633\\
          14 &    7.34 &    0.29 &    8.67 &    0.19 &   0.524\\
          20 &   11.88 &    0.06 &   10.40 &    0.11 &   2.545\\

\end{tabular}

\ec
\end{table}

\begin{figure}
\bc
\resizebox{8cm}{!}{\includegraphics{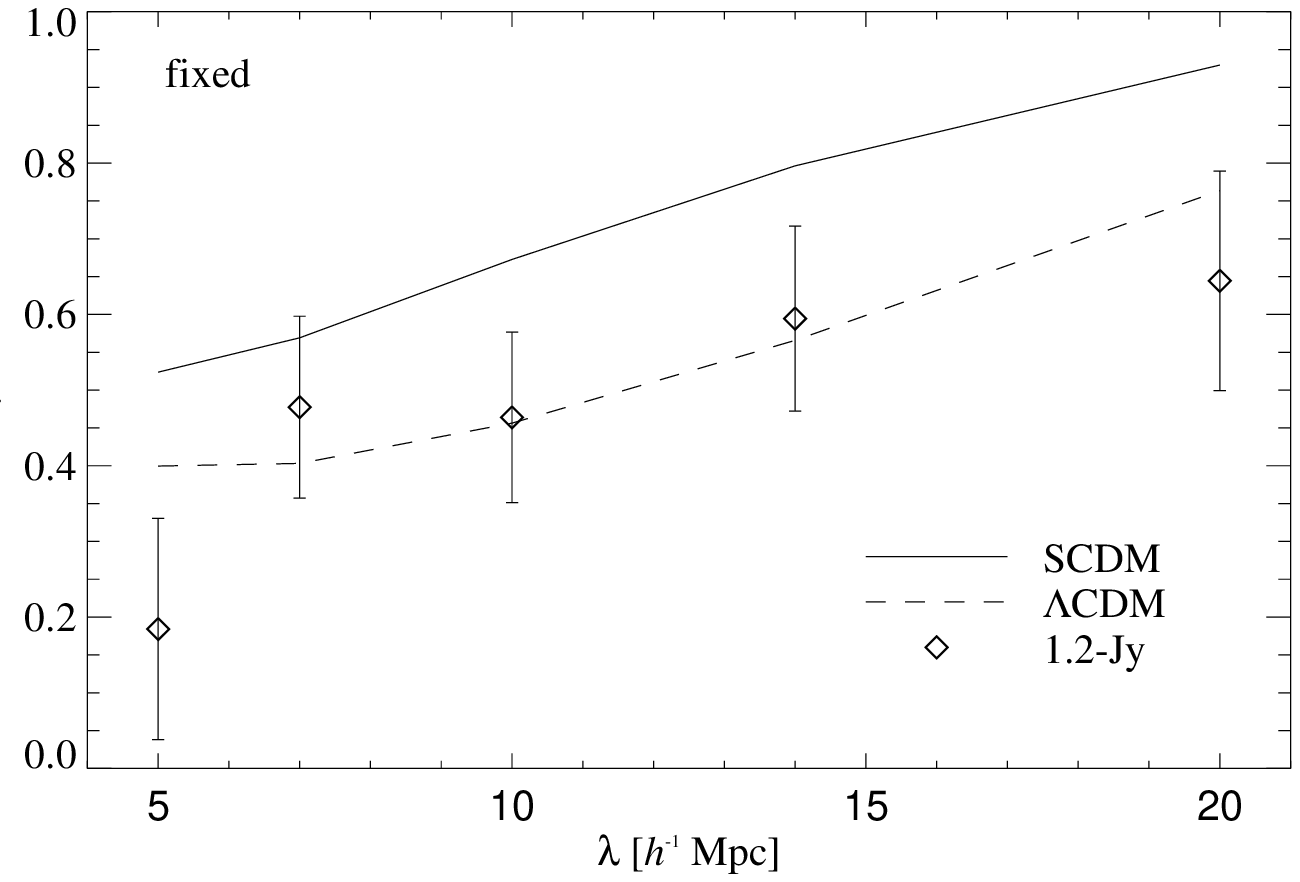}}
\resizebox{8cm}{!}{\includegraphics{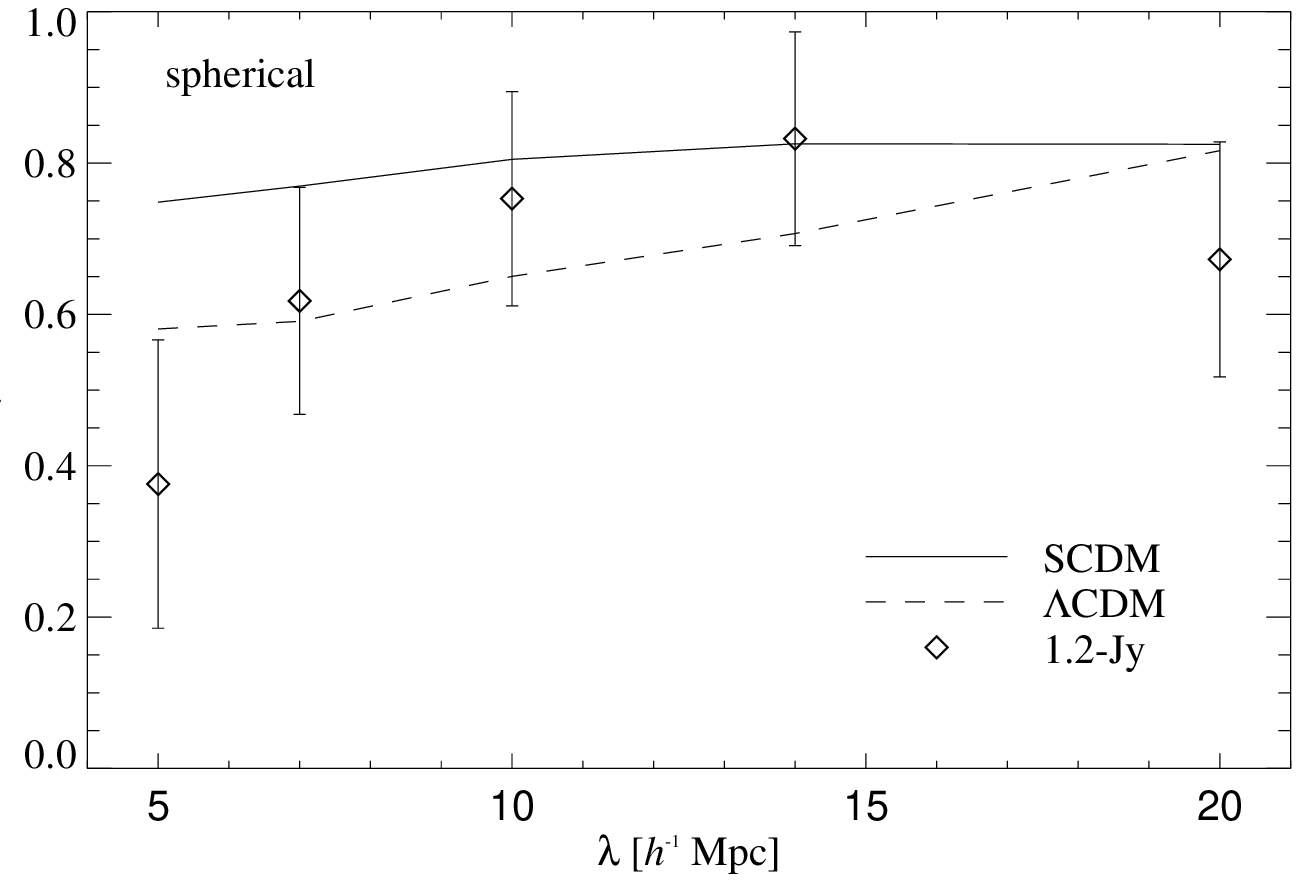}}
\resizebox{8cm}{!}{\includegraphics{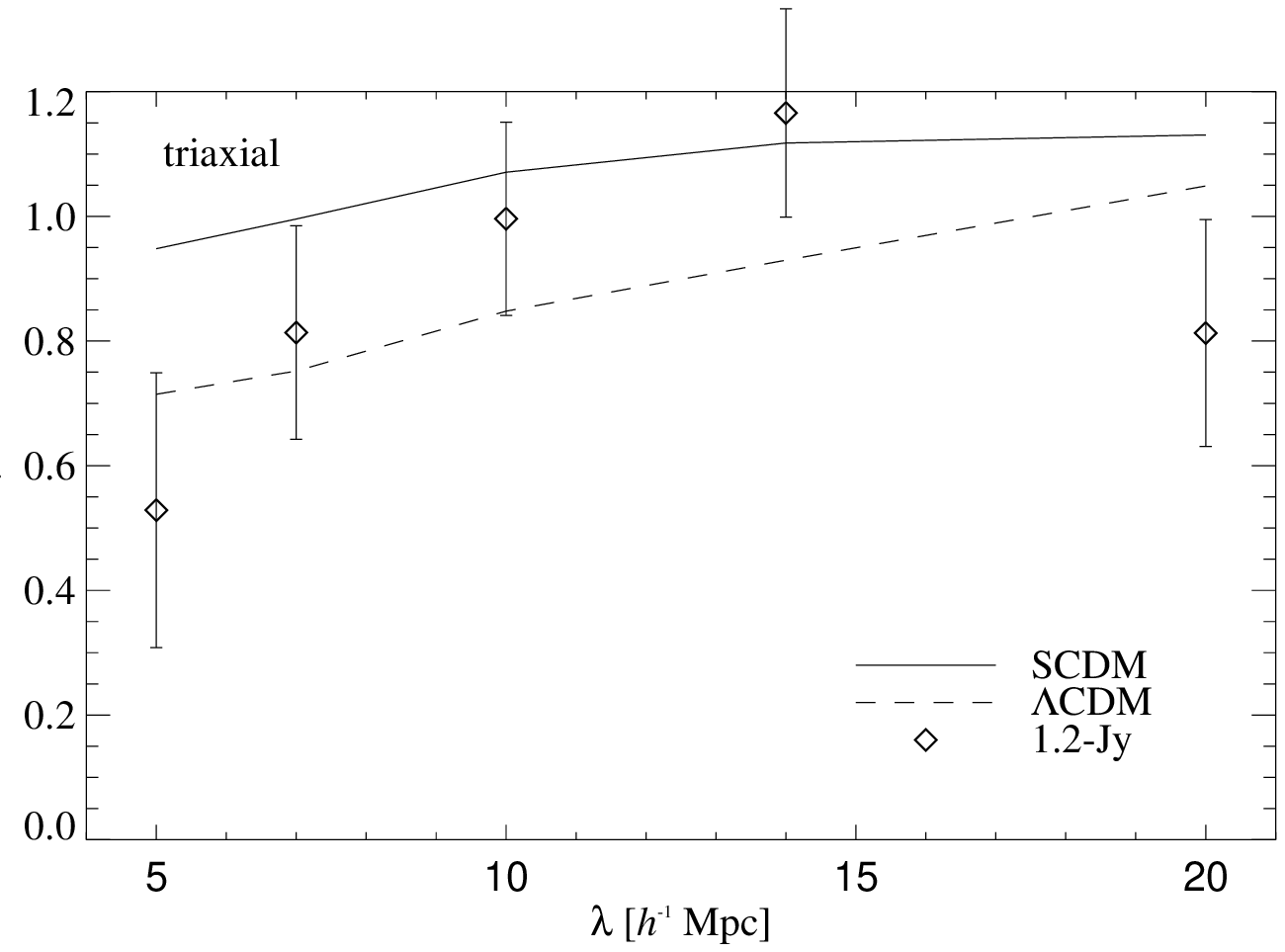}}
\caption{Measured genus amplitudes for the SCDM and $\Lambda$CDM
mock catalogues and the 1.2-Jy survey. 
Each panel shows results for 
one of the three different smoothing schemes employed. The mean of
the SCDM mock catalogues is displayed as solid line, the $\Lambda$CDM
result is the dashed curve, and the diamonds mark the 1.2-Jy
measurements. The attached error bars indicate the rms scatter of the
SCDM sample. 
Note that we have normalized 
the first principal component such that the
measured amplitude $N$ corresponds to that of a best-fitting Gaussian
curve (equation \protect\ref{GeRP}). 
\label{figxyz}
}
\ec
\end{figure}

All the measured probabilities $p$ are so high, that the 1.2-Jy genus
data are consistent with being drawn from either of the two CDM models,
i.e.\ the genus test cannot rule out the SCDM or $\Lambda$CDM
model with high significance when only one smoothing scale is considered.
However, we can still make a statement about their relative
likelihoods given the measured 1.2-Jy data.
For the fixed smoothing, the $\Lambda$CDM model is more likely than
the SCDM model for all smoothing scales. This preference of
$\Lambda$CDM is also found with the adaptive smoothing, however at a
weaker level since the results actually favour SCDM at $10\lu$ and $14\lu$.

\begin{table}
\bc
\caption{Comparison of the SCDM and $\Lambda$CDM models 
with the 1.2-Jy survey
using a combination of genus amplitudes for different smoothing scales.
Listed are the fractions $p_{\rm Q}$, $p_{\rm R}$ of mock catalogues that fit the
ensemble mean worse than the 1.2-Jy measurements. The $p_{\rm Q}$
statistic is just a sum over the $\chi^2$-deviations at individual
smoothing scales, and the  $p_{\rm R}$ statistic is based on a fit of the
run of genus amplitude with smoothing scale.
\label{tabcomp}
}
\begin{tabular}{c|c|c|c}
 & & SCDM & $\Lambda$CDM \vspace*{0.2cm} \\
 & $p_{\rm Q}$ & 0.012 & 0.522 \\
\raisebox{1.5ex}[0pt]{fixed} & $p_{\rm R}$ & 0.002 & 0.696\vspace*{0.14cm}\\
 & $p_{\rm Q}$ & 0.330 & 0.590 \\
\raisebox{1.5ex}[0pt]{spherical} & $p_{\rm R}$ & 0.242 & 0.992\vspace*{0.14cm}\\
 & $p_{\rm Q}$ & 0.174 & 0.384 \\
\raisebox{1.5ex}[0pt]{triaxial} & $p_{\rm R}$ & 0.132 & 0.944\vspace*{0.14cm}\\
\end{tabular}
\ec
\end{table}

To improve the discriminative power of the genus test we can  
try to combine the
measurements corresponding to different smoothing scales.
As we have demonstrated above, the genus amplitude (i.e. the first
principal component $h_1$) is most
sensitive to differences between the models, essentially because it
measures the shape of the power spectrum. Hence  we will focus
on it in the following.

In figure \ref{figxyz}
we show the average genus amplitudes of the SCDM and  $\Lambda$CDM
mock catalogues together with the measurements for the 1.2-Jy survey.
The models follow approximately parallel lines, albeit at different
heights, when we plot dimensionless genus densities $(2\pi)^2\lambda^3
N$.
To quantify the overall agreement between the 1.2-Jy survey 
and the models we
consider two simple statistics. First we compute a 
$\chi^2$-like quantity by adding up the amplitude measurements
corresponding to the different smoothing scales, viz.
\be
Q=\sum_i \frac{(A^{\rm m}_i-\ol{A}_i)^2}{\sigma_i^2},
\ee
where $A^{\rm m}_i$ is the measured 1.2-Jy amplitude at smoothing scale
$i$, and $\ol{A}_i$ and
$\sigma_i$ give the mean and dispersion of one of the two mock ensembles.
Because 
the survey volumes
corresponding to different smoothing scales are not 
independent,
the quantity $Q$ cannot be expected to follow a $\chi^2$
distribution. Instead, we calibrate its distribution with the mock
ensembles themselves. 
In particular, we compute the fraction $p_{\rm Q}$ 
of mock catalogues that give a
higher value for $Q$ than 1.2-Jy. This quantity can be interpreted as
an exclusion probability.

The above test makes no assumption about the run of genus amplitude with
smoothing scale. We can further strengthen the constraints 
by assuming that
the power spectrum of the 1.2-Jy survey 
is reasonably close to a CDM spectrum. Then the
measured genus amplitudes should lead to an approximately straight
line as well.
Using linear regression we 
therefore fit a straight line to the
measured amplitudes in the $(2\pi)^3\lambda^3 N$--$\lambda$ diagram and consider
the distribution of the amplitude of this fit at the intermediate 
smoothing scale
of $\lambda=12.5\lu$. 
Again we calibrate this statistic with the mock
catalogues. In particular, we compute for the measured 1.2-Jy value the
fraction $p_{\rm R}$ of mock catalogues that deviate more from the
mean than 1.2-Jy itself.

Table \ref{tabcomp} summarizes the results of these tests. As already 
suggested
by figure \ref{figxyz} the fixed smoothing technique clearly
favours the $\Lambda$CDM model, and rules out the SCDM model with high
significance. In fact, the combined genus test excludes the SCDM
model at a 99 per cent confidence level. 

The adaptive smoothing
techniques, however, disappoint the hopes for stronger
constraints.
While they also clearly favour the $\Lambda$CDM
model, they cannot exclude SCDM with reasonable significance. 
This is
mainly due to 
1.2-Jy measurements that are quite high at intermediate smoothing
scales and which fit the SCDM model better than $\Lambda$CDM in this regime.
This may well be a fluctuation
due to cosmic variance.
We also want to argue that this should not be viewed as a failure of
the adaptive smoothing techniques.
As we have shown, adaptive smoothing
does increase the signal-to-noise ratio 
of the measured genus curves. Hence it
is able to measure more properties of the examined density fields. 
The results we obtain just mean 
that these additional properties of the 1.2-Jy density field
agree with both models and are not able to discriminate more strongly
between them.

In summary the genus statistic of the 1.2-Jy survey is consistent
with the
$\Lambda$CDM model, while the SCDM cosmology 
is ruled out with high significance. 
This is in accordance with the expectation that a
model with power spectrum shape $\Gamma\approx 0.2$ should do
substantially better than SCDM, since the observed galaxy distribution
has been repeatedly shown to exhibit more large-scale power than SCDM.

\section{Conclusions}

In this work we have examined the genus statistics of N-body
simulations down to the smallest scales examined so far. With the
conventional smoothing technique of a spatially fixed kernel we
showed that the genus curves of CDM cosmologies 
retain their random phase shape far into
the non-linear regime. However, the genus amplitude is strongly
reduced by phase correlations in the density field on scales below
$10\lu$. At $2\lu$ the suppression reaches a factor of 4. 
While it is not obvious at the moment how this amplitude drop is
related to more traditional measures of higher order correlations, it
might be an interesting quantity to characterize non-linearity in
future investigations.
The genus in the fixed
smoothing regime fails to show strong differences 
between the $\tau$CDM, $\Lambda$CDM, and
OCDM models. This suggests that the genus is only sensitive to the
shape of the power spectrum in this regime.

We have shown that an adaptive smoothing
is required
to use 
the smallest resolved mass scales of the \virgo
simulations. 
Because of that we have computed
genus curves with a novel adaptive smoothing technique
for the fully sampled simulations. 
With the adaptive scheme 
we can clearly separate the four
models at the smallest scales we examined,
i.e. the `degeneracy'
between the three models 
$\tau$CDM, $\Lambda$CDM, and OCDM 
can be lifted. In addition, on these scales the genus statistics show
very strong departures from `quasi-Gaussian' behaviour.

We have also performed a large Monte-Carlo experiment in order to
establish the statistical properties of the genus statistic when it
is applied to a redshift survey like the \iras 1.2-Jy catalogue. For this
purpose we extracted a large number of 1.2-Jy mock catalogues from the
simulations.

We found that the genus statistic of the 1.2-Jy survey is well consistent
with the $\Lambda$CDM
simulation, while the SCDM model is ruled out 
at a 99 per cent confidence level.
We have not explicitly examined
$\tau$CDM and OCDM, since we expect them to show 
at most marginal
differences from $\Lambda$CDM at the resolution of the 1.2-Jy data.

In this work we also proposed two variants of adaptive smoothing
techniques for flux limited redshift surveys. 
We demonstrated that these techniques
can improve the signal-to-noise ratio of the measured genus curves. 
Hence they
are able to 
extract more topological information from
a given redshift survey.
Using the
1.2-Jy catalogue, however,
we could not achieve
a stronger discrimination between the SCDM and $\Lambda$CDM models.
Since adaptive smoothing is sensitive to additional properties of the
density field, we conclude that these properties of the 1.2-Jy survey
are
consistent with both models.
We remain convinced that adaptive smoothing will exhibit
a clear advantage, if redshift surveys are used that allow a 
reconstruction of the density field on
strongly clustered scales.
Since very large redshift surveys like the Sloan survey are underway,
this regime will be accessible in the next few years.

Due to its sensitivity to higher order correlations, the 
genus statistic remains a useful tool to test the random phase
hypothesis, and to compare theoretical models with observations.
In this work we confirm that the topology of the 1.2-Jy redshift survey is
consistent with current models of cold dark
matter universes that grow structure out of random phase initial conditions.
A particular advantage of the genus is that 
these results should be largely independent of a possible bias
between the galaxy density and the mass density,
at least if this bias relation is monotonic.

\bibliography{paper}

\appendix

\section{Distribution of errors}

\begin{figure}
\bc
\resizebox{8cm}{!}{\includegraphics{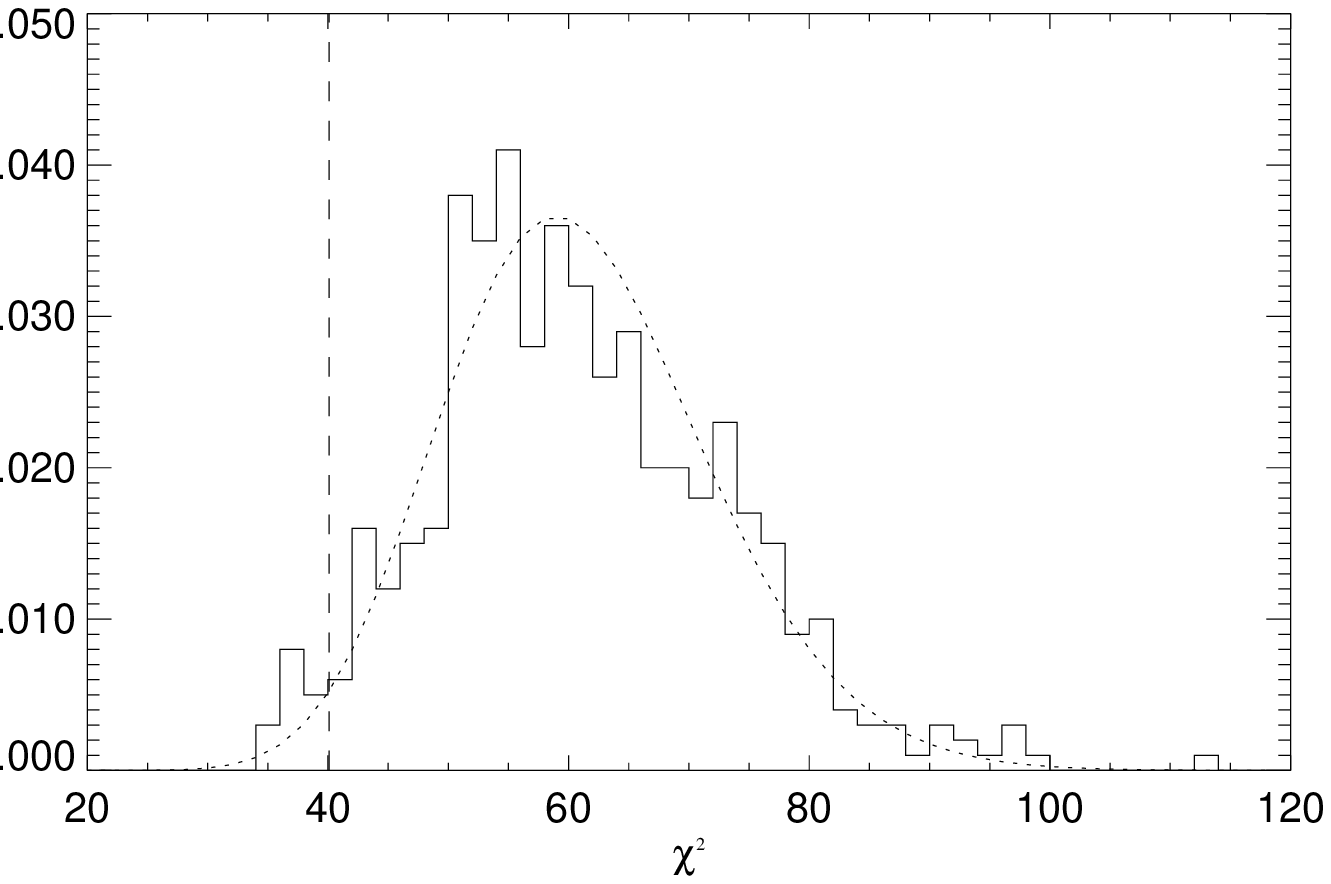}}
\resizebox{8cm}{!}{\includegraphics{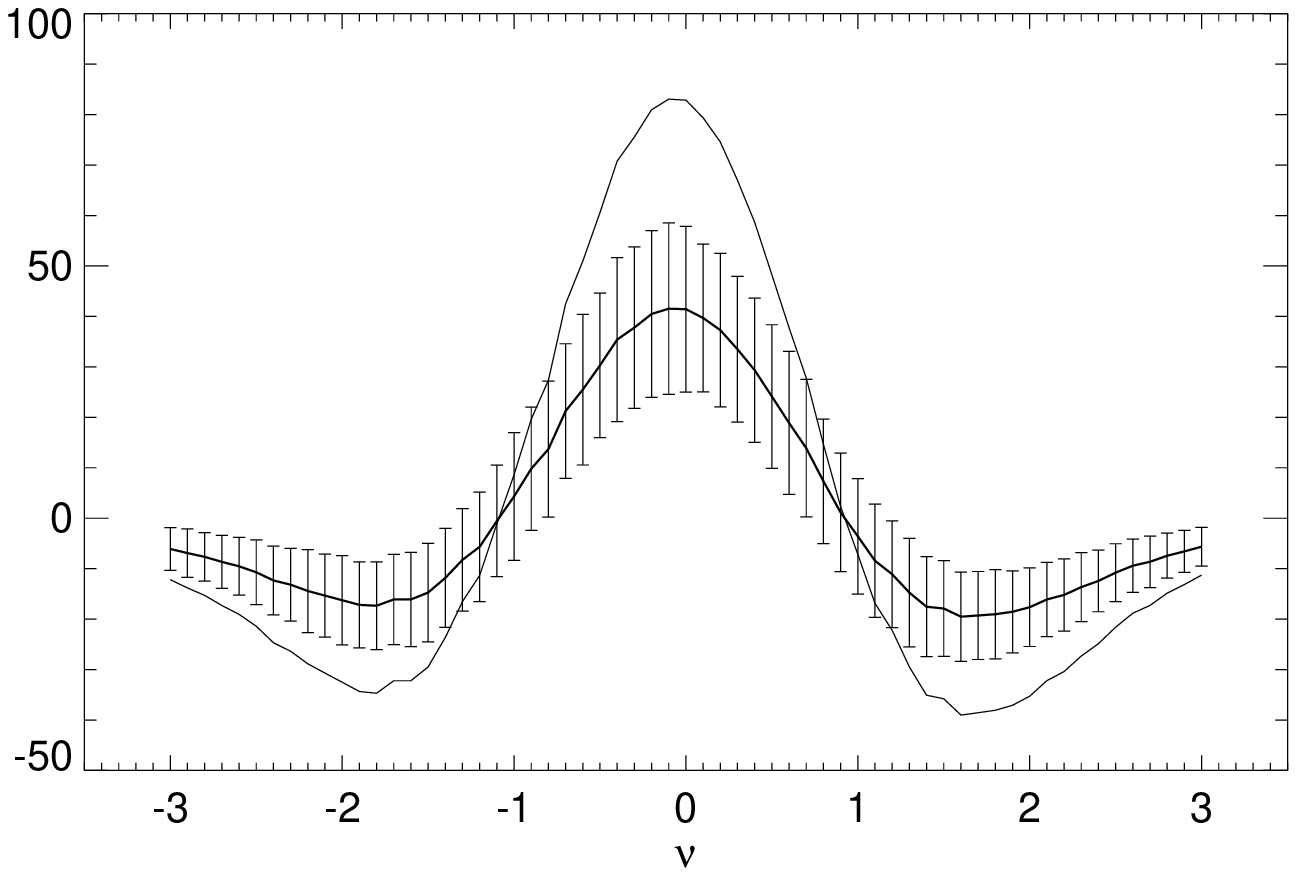}}
\caption{The top panel shows the 
distribution of $\chi^2$ defined by equation (\ref{AAA001})
for $k=61$ measured points of the genus curve (SCDM mock catalogues,
fixed smoothing of $7\lu$).
The dotted curve is the theoretical $\chi^2$ distribution for 61
degrees of freedom.
The dashed line marks the $\chi^2$-value
obtained if one checks the fit of a genus curve that is everywhere
twice the mean, i.e. $\vec{g}=2\ol{\vec{g}}$. The resulting low value of
$\chi^2$ suggests a perfect fit; yet this curve is
highly inconsistent with the mock ensemble, as we demonstrate 
in the bottom panel. Here the thick line gives $\ol{\vec{g}}$, while the
thin line shows $\vec{g}=2\ol{\vec{g}}$. The error bars are the rms
deviations of the genus at individual points $\nu_i$ of the curve.
\label{figchi2dist}
}
\ec
\end{figure}

\begin{figure}
\bc
\resizebox{8cm}{!}{\includegraphics{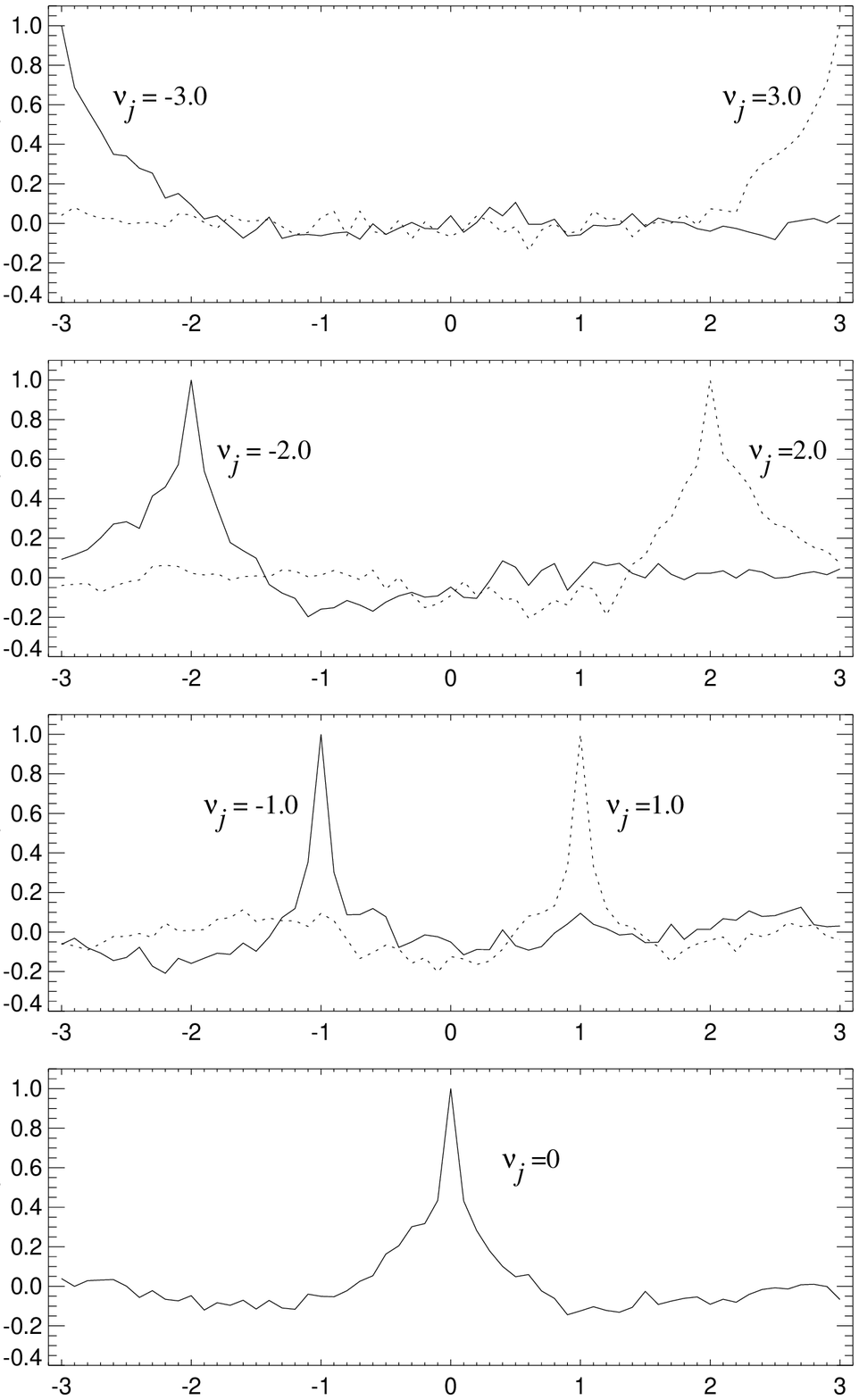}}
\caption{Pearson's correlation coefficient for
the genus measurement.
The labelled curves show the correlation coefficient for a number of different
points on the genus curve.
Adjacent points are quite strongly correlated over a
range $\Delta\nu\approx 1$.
The result shown here is for the SCDM catalogues with 
fixed smoothing of $5\lu$; it is very similar for all other
smoothing scales we consider in this work.
\label{figPearson}
}
\ec
\end{figure}

In section \ref{sec71} we defined a multivariate analysis of the
measured genus curves 
using the full $61\times 61$ covariance matrix of equation (\ref{chi2matrix}).
Here we demonstrate that this method fails in practice, and we
examine reasons for this failure.

In figure \ref{figchi2dist} 
we show the distribution of $\chi^2$ as defined in equation (\ref{AAA001})
for the SCDM mock catalogues, smoothed with $\lambda=7\lu$. 
The dotted line shows a $\chi^2$ distribution for $k=61$
degrees of freedom, which is apparently able to fit
the observed distribution reasonably well. 
However, this does not necessarily mean
that the distribution of errors is in fact well approximated by a
multivariate Gaussian. A first hint that a problem is lurking
here may be obtained by computing the $\chi^2$ value for a genus curve
with $\vec{g}=2\ol{\vec{g}}$, i.e. one that deviates by 100 per cent from
the mean. For this curve the $\chi^2$ comes out as 40.1, suggesting a perfect
fit, although the curve $\vec{g}$ is actually discrepant at individual
points on the genus curve with high significance level, as is shown in
the bottom panel of figure \ref{figchi2dist}.

Clearly, this peculiar result needs to be understood. In figure 
\ref{figPearson} we show Pearson's correlation coefficient for a
number of places on the genus curve. In general, adjacent points on
the genus curve are correlated over roughly a 
range $\Delta\nu\approx 1$.
The main
effect of these correlations 
is to
introduce negative values 
in the inverse 
$V_{ij}^{-1}$ of the 
covariance matrix 
in the elements just off the diagonal.
These terms reduce the weight of deviations from the
mean if adjacent points exhibit deviations of the same sign.

As a result, coherent deviations (as in $\vec{g}=2\ol{\vec{g}}$) 
are hardly penalized
at all. This partly explains the low $\chi^2$ we obtained for the test curve
$\vec{g}=2\ol{\vec{g}}$; 
yet it does not fully account for why the test fails so badly.
We think that there are two reasons for this. 
First, 
the statistical properties of the genus
errors cannot be fully described by the covariance matrix alone,
because the distribution of errors is not consistent with a
multivariate Gaussian. This will be shown below. Second, the
noise in the covariance matrix compromises its inversion; although 
the inversion is mathematically possible and stable, 
the result is not
necessarily meaningful, because it is strongly affected by the noise.
The principal components analysis developed
in section \ref{sec72} provides a solution to this problem.


\begin{figure}
\bc
\resizebox{8cm}{!}{\includegraphics{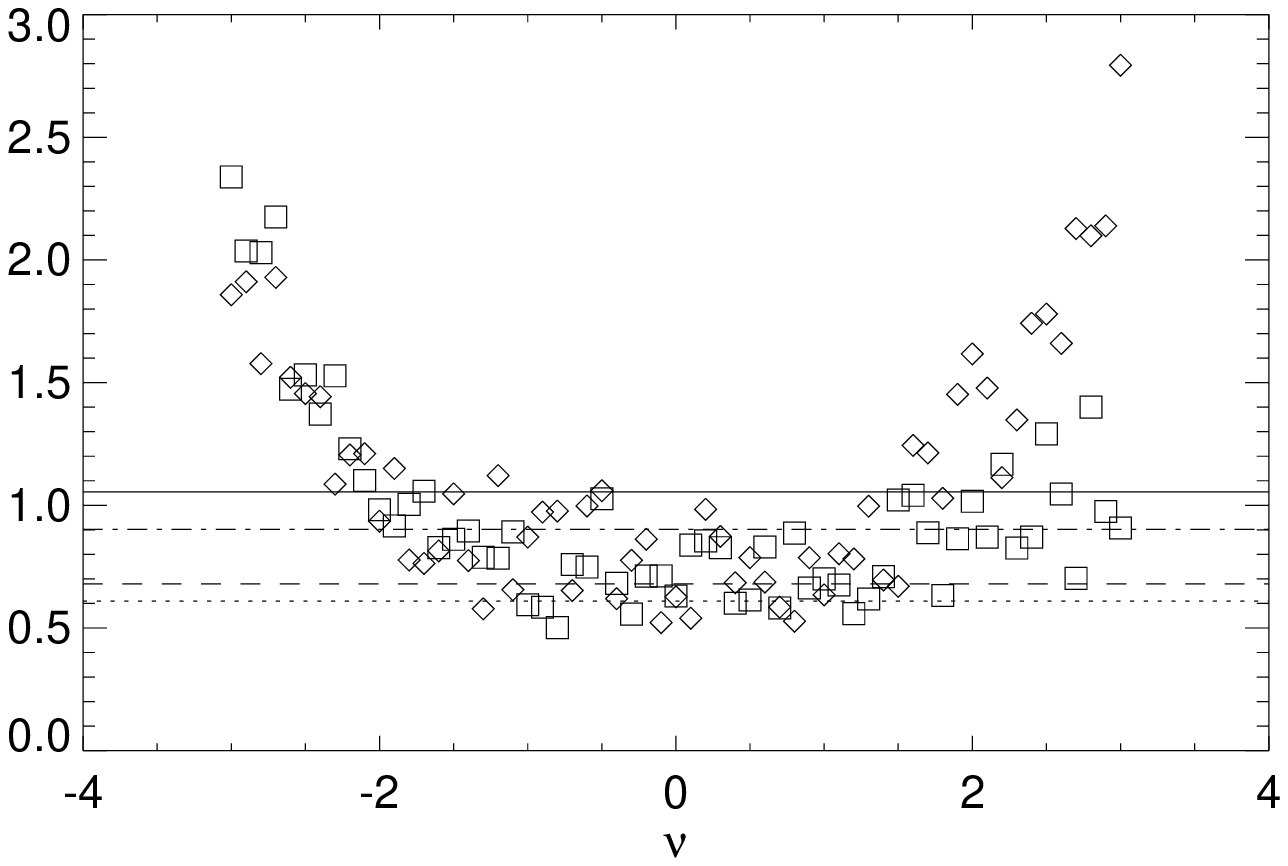}}
\caption{Kolmogorov-Smirnov test of normality of the genus
measurements.
The symbols give the test measure $\sqrt{n}D_n$ for 61 points on the
genus curve of the SCDM mock sample. Diamonds refer to fixed
smoothing with $5\lu$, boxes are for adaptive smoothing.
The horizontal lines delimit probability regions for various exclusion
levels. If the distribution were normal, a point should be found with
probability 50\% below the lowest line, with 68\% below the dashed
line, with 95\% below the dot-dashed line, and with 99\% below the
solid line. We obtain a qualitatively similar result for other
smoothing scales and mock ensembles.
\label{figKStest}
}
\ec
\end{figure}

We now demonstrate 
that the distribution of the genus measurements is in
fact not a multivariate Gaussian. It is sufficient to show that even the
distribution of just the genus at a single value of $\nu_i$ is not normal.
For this purpose we employ a Kolmogorov-Smirnov (KS) test.
Based on the $n=500$ genus measurements for a mock ensemble 
we can 
estimate the mean 
and the variance at individual points on the genus curve
from the sample itself.
We can then evaluate the KS test 
\be
D_n={\rm sup}\left| S_n(g)-F(g) \right|,
\ee
where
$S_n(g)$ denotes the cumulative distribution function of the measurements
and $F(g)$ is
the cumulative probability distribution function 
of the presumed Gaussian.

In general the KS test is fully distribution-free only if the test 
distribution $F$ is known beforehand. However, here we estimate the
parameters of the normal distribution from the sample itself. Because
mean and variance are only scale factors of the normal distribution,
the KS test remains applicable, although the distribution of $D_n$ is
changed \cite{Ke70}. 
We calibrate the latter with a Monte-Carlo experiment.

In figure \ref{figKStest}
we show measurements of $\sqrt{n}D_n$ for the SCDM mock ensemble, 
smoothed at $5\lu$ with the spherical and, alternatively, the 
triaxially adaptive method.
If the distribution of the $g_i$
were Gaussian, in half of the cases we should measure
$\sqrt{n}D_n>0.609$, and in only 1 per cent 
of the cases $\sqrt{n}D_n>1.055$.
Hence the KS test shows that the genus distribution is 
inconsistent with a Gaussian at low and high values of $\nu$.
Close to $\nu=0$ the points are consistent with a normal
distribution, although there seems to be a 
lack of points with low values of
$\sqrt{n}D_n$.

We think that it is not too surprising that the distribution of errors is
not well described by 
a multivariate Gaussian. For example, sampling fluctuations
can affect the whole volume-fraction/density-threshold relation and
thus result in coherent shifts of parts of the genus curve along the
x-axis. Similarly, the small survey volumes examined here can lead to
large irregular fluctuations in the genus due to particular
density features of the patch under examination. As a result, we
expect that the genus curve exhibits complicated higher order
correlations which make it rather difficult to take full advantage of
its information content.

\end{document}